\documentclass[useAMS,usenatbib]{mn2e}
\usepackage{epsfig}
\usepackage{subfigure}
\usepackage{amssymb,amscd}

\title[Angular momentum in BEC-CDM haloes]
  {Angular Momentum and Vortex Formation in Bose-Einstein-Condensed Cold Dark Matter Haloes}
\author[T. Rindler-Daller, P.R. Shapiro]
  {Tanja Rindler-Daller$^{1,2}$\thanks{daller@astro.as.utexas.edu} and Paul R.~Shapiro$^{1}$\thanks{shapiro@astro.as.utexas.edu} \\
  $^{1}$ Department of Astronomy and Texas Cosmology Center,
  The University of Texas at Austin, Austin, TX 78712, USA\\
  $^{2}$ Institut f\"ur Theoretische Physik, Universit\"at zu K\"oln, 50937 Cologne, Germany}
\date{December 23, 2011}

\pagerange{\pageref{firstpage}--\pageref{lastpage}} \pubyear{}

\def\LaTeX{L\kern-.36em\raise.3ex\hbox{a}\kern-.15em
    T\kern-.1667em\lower.7ex\hbox{E}\kern-.125emX}

\newcommand{\gpf}{\mathcal{E}}
\newcommand{\mb}{\mathbf}
\newcommand{\p}{\partial}
\newcommand{\tb}{\textbf}
\newcommand{\mc}{\mathcal}
\newcommand{\tx}{\textit}
\newcommand{\beq}{\begin{equation}}
\newcommand{\eeq}{\end{equation}}
\newcommand{\bdi}{\begin{displaymath}}
\newcommand{\edi}{\end{displaymath}}
\newcommand{\glf}{\mathcal{G}_f}
\newcommand{\rotf}{\mathcal{R}_f}
\newcommand{\f}{\frac}

\begin{document}

\label{firstpage}

\maketitle

\begin{abstract}

Various extensions of the standard model of particle physics predict
the existence of very light bosons, with masses ranging from about
$10^{-5}$ eV for the QCD axion down to $10^{-33}$ eV for ultra-light
particles. These particles could be responsible for all or part of
the cold dark matter (CDM) in the Universe. For such particles to
serve as CDM, their phase-space density must be high enough to form
a Bose-Einstein condensate (BEC). The fluid-like nature of BEC-CDM
dynamics differs from that of standard collisionless CDM, however,
so different signature effects on galactic haloes may allow
observations to distinguish them. Standard CDM has problems with
galaxy observations on small scales; cuspy central density profiles
of haloes and the overabundance of subhaloes seem to conflict with
observations of dwarf galaxies. It has been suggested that BEC-CDM
can overcome these shortcomings for a large range of particle mass
$m$ and self-interaction coupling strength $g$. For
quantum-coherence to influence structure on the scale of galactic
haloes of radius $R$ and mass $M$, either the de-Broglie wavelength
$\lambda_{deB} \lesssim R$, which requires $m \gtrsim m_H \cong
10^{-25}(R/100~\rm{kpc})^{-1/2}(M/10^{12}~ M_{\odot})^{-1/2}$ eV, or
else $\lambda_{deB} \ll R$ but gravity is balanced by
self-interaction, which requires $m \gg m_H$ \tx{and} $g \gg g_H
\cong 2 \cdot 10^{-64} (R/100~
   \rm{kpc})(M/10^{12}~M_{\odot})^{-1}$ eV
   cm$^3$.
 Here we study the largely-neglected effects of angular momentum on BEC
haloes. Dimensionless spin parameters $\lambda \simeq 0.05$ are
expected from tidal-torquing by large-scale structure formation,
just as for standard CDM. Since laboratory BECs develop quantum
vortices if rotated rapidly enough, we ask whether this amount of
angular momentum is sufficient to form vortices in BEC haloes, which
would affect their structure with potentially observable
consequences. The minimum angular momentum required for a halo to
sustain a vortex, $L_{QM}$, corresponds to $\hbar$ per particle, or
$\hbar M/m$. For $\lambda = 0.05$, this requires $m \geq 9.5 m_H$,
close enough to the particle mass required to influence structure on
galactic scales that BEC haloes may be subject to vortex formation.
While this is a necessary condition, it is not sufficient. To
determine if and when quantum vortices will form in BEC halos with a
given $\lambda$-value, we study the equilibrium of self-gravitating,
rotating, virialized BEC haloes which satisfy the
Gross-Pitaevskii-Poisson equations, and calculate under what
conditions vortices are energetically favoured, in two limits:
either just enough angular momentum for one vortex or a significant
excess of angular momentum. For $\lambda = 0.05$, vortex formation
is energetically favoured for $L/L_{QM} \geq 1$ as long as \tx{both}
$m/m_H \geq 9.5$ \tx{and} $g/g_H \geq 68.0$. Hence, vortices are
expected for a wide range of BEC parameters. However, vortices
cannot form for vanishing self-interaction (i.e. when $\lambda_{deB}
\lesssim R$), and a range of particle parameters also remains even
for BEC haloes supported by self-interaction, for which vortices
will \tx{not} form. Such BEC haloes can be modelled by compressible,
($n=1$)-polytropic, irrotational Riemann-S ellipsoids.

\end{abstract}


\section{Introduction}

In the last decades, astronomical observations have provided a range
of supporting evidence that about 23 \% of the energy density of the
Universe is comprised of non-baryonic dark matter. Its particle
nature is still unknown, but all candidates find their justification
in extensions of the standard model of particle physics. Based on
observations, it is now believed that dark matter is
weakly-interacting and non-relativistic, hence behaving like a cold
gas, and generally termed cold dark matter (CDM).

The standard form of CDM is often supposed to be a relic of the Big
Bang in the form of weakly-interacting, massive particles (WIMPs),
in particular, the lightest supersymmetric particles, the most
popular of which is the neutralino, with particle mass of the order
of 100 GeV. Efforts are underway to measure the presence of those
particles, but no direct detection has yet been reported. On the
other hand, as far as the dynamics of galaxies and large-scale
structure is concerned, standard CDM is modelled as a cold,
collisionless gas with vanishing pressure. While many observational
properties of galaxies can be reproduced, some crucial issues are
subject to controversy. Simulations of standard CDM structure
formation predict a universal halo density profile, which has a cusp
going like $r^{-1}$ in the center. This seems to be in conflict with
the observational properties of low-surface brightness (LSB) and
dwarf galaxies. Moreover, standard CDM predicts the hierarchical
merger of smaller structures into larger structures over time, with
an overly abundant population of subhaloes inside a halo like the
Local Group for which there does not seem to be an observational
counterpart. Given the above shortcomings, and the null results of
direct detection experiments to date, we are still free to consider
other candidates for the CDM paradigm.

Another major candidate for dark matter is the QCD axion, the
pseudo-Nambu-Goldstone boson of the Peccei-Quinn phase transition,
proposed as a dynamical solution to the CP-problem of the strong
interactions. For the axion to be CDM, it has to be very light, $m
\sim 10^{-5}$ eV, and direct detection experiments rely on its
electrodynamical coupling, employing the haloscope idea of
\cite{sikivie} in the ADMX experiment (\cite{asztalos}), and
axion-photon oscillations in the presence of high magnetic fields
using lasers (so-called light-shining-through-wall experiments),
e.g. \cite{chou}.

In addition to the QCD axion, multidimensional cosmological and
string theories generically predict the existence of even much
lighter bosonic particles down to masses of the order of $10^{-33}$
eV (see e.g. \cite{HW, GZ}), which could form all or part of the
dark matter\footnote{While those particles all the way up to the QCD
axion are often summarized as 'axion-like' particles, we will avoid
this term here, since the definition of what is 'axion-like' may
sometimes encompass particles in the keV mass range. Instead, we
will refer to them as ultra-light bosons.}, e.g. \cite{carroll, ADD,
axiverse}. For such particles to serve as CDM, their phase-space
density must be high enough to form a Bose-Einstein condensate
(BEC), i.e. a macroscopic occupancy of the many-body ground state.
Generally, for a system to undergo Bose-Einstein condensation, the
phase-space density $n \lambda_{deB}^3$ must exceed a number of
order one, where $n$ is the number density and $\lambda_{deB}$ is
the de-Broglie wavelength of the particles. While for axions
$n\lambda_{deB}^3 \gg 1$, ultra-light bosons even fulfill
$n\lambda_{deB}^3 \ggg 1$. It has been shown in models by \cite{SM},
\cite{urena2} and \cite{lundgren} that ultra-light particles are
able to undergo a phase transition in the early universe to a BEC at
a temperature which is well above that of the time of recombination.
These bosons are thereby able to form a non-relativistic,
quantum-degenerate gas, in contrast to the (fermionic) neutrinos
which remain relativistic. While numerous searches for the QCD axion
are currently pursued, the ultra-light bosons seem to be out of
reach for direct detection. However, low-temperature ultra-precision
experiments promise new possibilities to search directly even for
these ultra-light particles, for a review see e.g. \cite{frontier}.
An interesting, but very challenging experimental proposal for
detecting such particles, by exploiting their supposed coupling to
gluons, has been presented recently by \cite{GR}. Nevertheless, this
kind of dark matter will best be traced in the near future through
the signature of its dynamical differences from standard CDM, an
aspect of which we will study in this paper. A full account of all
the existing literature is beyond the scope of this paper, but
earlier investigations of models for light bosonic dark matter
include e.g. the works of \cite{IS, baldeschi, khlopov, MPS, press,
widrow, sin, schunck, lee, MFS} and \cite{MU} to name a few. Most of
this literature has been restricted to free particles without
self-interaction.

We will see that BEC cold dark matter (henceforth also BEC-CDM)
obeys quantum-mechanical \tx{fluid equations}. Therefore,
small-scale structure can be very different for BEC dark matter from
that of collisionless standard CDM. The behaviour of
\tx{self-interacting}\footnote{This is different from the kind of
self-interacting, cold dark matter particles referred to elsewhere
in the literature as SIDM, suggested by \cite{spst}, which we have
studied in \cite{AS} and \cite{KS}. In SIDM, the particle
self-interaction results in 2-body elastic scattering which adds
'collisionality' to the otherwise collisionless CDM gas, but does
not make a BEC or exhibit any form of macroscopic quantum coherence.
The reader should henceforth avoid any confusion between the BEC-CDM
discussed here and SIDM.} BEC-CDM as a superfluid makes possible
entirely new phenomena like the formation of \tx{quantum vortices}.
Astronomical observations may thus provide a means to diagnose
different dynamical effects, thereby allowing us to constrain (or
rule out) this form of dark matter. In fact, BEC-CDM has previously
been invoked to overcome the above mentioned shortcomings of
standard CDM: One prime motivation for considering BECs for CDM has
been their ability to produce galactic haloes with constant density
cores, see e.g. \cite{goodman} ("repulsive DM") and \cite{peebles}
("fluid DM"), since the corresponding profiles may then better agree
with observed rotation curves of dwarf and LSB galaxies as has been
shown in the papers of \cite{arbey, BH} and B\"ohmer, Martins,
Salucci (private communication). The problem of overabundance of
subhaloes in standard CDM is naturally resolved in some BEC dark
matter models, since the uncertainty principle prevents the
formation of gravitationally bound isolated structures below a
certain length scale, which depends on the dark matter particle
mass, see e.g. \cite{hu} ("fuzzy DM"). However, the formation of
large-scale structures involving BEC dark matter has not yet been
studied in the same depth as for standard CDM and many issues are
still unresolved. Some recent results on structure formation studies
involving BEC-CDM can be found in the works of \cite{urena,
fukuyama, woo, MF, harko} and
\cite{chavanis}\footnote{\cite{slepian} recently suggested that
astronomical constraints can be used to rule out strongly-repulsive
bosonic dark matter. However, their analysis does not apply to the
case considered here. They assume that the bosons are in
thermodynamic equilibrium in isothermal haloes at the halo virial
temperature, with a condensate core surrounded by a non-condensate
envelope. As they themselves point out, their assumption breaks down
when 2-body collisions are not frequent enough, which is the case
here (see also \tx{Section 5}).}.

We will see in the next section that for quantum-coherence to
influence structure on the scale of galactic haloes of radius $R$,
either the de-Broglie wavelength of the dark matter particle is of
that same order,
 \bdi
 \lambda_{deB} \lesssim R,
  \edi
  requiring the particle mass to be of the order
   \bdi
    m \gtrsim m_H \cong 10^{-25}(R/100 ~\rm{kpc})^{-1/2}(M/10^{12}~ M_{\odot})^{-1/2} ~\rm{eV},
     \edi
or else
 \bdi
  \lambda_{deB} \ll R
  \edi
  and the halo is supported against gravity by
self-interaction pressure, which requires \tx{both}
 \bdi
 m \gg m_H
  \edi
\tx{and}
 \bdi
  g \gg g_H \cong 2 \cdot 10^{-64} (R/100~\rm{kpc})(M/10^{12}~M_{\odot})^{-1}~
  \rm{eV}
  \rm{cm}^3,
   \edi
 and we choose units where $c=1$.

However, previous literature on BEC-CDM has mostly neglected an
important aspect of halo physics, namely angular momentum. In the
early phases of halo collapse, tidal torques caused by large-scale
structure give a halo most of its angular momentum. Cosmological
N-body simulations of the standard CDM universe show that haloes
form with a net angular momentum such that the dimensionless ratio
 \beq \label{lam}
  \lambda = \f{L |E|^{1/2}}{GM^{5/2}},
   \eeq
which expresses their degree of rotational support, has typical
values in the range $[0.01,0.1]$ with median value $\simeq 0.05$.
These values can be found in \cite{BE}, and more recently in
\cite{antonucci}, which seem to be confirmed by observations, see
\cite{hernandez}. In the above expression, $L$ is the total angular
momentum, $E$ is the total energy, and $M$ is the total mass of the
halo. The quantity $\lambda^2$ corresponds roughly to the so-called
$t$-parameter, $t \equiv T/|W|$, where $T$ is the rotational kinetic
energy and $W$ is the gravitational potential energy. Tidal-torque
theory can successfully account for the $\lambda$-values found in
$N$-body simulations of structure formation, see e.g.
\cite{porciani,porciani2}. We shall here be interested in the case
where the BEC nature of CDM affects small-scale structure and the
internal dynamics of galactic haloes, while large-scale structure
formation follows that of standard CDM to a great extent. In this
paper, therefore we will adopt the above range of $\lambda$-values
for BEC-CDM haloes, too. The general problem of \tx{acquisition} of
angular momentum by BEC haloes is worth further consideration in a
future work, but in the context of this work, we will content
ourselves with the convenience of the above prescription.

Once haloes rotate, additional effects come into play. Halo shapes
and profiles will differ from those which result when angular
momentum is zero, and, since self-interacting BEC-CDM behaves as an
irrotational superfluid, these will differ from the case of
collisionless CDM. In addition, it is known that laboratory BECs can
develop quantum vortices when rotated with a sufficient angular
velocity (see e.g. \cite{mad2}), thereby changing the density
profile in a characteristic way. The question arises as to whether
this may be also possible for BEC haloes and we will attempt to
answer that question here.

Several authors have previously pursued related questions. \cite{SM}
postulated vortices in galactic haloes comprised of ultra-light
bosonic particles by comparing the critical angular velocity for
vortex formation with the rotation rate of M31. However, the formula
quoted there without derivation applies actually to laboratory
superfluids with strong self-interaction but without self-gravity.
Subsequently, \cite{YM} have studied the influence of a vortex
lattice on the velocity profile of a spherical galactic halo
composed of ultra-light bosons.

On the other hand, \cite{DS} have argued that certain fine-structure
in the observed inner mass distribution of the Milky Way can be
explained only if the infalling dark matter particles had a net
overall rotation, causing a 'tricusp' caustic ring in the
catastrophe structure of dark matter in that case. For standard,
non-interacting CDM models, however, one expects infall to be
\textit{irrotational}. According to \cite{SY}, axionic dark matter,
as a BEC, may be able to form a vortex lattice with high enough
vorticity so as to mimic a net rotational component, producing the
above structure. As such, the authors suggest, Milky-Way
observations may already have detected the signature of axionic dark
matter. We will come back to this claim in \tx{Section 5}.

This paper will extend the analysis which we presented in \cite{RS}.
We showed there that vortices are favoured to form for a wide range
of possible particle mass and self-interaction strength, by
calculating for the first time the critical angular velocity for
vortex creation for a simple model of BEC-CDM galactic haloes and
comparing the result with the angular velocity expected from
cosmological N-body simulations of standard CDM haloes. \cite{kain}
subsequently studied the formation of a vortex in spherical
(non-rotating) haloes in the Thomas-Fermi regime of strong
self-interaction. The major drawback of their approach is the fact
that it starts from haloes which have no angular momentum. The
physical mechanism by which vortices shall form is thus absent from
the beginning. In fact, the amount of angular momentum of a
singly-quantized, axisymmetric vortex in the center of a halo is
given by
 \beq \label{lqm}
  L_{QM} \equiv N\hbar,
  \eeq
and this angular momentum has to be provided by halo spinning. We
will report here our study of the equilibrium of rotating,
self-gravitating, virialized haloes, and the conditions for which
vortex formation is favoured. For the latter, we will focus our
attention on two extreme cases, one in which haloes have just enough
angular momentum to support a single vortex, i.e. their angular
momentum is $L = L_{QM}$, and the other case in which haloes have
much excess angular momentum, such that $L \gg L_{QM}$. Haloes will
be modelled in each case as rotating, ellipsoidal bodies. While the
main conclusion of \cite{RS} will not change, this paper will
significantly improve on the analytic modeling of BEC haloes
presented there and thereby establish better constraints on the BEC
dark matter particle mass and self-interaction required for vortex
formation to happen. We will employ an energy analysis of the full
equations of motion, thereby deriving analytically the conditions
under which vortex formation lowers the energy and is, hence,
favoured in haloes with the amount of angular momentum expected from
large-scale structure in the CDM universe.

The impact of vortices may be profound: We can expect vortices to
reside preferentially in the centers of dark matter haloes. Vortices
deplete the dark matter density in their core region once they have
formed. This in turn influences the dark matter density profile and
also the gravitational coupling to the baryons. If less dark matter
is around than without vortices, for example, then baryon cooling
and condensation, a prerequisite for star formation, may as a
consequence be reduced and delayed. It is thus interesting to ask
for which BEC-CDM models such vortices can be expected to form.
However, we stress that this work will be a 'dark matter only'-
analysis, and the baryonic component of galactic haloes will not be
considered.

This paper is organized as follows: In \tx{Section 2}, we review the
basic equations, the self-interaction regimes of BEC dark matter and
their consequences for halo density profiles. The underlying
equations we are going to use are the Gross-Pitaevskii equation for
the wave function of the Bose-Einstein-condensed halo, and the
Poisson equation by which the density is coupled to the halo
gravitational potential. The equation of motion can be restated in
terms of quantum-mechanical fluid equations. So, in \tx{Section 3}
we will study approximate figures of equilibrium models for BEC
haloes with angular momentum included: rotating haloes will be
approximated either by irrotational Riemann-S ellipsoids or by
Maclaurin spheroids. In \tx{Section 4}, we will present the energy
analysis to determine when vortex formation will occur in such BEC
haloes, whose net amount of angular momentum we will fix by values
for the $\lambda$-spin parameter, which are representative of those
of standard CDM. In the regime of strong self-interaction, we will
thereby establish bounds on the BEC dark matter particle mass and
self-interaction strength above which vortices will be favoured. The
implications of our results for halo models comprised of BEC-CDM
will be presented in \tx{Section 5}, which contains our conclusions
and discussion. Some further detailed derivations, tables and
frequently used relationships are deferred to three appendices.

\section{Fundamentals of BEC dark matter haloes}

\subsection{Basic equations}

We will describe self-gravitating BEC haloes by self-consistently
coupling the Gross-Pitaevskii (GP) equation of motion for the
complex scalar BEC wavefunction $\psi(\mathbf{r},t)$ (see e.g.
\cite{PS}) of the dark matter halo to the Poisson equation,
 \begin{equation} \label{gp}
 i\hbar \frac{\partial \psi}{\partial t} = -\frac{\hbar^2}{2m}\Delta \psi + (m\Phi +
 g|\psi|^2)\psi,
 \eeq
 \beq \label{poisson}
  \Delta \Phi = 4\pi G m |\psi|^2,
   \end{equation}
    where
$|\psi|^2(\mathbf{r},t) = n(\mb{r},t)$ is the number density of dark
matter particles of mass $m$ and $\Phi(\mb{r})$ is the gravitational
potential of the halo.
 We assume that \tx{all} $N$ particles comprising a given halo of volume $V$ are in the condensed state
  described by $\psi$, such that
   \beq \label{norm}
    \int_V |\psi|^2 = N.
     \eeq
This effectively amounts to assuming that the gas is at zero
temperature. Self-gravitating BEC matter has been considered before
as a possible candidate for making ultra-dense stars, called boson
stars, that can avoid collapse to a black hole (see e.g. \cite{kaup,
RB, CSW}). BEC haloes can be imagined as boson stars on galactic
scales, except that their densities are very low.
General-relativistic effects are thus usually considered to be
negligible on those scales. As such, we are free here to treat the
effects of gravity in the Newtonian limit, as is customary in the
literature on BEC haloes.

BEC-CDM, like standard CDM, is assumed to interact so weakly with
other matter and radiation, once its abundance is fixed in the early
universe, that we can neglect all other, non-gravitational
couplings. However, in contrast to standard CDM, BEC dark matter can
be \textit{self-interacting}. The BEC self-interaction has been
included in equ.(\ref{gp}) in the usual way in terms of an effective
interaction potential $g|\psi|^4/2$
  with coupling constant (or self-interaction strength) $g$.
The possibly complicated particle interactions are simplified this
way in the GP framework in the low-energy limit of a dilute gas:
disregarding higher than 2-body interactions, the cross section for
elastic scattering of indistinguishable particles becomes constant
in the low-energy limit,
 \beq \label{scattcross}
  \sigma_s = 8\pi a_s^2,
   \eeq
with the s-wave scattering length $a_s$. The coupling constant of
the effective interaction is then given by
 \beq \label{coupst}
 g = 4\pi \hbar^2 \f{a_s}{m}.
  \eeq
In this work, we will restrict our consideration to $g \geq 0$ ($a_s
\geq 0$).
   Condensates having attractive particle interactions $g < 0$ have been shown not to support vortices, but rather
   produce bright solitons with which we will not be
concerned here. We shall note that the above GP equation is strictly
valid for dilute quantum gases only, which means that $a_s$ must be
much smaller than the mean interparticle distance, i.e. $a_s \ll
n^{-1/3}$.

 The complex wave function
 of the condensate in the GP equation, a form of non-linear Schr\"odinger equation (\ref{gp}), can be decomposed
 into its amplitude and phase function,
 \beq \label{polar}
\psi(\mb{r},t) = |\psi|(\mb{r},t)e^{iS(\mb{r},t)} =
\sqrt{\f{\rho(\mb{r},t)}{m}}e^{iS(\mb{r},t)}
 \eeq
  with the corresponding dark matter halo mass density
   \beq \label{massdens}
  \rho(\mb{r},t) = m |\psi|^2.
   \eeq
    Inserting (\ref{polar}) into (\ref{gp}), the GP equation
    decouples into two equations for the real functions $|\psi|$ and
    $S$,
\beq \label{hd1} -\f{2m}{\hbar}|\psi|\f{\p S}{\p t} + \Delta |\psi|
- |\psi|(\nabla S)^2 - \f{2m}{\hbar^2}(m\Phi + g |\psi|^2)|\psi| = 0
 \eeq
  and
  \beq \label{hd2}
\f{\p |\psi|^2}{\p t} + \nabla \cdot \left[ |\psi|^2 \f{\hbar}{m}
\nabla S \right] = 0.
 \eeq
The associated quantum-mechanical current density,
  \beq \label{qmcurrent}
 \mb{j}(\mb{r},t) = \f{\hbar}{2i m}(\psi^* \nabla \psi - \psi \nabla
 \psi^*) = n(\mb{r},t)\f{\hbar}{m}\nabla S(\mb{r},t),
  \eeq
 can be expressed in terms of the bulk velocity $\mb{v}$ of the
 gas, if we write
   \beq \label{fluidvelo}
    \mb{v} = \f{\hbar}{m} \nabla S.
    \eeq
     The GP equation (\ref{gp}) can hence via equ.(\ref{hd1}) and (\ref{hd2}) be written as a system of quantum-mechanical
    hydrodynamic equations for the mass density $\rho$ and the
    velocity $\mb{v}$, easily recast in the form of an
  Euler-like equation of motion along with the continuity
  equation,
   \beq \label{fluid}
     \rho \f{\p \mb{v}}{\p t} + \rho (\mb{v} \cdot \nabla)\mb{v} = -\rho \nabla
     Q
    - \rho \nabla \Phi - \nabla P_{SI},
     \eeq
 \beq \label{hd3}
\f{\p \rho}{\p t} + \nabla \cdot (\rho \mb{v}) = 0,
 \eeq
  where we define
  \beq \label{qpot}
   Q = -\f{\hbar^2}{2m^2}\f{\Delta \sqrt{\rho}}{\sqrt{\rho}}.
    \eeq
    The term $Q$ gives rise to what is often called 'quantum
 pressure', an additional force on the rhs of equ.(\ref{fluid}) which basically stems from the quantum-mechanical
 uncertainty principle. The particle self-interaction, on the other hand, gives rise to a pressure
 of
 polytropic form
  \beq \label{selfpressure}
 P_{SI} = K_p \rho^{1+1/n} \equiv \f{g}{2m^2}\rho^2,
  \eeq
  where the polytropic index is $n=1$ (not to be confused with the number density $n(\mb{r},t)$), and where the
polytropic constant $K_p$ depends only on the dark matter particle
parameters. According to equations
(\ref{fluid})-(\ref{selfpressure}), BEC haloes as quantum gases have
fluid-like properties, in contrast to the collisionless nature of
standard CDM haloes.

The definition of the bulk velocity in equ.(\ref{fluidvelo}) implies
that for any smooth phase function $S$, the velocity flow of the
system is \tx{irrotational}, $\nabla \times \mb{v} = 0$, which is a
typical characteristic of superfluids. However, in the presence of a
vortex, the fluid velocity experiences a singularity and the
circulation around a contour enclosing the vortex is non-vanishing
and a multiple integer of the elementary circulation,
 \beq \label{circ}
\oint_{\mc{C}} \nabla S \cdot d\mb{l} = 2\pi d \f{\hbar}{m}
 \eeq
  where
$d\mb{l}$ is a unit tangent vector to the curve $\mc{C}$ encircling
the vortex with winding number $d$. The parameter $d$ is an integer
in order to ensure that the wave function, whose amplitude vanishes
along the vortex, is singly-valued. The genuine quantum character of
the system can be inferred by the fact that the above fluid
circulation condition has effectively become a 'quantization
condition'. For an axial-symmetric vortex, $S = d\phi$, the velocity
flow around it is
 \beq \label{vspecial}
 \mb{v} = \f{\hbar}{m}\f{d}{r}\mb{\hat \phi}.
  \eeq
 A wavefunction having such a vortex is an eigenstate of the angular
 momentum with $l_z = d\hbar$, such that the vortex carries a total
 angular momentum equal to
  \beq \label{vortexL}
  L_z = d N \hbar \equiv d L_{QM}.
   \eeq
   $L_{QM}$, as already defined in (\ref{lqm}), is the minimum angular momentum necessary to sustain a
   singly-quantized vortex and will constitute an important quantity
   of the analysis in \tx{Section 3} and \tx{4}.

The hydrodynamic equations described above make it possible for us
to apply familiar results for self-gravitating, classical fluids to
derive properties of quantum-mechanical BEC dark matter (e.g. the
classical figures of gravitational equilibrium, the properties of
vorticity-free flow, the conditions leading to gravitational
instability or vortex formation). In this way, we can also make
clear the distinct role played by the quantum pressure term in the
Euler equation in contrast to the roles of the polytropic
self-interaction pressure and gravity. In the future, this will also
be useful when we consider the dynamical origin of the virialized
structures which constitute the cosmological haloes in this model,
including the application of numerical hydrodynamic methods.

In this paper, however, we shall limit our treatment to the
equilibrium structure of the virialized haloes expected to emerge
during the cosmological formation of galaxies and large-scale
structure in a BEC-CDM-dominated universe. The dynamical formation
of such structures is presumed to result from the gravitational
instability of primordial density fluctuations, like those in the
standard CDM model, leading to the 'cosmic web' of filamentary
structure. In order to follow the development of vortices, predicted
here by a stability analysis, an initial-value problem involving the
fully time-dependent 3D evolution from linear perturbations to the
turn-around and nonlinear collapse of individual haloes must be
solved, including the tidal torques responsible for halo angular
momentum. This is a challenging task and beyond the scope of this
paper. Here, instead, we take the first step of determining whether
the virialized haloes in a BEC-CDM universe are subject to vortex
formation by applying an energy argument to the stationary
description of BEC-CDM haloes as rotating, self-gravitating figures
of equilibrium. If the energy of such a halo is lower in the
presence of a single, central vortex than without a vortex for some
values of the dark matter particle mass and self-interaction
coupling strength, for rotation rates consistent with the angular
momentum expected from large-scale structure formation, then the
halo is unstable to vortex formation for those parameters. Before we
turn our attention to rotating haloes, we set the stage by reviewing
two important limiting cases which give insight into how the
properties of the DM particle can lead to quantum-coherence on
galactic scales.

\subsection{BEC dark matter halo regimes}

In this paper, we will study bound, isolated systems whose
equilibrium structure must satisfy the Virial theorem (see
relationship (\ref{virial}) in the next subsection)\footnote{The
dynamical evolution which leads to the formation of the virialized
haloes in a BEC-CDM cosmology is beyond the scope of this paper,
which focuses on the haloes themselves. For some discussion of this
dynamical evolution, the reader is referred to \cite{alcubierre, GU}
and \cite{SC}.}. BEC haloes can be stabilized against gravitational
collapse by internal pressure or rotation. In addition, BEC haloes
can be prevented from collapse by the quantum-mechanical uncertainty
principle, which is expressed by the quantum-kinetic term in
(\ref{gp}). In order to gain an idea for the length scales involved,
we may simply check the importance of terms of the time-independent
GP equation in different limiting cases\footnote{A Jeans instability
analysis in these limiting cases has been recently presented in
\cite{chavanis1}.}. Obviously, the corresponding characteristic
length scales will be determined by the respective particle or halo
parameters which are involved in the terms under consideration. In
the following, let $R$ denote the size, $M$ the total mass and $\bar
\rho$ the mean density of a given halo.

\subsubsection{Non-interacting DM particles: quantum pressure versus gravity}

For quantum-coherence to be relevant on the scale of a halo, the
dark matter particle de-Broglie wavelength
 \beq \label{deB}
    \lambda_{deB} = \f{h}{m v},
 \eeq
should be of the order of the size of the system
  \beq \label{deBconstraint}
   \lambda_{deB} \lesssim R.
    \eeq
Using $v_{vir} \simeq v_{circ} = (G M/R)^{1/2}$ in the expression
for $\lambda_{deB}$, the particle mass must be of the order of
 \beq
  m \simeq \f{h}{(G M R)^{1/2}} \simeq \f{h}{R^2(G \bar \rho)^{1/2}}.
   \eeq
For galactic virial velocities of the order of $v_{vir} \sim 10 -
200$ km/s, it is clear that it is the particle mass which must be
small enough in order for the de-Broglie wavelength to be comparable
to the extent of the halo.

The quantum-kinetic term can stabilize a system against
gravitational collapse,
 \beq \label{secondlength}
  0 = -\f{\hbar^2}{2m}\Delta \psi + m\Phi \psi
   \eeq
above a length scale of
 \beq \label{jeans1}
 l_{QP} = \sqrt{\f{\hbar}{m(2G \bar \rho)^{1/2}}},
  \eeq
 which plays the role of a quantum Jeans length (see also \cite{khlopov}; \cite{hu}).
 In BEC-CDM without self-interaction,
 $l _{QP}$ is the smallest scale above which bound structures can
 form.
 More precisely, it can be shown that the density profile of a bound system like a halo,
 can only be determined numerically, and is given for instance in
 \cite{MPS}. It has a flat core and falls off as $r^{-4}$ at
 infinity. Although the total mass $M = N m$ is conserved and hence finite, the
 system has no finite size (i.e. no compact support) in this regime,
 and so one may calculate a radius which includes 99 \% of the mass,
  \beq \label{fuzzysize}
   R_{99} = 9.9 \f{\hbar^2}{G M m^2},
    \eeq
 \cite{MPS}. Indeed, this radius which is about 5 times as large as (\ref{jeans1}), is just the more accurate form of
 $l_{QP}$, which can be seen by expressing the density in (\ref{jeans1}) by using the total
 mass and radius, instead. Now, if we require $\lambda_{deB}$ to be
 of the order of
  \beq \label{deBconstraint2}
   l_{QP} \lesssim \lambda_{deB} \lesssim R,
    \eeq
the particle mass has to satisfy $m \lesssim \sqrt{18 \pi^3}m_H$,
where we have defined
   \beq \label{mmin}
    m_H \equiv \f{\hbar}{R^2(\pi G \bar \rho)^{1/2}} = \f{2\hbar}{\sqrt{3G}}(R
    M)^{-1/2},
  \eeq
depending on halo properties. This mass of a \tx{non-interacting} DM
particle is hence characteristic for making the dark matter of a
halo quantum-coherent on a substantial fraction of its size.

There is, however, yet another meaning for this definition, which is
of importance in the context of our paper: If a halo of mass $M$ and
angular momentum $L = L_{QM} \equiv N\hbar$ rotates uniformly with
an angular velocity $\Omega_{QM}$, then
 \beq \label{solidang}
    L_{QM} = MR^2\Omega_{QM}
    \eeq
  implies
   \beq \label{oqm}
    \Omega_{QM} = \f{\hbar}{mR^2}.
    \eeq
 On the other hand, the characteristic gravitational
angular frequency is usually defined as
 \beq \label{freefall}
 \Omega_G \equiv \sqrt{\pi G \bar \rho}.
  \eeq
  It is basically the inverse free-fall time of a
  self-gravitating body, i.e. this angular velocity corresponds to a rotation, which supports the halo
  against gravitational collapse. Now, the dark matter particle mass assumes the form (\ref{mmin})
\tx{if}
 $\Omega_G = \Omega_{QM}$, i.e. if the halo's angular velocity due
 to gravitational stability is sufficient to have each particle
 contribute an amount of $\hbar$ to the total angular momentum.

This regime, where self-interaction is neglected, is denoted as
\tx{Regime I} in Table \ref{tab1}.

\subsubsection{Strongly-interacting DM particles: self-interaction pressure versus gravity}

In the regime where the particle self-interaction dominates, the
so-called \tx{Thomas-Fermi (TF) regime}\footnote{The Thomas-Fermi
energy functional appeared originally in atomic physics
 as an energy depending only on the density of the system. Since the GP energy, too, depends
 on density only in the above limit case, its formal analogy has been
 given the same name.}, it is only the self-interaction which balances
gravity,
 \beq \label{thirdlength}
  0 = m\Phi \psi + g|\psi|^2\psi.
   \eeq
 A system is then gravitationally bound above a length scale of
  \beq \label{jeans2}
 l_{SI} = \f{1}{m}\sqrt{\f{g}{G}}.
  \eeq
More precisely, the GP equation of state is \tx{exactly} an
($n=1$)-polytropic law in this case, and equ.(\ref{fluid}) reduces
to a Lane-Emden
 equation of motion. Solving (\ref{stat}) with (\ref{statdec}), or equivalently
(\ref{fluid}) in this regime, leads to the following general result
for the mass density profile,
  \beq
  \rho(\mb{r}) = \f{m}{g}\left(\mu - m\Phi -
  \f{m}{2}\mb{v}^2\right),
  \eeq
 which can be determined, once $\Phi$ and $\mb{v}$ are known. For spherical haloes in the static case, the density
 profile is given by
  \beq \label{tfprofile}
   \rho^S(\mb{r}) = \rho_c^S \sin \left(\sqrt{\f{2\pi G}{K_p}} r\right)/\left(\sqrt{\f{2\pi G}{K_p}}r\right),
    \eeq
 with the corresponding ratio of central to mean density as
  \beq \label{kreuz}
   \rho_c^S/\bar \rho^S = \pi^2/3 \simeq 3.29.
    \eeq
     A spherical BEC halo is thus an ($n=1$)-polytrope with radius
  \beq \label{onesphere}
   R_0 = \pi \sqrt{\f{K_p}{2\pi G}} = \pi
   \sqrt{\f{g}{4\pi G m^2}}.
    \eeq
This result, derived in \cite{chandrasek} for stellar bodies, has
been re-derived in the context of long-range-interacting atomic BEC
gases in \cite{kurizki} and in the context of dark matter, for
instance in the works of \cite{goodman}, \cite{peebles} and
\cite{BH}. The Thomas-Fermi regime will be studied in more detail in
\tx{Section 3}. Since this limit corresponds to the BEC becoming an
$(n=1)$-polytrope, we will refer to this case also as the
\tx{polytropic regime}. The convenient simplifications, which come
by disregarding the complicated quantum pressure term in
(\ref{fluid}) in this regime, have been heavily exploited in much of
the literature on the subject. However, this comes with a price.
According to equ.(\ref{onesphere}), the size of the polytrope is
fixed by the BEC-CDM particle mass and coupling strength in the
combination $g/m^2$. While we shall continue to refer to this size
as the halo size, it shall be understood that this refers
\tx{either} to the actual halo size \tx{or} to the size of a
virialized core region within a larger halo. The TF regime
considered here is denoted as \tx{Regime II} in Table 1.

\subsubsection{Validity of the Thomas-Fermi regime: quantum pressure
versus self-interaction}

To determine whether a halo is in \tx{Regime I} or \tx{II}, we must
compare the quantum pressure and self-interaction pressure terms to
each other. When the length scale for density variation in the BEC
fluid is such that the two terms are of equal magnitude, that length
scale is sometimes called the 'healing length'. The reason for this
name is because it is the characteristic distance over which the
wavefunction approaches the background value of a smooth density
distribution, subjected to a localized perturbation. Balancing
quantum-kinetic term and self-interaction,
 \beq \label{firstlength}
  0 = -\f{\hbar^2}{2m}\Delta \psi + g|\psi|^2\psi,
  \eeq
 results thus in the following expression for the healing length,
  \beq \label{healing}
   \xi = \f{\hbar}{\sqrt{2 \bar \rho g}}.
   \eeq
We will see below that if the system's characteristic size is much
larger than this, $R \gg \xi$,
 than\footnote{Since the diluteness condition implies $n^{-1/3} \ll
\xi$, one should consider the limit $R/\xi \to \infty$, rather than
$\xi \to 0$.} the quantum pressure can be neglected compared to the
self-interaction term and we are in the Thomas-Fermi
 regime (\tx{Regime II}). For a given mean halo density $\bar \rho$, we see that
$\xi$ grows with decreasing coupling $g$, approaching the opposite
regime of non-interacting particles. On the other hand, if $g$ is
fixed, the healing length is smaller for haloes of higher mean
density. Let us define the characteristic coupling strength for
which $R = \xi$, therefore, as $g_H$, given by
  \beq \label{gmin}
 g_{H} \equiv \f{\hbar^2}{2\bar \rho R^2} = \f{2}{3}\pi \hbar^2 \f{R}{M},
  \eeq
  where $g_H$ is
  determined by the mean density and radius (or mass and mean radius) of the
  halo.
In later sections, it will be convenient to describe dimensionless
BEC particle parameters, and so already here we define the
quantities
  \beq \label{becdef}
   y \equiv \f{m}{m_H} \mbox{ and } x \equiv \sqrt{\f{g}{g_H}} =
   \f{R}{\xi}.
    \eeq
 Thus, the BEC self-interaction regime can also be determined by the ratio
 of $g/g_H$.

Let us investigate this in more detail: In the TF-regime, the
quantum-mechanical pressure in (\ref{fluid}) is neglected. In order
to see for which part of the BEC parameter space this regime can be
expected to be a good approximation, we estimate the respective
forces associated with the quantum-pressure term and the
self-interaction term in (\ref{fluid}), respectively. Both depend on
the halo density profile's varying on a length scale of order $R$.
Using (\ref{qpot}), the force due to quantum pressure is estimated
by
 \beq \label{line1}
  -\rho \nabla Q \sim \f{\hbar^2}{2m^2}\f{\rho}{R^3},
   \eeq
 while the force due to the self-interaction is of order
  \beq \label{line2}
   -\nabla P_{SI} \sim -\f{g\rho^2}{2m^2R}.
    \eeq
The ratio of the two becomes
 \beq \label{line3}
  \f{|-\rho \nabla Q|}{|-\nabla P_{SI}|} \sim \f{\hbar^2}{g\rho R^2}
  \sim
  2 \f{g_H}{g}
  \eeq
  by using the definition of $g_H$. We see
  that for the above ratio to be much smaller than one, we require that
   \beq \label{gcond}
   \f{g}{g_H} \gg 2,
   \eeq
    which is thus the condition for the TF regime to be valid. If we go
 back to the characteristic size of a spherical halo in (\ref{onesphere}), we can furthermore see
 that, if we set $v = v_{vir}$ in (\ref{deB}), we can rewrite
  \beq \label{relate}
   R_0 \simeq
   \f{\sqrt{3}\pi^{1/4}}{12}\left(\f{g}{g_H}\right)^{1/2}\lambda_{deB},
    \eeq
from which follows $R_0/\lambda_{deB} \gg 1 $ \tx{if} $(g/g_H)^{1/2}
=R_0/\xi \gg 1$. Therefore, in contrast to \tx{Regime I},
$\lambda_{deB}$ is much smaller than the size of the halo. However,
using (\ref{becdef}), we see that $\lambda_{deB} \simeq 4.3 \xi$ in
this regime, so the de-Broglie wavelength is still larger than the
healing length. In \tx{Section 3}, we will apply the Virial theorem
to ellipsoidal BEC haloes in the TF regime, and we will see that
this enforces the particle parameters $m$ and $g$ to be related
according to (\ref{npoly})-(\ref{npolydim}) and
(\ref{compR})-(\ref{compRdim}), respectively. In the spherical case,
the dimensionless quantities (\ref{becdef}) are then related as
 \beq \label{sphervirial}
  \f{m}{m_H} = \f{\sqrt{15}}{4}\sqrt{\f{g}{g_H}}.
  \eeq
Therefore, it follows that in the TF regime we also have $m/m_H \gg
1$. In other words, $\lambda_{deB}$ can be much smaller than $R$ and
hence the particle mass $m$ can be much larger than $m_H$, but only
if the self-interaction is high enough.

The above characteristic units for the DM particle mass and coupling
strength can be written in fiducial values as
  \beq \label{mfidu}
  m_H = 1.066 \cdot 10^{-25}\left(\f{R}{100 ~\rm{kpc}}\right)^{-1/2}\left(\f{M}{10^{12}~ M_{\odot}}\right)^{-1/2}
  \rm{eV},
   \eeq
    and
     \beq \label{gfidu}
   g_H = 2.252 \cdot 10^{-64} \left(\f{R}{100~
   \rm{kpc}}\right)\left(\f{M}{10^{12}~M_{\odot}}\right)^{-1}
   \rm{eV} ~\rm{cm}^3,
 \eeq
 in units where $c=1$ and $[R] = $ kpc, $[M] = M_{\odot}$.
  For haloes with a size between the Milky Way ($M=10^{12}M_{\odot}, R=100$
kpc) and a typical
 dwarf galaxy ($M=10^{10}M_{\odot}, R=10$ kpc), we see that the
 above quantities are in the range of
 \beq \label{ranges}
   m_H \simeq 10^{-25..-24} ~\rm{eV},~~ g_H \simeq 2\cdot
   10^{-64..-63}~ \rm{eV}~\rm{cm}^3.
 \eeq

The two BEC dark matter regimes are schematically described in Table
\ref{tab1}, where also some common names are listed. The latter,
however, have appeared in more recent papers, even though both
regimes have been studied already in earlier literature on
self-gravitating bosonic matter.

To re-iterate, both regimes revisited in this section exhibit
non-classical behaviour: in \tx{Regime I}, $\nabla Q$ in
equ.(\ref{fluid}) causes the Heisenberg uncertainty principle to act
on galactic scales. In \tx{Regime II}, on the other hand, $\nabla
P_{SI}$ results in $(n=1)$-polytropic solutions. However, the
physical nature of this pressure and the so-caused "internal energy"
stems from the repulsive quantum-mechanical 2-body scattering at
vanishing temperature.

\begin{table}
\caption{BEC dark matter regimes for a halo of (mean) radius $R$ and
mass $M$} \label{tab1}
\begin{center}
\begin{tabular}{l|l|l}
       & $\mb{I}$      & $\mb{II}$  \\
\hline
          & quantum $\gg$ self-interaction &  self-interaction $\gg$ quantum \\
          & pressure   & pressure   \\
          & $\xi \to R$, $g \to 0$,     & $\xi \ll R$, $g \gg g_H$,  \\
          & $m \lesssim m_H = f(R,M)$          & $m \gg m_H$, $g/m^2 = f(R)$ \\
\hline
          & $R \gtrsim \lambda_{deB} \gtrsim l_{QP}$ & $R = R_0 \simeq l_{SI} \gg \lambda_{deB}$ \\
          & $\rightarrow$ grav. bound,  & $\rightarrow$
          grav. bound, \\
          & halo core $\sim \lambda_{deB}$  & halo core $\sim R_0$\\
          & no vortices                   & vortices favoured
          \\
          & "fuzzy dark matter"  & "repulsive/fluid dark
          matter"  \\
          & (free scalar-field DM)  & (scalar-field DM with strong, \\
          &                         &  positive self-interaction) \\
 \hline
\end{tabular}
\end{center}
\end{table}

\subsection{Stationary systems and virial equilibrium}

Since we are going to study the energetic stability of vortices in
this paper, we will restrict ourselves to some simplified models in
the forthcoming analysis: First, we will consider stationary systems
and their corresponding energy, i.e. we will restrict to
time-independent systems. Our second major assumption will be that
haloes rotate with a \tx{constant} angular velocity $\mb{\Omega}$
about their rotation axis. This will allow us later to move into the
frame co-rotating with the halo and to study stable structures in
that frame.

Stationary states are described by wavefunctions of the form
$\psi(\mb{r},t) = \psi_s(\mb{r})e^{-i\mu t/\hbar}$, where $\mu$, the
GP chemical potential, is fixed by the conservation of particle
number. For these states, the mass density $\rho = m|\psi_s|^2$ and,
hence, the gravitational potential $\Phi$ are time-independent,
while the wavefunction evolves harmonically in time. Inserting this
$\psi$ into (\ref{gp}) results in the time-independent GP equation
with eigenvalues $\mu$ (see also \tx{Appendix C}),
 \beq \label{stat}
 \left(-\f{\hbar^2}{2m}\Delta + g|\psi_s|^2 + m \Phi\right)\psi_s = \mu
 \psi_s.
  \eeq
 The time-independent part $\psi_s(\mb{r})$ itself can be decomposed as
   \beq \label{statdec}
   \psi_s(\mb{r}) = |\psi_s|(\mb{r})e^{iS_s(\mb{r})}
    \eeq
   with amplitude and phase both depending on position only. We
   will omit the subscript 's' in the forthcoming analysis.
 Systems obeying (\ref{stat}) can
 be studied via the corresponding GP energy functional, which is given by
 \begin{equation} \label{energie}
   \mathcal{E}[\psi] = \int_V \left[\frac{\hbar^2}{2m}
 |\nabla \psi|^2 + \frac{m}{2}\Phi |\psi|^2 +
 \frac{g}{2}|\psi|^4\right]d^3\mb{r}.
 \end{equation}
 On the other hand, equ.(\ref{stat}) could have been also obtained by variation of
(\ref{energie}) with respect to $\psi$ (or its complex conjugate
$\psi^*$), under the constraint (\ref{norm}),
 \beq \label{variation}
  \f{\delta \gpf[\psi]}{\delta \psi^*} - \mu \f{\delta}{\delta
  \psi^*} \int |\psi|^2 d^3\mb{r} = 0,
   \eeq
 with $\mu$ playing the role of the Lagrange multiplier. Inserting (\ref{statdec}) into (\ref{energie}), the total energy
  can be written in a very instructive way as
  \beq \label{sumenerg}
   E = K + W + U_{SI},
    \eeq
 where the \tx{total} kinetic energy term $K$ is given by
  \bdi
   K \equiv \int_V \f{\hbar^2}{2m}|\nabla \psi|^2 d^3\mb{r} =
    \edi
   \beq \label{kname}
    = \int_V
   \f{\hbar^2}{2m^2}(\nabla \sqrt{\rho})^2d^3\mb{r} + \int_V
   \f{\rho}{2}\mb{v}^2d^3\mb{r} \equiv K_Q + T,
    \eeq
 with $K_Q$ accounting for the quantum-kinetic energy and $T$ for the
 bulk kinetic energy (rotational, internal motion, ect.) of the body. $K_Q$ has \tx{no} classic
 counterpart, and is absent in the classical figures of equilibrium studied in
 \cite{chandra} or in \cite{LRS}. Also, $K_Q$ is \tx{neglected} whenever the Thomas-Fermi regime is employed.
  The other terms in (\ref{sumenerg}) are simply the gravitational
 potential energy
  \beq \label{wname}
   W \equiv \int_V \f{\rho}{2}\Phi d^3\mb{r}
    \eeq
 and the internal energy
  \beq \label{internal}
   U_{SI} \equiv \int_V \f{g}{2m^2}\rho^2 d^3\mb{r},
    \eeq
 which is determined by the particle interactions. We have defined the latter essentially as
 $U_{SI} = \int P_{SI} dV$ with
 $P_{SI}$ the pressure due to self-interaction (\ref{selfpressure}), which, in
general, is \tx{not}
  the total pressure given the presence of $K_Q$. While we call $U_{SI}$ the internal energy,
  it shall be kept in mind that its origin is not due to thermal processes but solely due to the repulsive 2-body
  elastic scattering.
  The above energy
 contributions are those which enter the scalar Virial theorem of a
 rotating, \tx{isolated} BEC halo under self-gravity,
  \beq \label{virial}
 2K + W + 3U_{SI} = 0.
  \eeq

\section{BEC haloes with angular momentum}

In the previous section, we have reviewed halo density profiles in
the absence of rotation and angular momentum. However, we expect and
assume galactic haloes to have undergone
 tidal torquing in the early phases of their collapse, and as such, we will model them as rotating,
 \tx{non-spherical} bodies. In case of a (non-rotating) sphere
 ($\Omega=0$), the BEC wavefunction were real and positive.
However, for non-vanishing angular velocities, $\Omega \not= 0$,
superfluid currents arise. As a result, the wave function is complex
and therefore has a non-trivial phase or velocity flow,
respectively, even without vortices (for laboratory BEC examples see
for instance the systems studied in \cite{fetter} and
\cite{recati}). This means, in particular, that such a system has a
non-vanishing, bulk angular momentum \tx{prior} to vortex formation,
which we will want to associate with that provided by tidal
torquing, and characterized by the $\lambda$-spin parameter defined
in equ.(\ref{lam}).

So, in modelling dark matter haloes as BEC wave functions, $\psi =
|\psi|e^{iS}$, their velocity information is contained in the (real)
phase function $S(t,\mb{r})$. As long as no defect structures like
vortices appear in the flow, the condensate phase is a smooth
function in which case we will use the notation $S_0$ in the rest of
the paper. The fluid velocity is generally given by $\mb{v} = \hbar
\nabla S_0/m$, see equ.(\ref{fluidvelo}), while in a frame rotating
rigidly with a constant angular velocity $\mb{\Omega}$, it is given
by $\mb{v}' = \mb{v} - \mb{\Omega} \times \mb{r}$. In \tx{Section
4}, we will study the energetic stability of vortices, and it will
turn out to be convenient to move into a frame which rotates with
the figure, i.e. at an angular velocity $\mb{\Omega}$. It is this
co-rotating frame we shall consider, assuming that the haloes rotate
about the $z$-axis, such that $\mb{\Omega} = (0,0,\Omega)$.

As we have pointed out in \tx{Section 2}, the (vortex-free) density
profile of a BEC halo has a flat core, in contrast to the
$r^{-1}$-central cusps of standard CDM. However, that equ.
(\ref{gp}) - with or without rotation - favours a flat core over a
cusp can be also seen as follows: The equilibrium matter
distribution of the BEC halo in a frame rotating rigidly with
constant angular velocity $\mb{\Omega}$ is given by
 \bdi
 -\f{\hbar^2}{2m}\Delta' |\psi'| + \f{\hbar^2}{2m}|\psi'|(\nabla S')^2 +
  \edi
   \beq \label{static}
    + (m \Phi' + g|\psi'|^2 - \mu')|\psi'|
  - \hbar |\psi'|\nabla S'\cdot (\mb{\Omega}\times \mb{r}) = 0, \eeq
 (see (\ref{hd1}) and (\ref{stat})) where primes will in the following denote quantities as measured in the
 rotating frame, and $\mu' = const$.
Inserting a cuspy test function, $|\psi'|^2 = (r')^{-\alpha}/m$ with
constant exponent $\alpha > 0$, results in
 \bdi
  m\Phi' + \f{g}{m}(r')^{-\alpha} + \f{\hbar^2}{2m}(\nabla S')^2 - \hbar \nabla S'\cdot (\mb{\Omega}\times \mb{r})
  \edi
   \beq \label{gprot}
    = \mu' +
  \f{\hbar^2}{2m}\f{\alpha}{2}\left(\f{\alpha}{2}+1\right)(r')^{-2}.
   \eeq
Calculating the associated gravitational potential $\Phi'$ to the
test function by requiring that $\nabla' \Phi' = 0$ at $r'=0$ (i.e.
no net gravitational force at the centre) requires $\alpha < 1$.
However, if $\alpha > 0$ and since $\mu' = const.$, one can easily
show that the resulting velocity flows $\mb{v}'$ would diverge in
the centre, regardless of the absence of a vortex, which would
contradict the irrotationality constraint. In fact, in the next
section, we will study figures of revolution as halo models, whose
velocity fields \tx{prior} to vortex formation are smooth in either
frame of reference.

\subsection{BEC haloes as Maclaurin spheroids}

\subsubsection{General properties}

For the sake of computational simplicity and in order to gain
analytical insight, we start our analysis by considering a simple
halo model, the densities and potentials of which are given by
homogeneous Maclaurin spheroids, rotating uniformly with angular
velocity $\mb{\Omega} = (0,0,\Omega)$. Although the BEC fluid is
compressible, as described in \tx{Section 2}, the spherical
polytrope solution shows that the density only varies by a factor of
about three from the centre to the surface, see (\ref{kreuz}).
Hence, the assumption of a uniform density as a first approximation
is more justified for BEC than for standard CDM haloes. However,
homogeneous, i.e. incompressible Maclaurin spheroids are only
approximate solutions of the GP equation (\ref{stat}). In order for
our model to be in global virial equilibrium, equ.(\ref{virial}), in
accordance with the equations of motion, we will have to find a
constraint on the underlying DM particle parameters, as will be
shown in the next subsection.

We denote the mass density as $\rho_0$, and the semi-axes
$(a_1,a_2,a_3)$, lying along $(x,y,z)$, are such that $a_1=a_2
\equiv a > a_3 \equiv c$. The total volume of a spheroid is $V =
4\pi a^2c/3$, and we will make use of its mean radius, defined as $R
= (a^2c)^{1/3}$. The gravitational potential inside the spheroidal
body is given by
 \bdi
  \Phi_0(r,z) = \pi G \rho_0 \times
   \edi
   \beq \label{macpotential}
    \times \left[\left(A_1(e)r^2 + A_3(e)z^2\right) -
  2\sqrt{1-e^2}a^2\f{\arcsin(e)}{e}\right]
  \eeq
 in cylindrical coordinates
 $(r,z)$ (see \cite{chandra} or \cite{BT} for definitions and formulas, but note the difference in the sign convention
 of $\Phi_0$ between the two references. We choose the convention in \cite{BT}). The functions $A_1$ and $A_3$ depend on the
 eccentricity $e = \sqrt{1-(c/a)^2}$ of the figure. Their general definition can be found in \tx{Appendix A}. Here,
 they reduce to functions in closed-form,
 \beq \label{A1}
 A_1(e) = A_2(e) = \f{\sqrt{1-e^2}}{e^3}\arcsin(e) -
 \f{1-e^2}{e^2},
  \eeq
  \beq \label{A3}
  A_3(e) = \f{2}{e^2} -
 \f{2\sqrt{1-e^2}}{e^3}\arcsin(e),
  \eeq
   and both are positive for $e \in [0,1)$.
   A family of spheroids is characterized by the following relationship
between angular velocity and eccentricity, the Maclaurin formula
(see e.g. \cite{BT})
 \beq \label{macomega}
  \tilde \Omega \equiv \f{\Omega}{\Omega_{G}} = \sqrt{2}(A_1(e) -
  (1-e^2)A_3(e))^{1/2}
  \eeq
   with $\Omega_G$ in (\ref{freefall}).
   Relation (\ref{macomega})
   parametrizes the rotation via $e$, and $e=0$ for $\Omega = 0$. The figure is stable
   only for values of $e$ below $0.9529$ (see \cite{chandra}).
For our analysis, we will also need the gravitational potential and
rotational kinetic energy of the homogeneous Maclaurin spheroid,
 \beq \label{macpot}
  W = -\f{3}{5}\f{GM^2}{R}(1-e^2)^{1/6}\f{\arcsin(e)}{e}
   \eeq
   and
  \beq
    T =
  \f{4\pi}{15}(1-e^2)^{-1/3}\rho_0 \Omega^2 R^5,
  \eeq
  respectively,
   along with its angular momentum
   \beq \label{macL}
   L = |\mb{L}| = \sqrt{\f{4}{5}(1-e^2)^{-1/3}T M R^2}
    \eeq
 (see again \cite{chandra} and \cite{BT} for reference).
The Maclaurin spheroid experiences rigid rotation only, therefore
its velocity in the two frames is simply given by
 \beq \label{macvelo}
  \mb{v} = \mb{\Omega} \times \mb{r} = \sqrt{2}\Omega_G (A_1 - (1-e^2)A_3)^{1/2}(-y, x,
  0),
   \eeq
   and $\mb{v}' = 0$.
   An illustrative plot of a velocity field in the rest frame can be found in Fig.1.

\begin{figure*}
      \centering\includegraphics[angle=270,width=6cm]{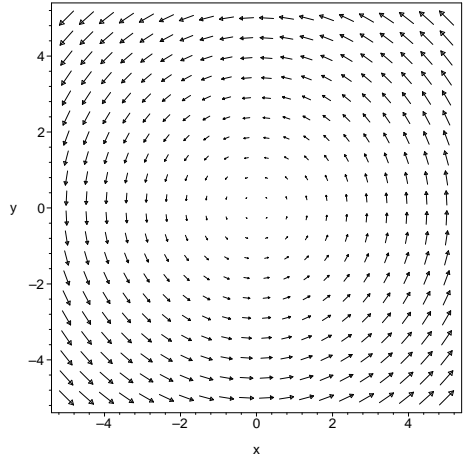}
 \caption{Illustrative velocity field of a Maclaurin spheroid in the
    rest frame having $\lambda = 0.05$.}
 \label{fig1}
\end{figure*}

\subsubsection{Comparison to CDM $\lambda$-spin parameter and virial
constraint}

In this subsection, we aim to express the spin parameter $\lambda$
in equ.(\ref{lam}) as a function of the halo's eccentricity only,
i.e. its shape. To this end, we first rewrite
 \beq \label{lam2}
  \lambda = \f{L|W|^{1/2}}{GM^{5/2}}\left|\f{E}{W}\right|^{1/2}.
  \eeq
  According to equations (\ref{macpot}) and (\ref{macL}),
  \bdi
   \f{L|W|^{1/2}}{GM^{5/2}} =
    \edi
   \beq \label{lam3}
   = \sqrt{\f{4}{5}M(1-e^2)^{-1/3}R^2t}\f{|W|}{GM^{5/2}} = \f{6}{5\sqrt{5}}\f{\arcsin
   e}{e}\sqrt{t(e)},
   \eeq
  with the $t$-parameter $t \equiv T/|W|$, a measure of rotational support,
  given by
 \beq \label{tpam}
  t(e) = \f{3}{2e^2} - 1 - \f{3\sqrt{1-e^2}}{2e\arcsin(e)}.
   \eeq
 This relation also applies to compressible Maclaurin spheroids as has been shown in \cite{LRS}.
The Virial theorem in (\ref{virial}) allows us to write
 \beq \label{hilf}
\left|\f{E}{W}\right|^{1/2} =
t^{1/2}\left(1+\f{2U_{SI}}{T}\right)^{1/2}, \eeq
 where we used the fact that, for a homogeneous body, $K_Q=0$. Using
 now $U_{SI}=P_0V$ with \tx{constant} pressure $P_0$ and $V=4\pi R^3/3$
according to the definition of the spheroid's mean
 radius $R$, we get for the ratio
  \beq \label{vir}
   \f{2U_{SI}}{T} = \f{2P_0V}{|W|t} =
  \f{2}{t}\f{5}{3}\f{e}{\arcsin(e)}(1-e^2)^{-1/6}\f{g}{2m^2}{R}{GV}.
   \eeq
It remains to establish a connection between the mean radius $R$ of
the halo and the dark matter particle parameters. The 'classical',
homogeneous Maclaurin spheroid, which formally is an
$(n=0)$-polytrope, has a pressure profile and its internal energy
vanishes. A BEC system, on the other hand, possesses a formal
"internal energy" due to the particle self-interaction
(\ref{internal}), and its equation of state is such that the
pressure is constant in case of constant density. This apparent
inconsistency is due to the fact that we model the BEC halo as a
homogeneous system. In order for the BEC halo to fulfil the virial
theorem globally, we thus set the BEC pressure equal to the
\tx{average} pressure of the classical spheroid, i.e. the DM
particle parameters must be such that the resulting halo size
corresponds to the one given by the classical spheroid. Using the
pressure distribution of a homogeneous, ellipsoidal equilibrium
configuration,
 \beq
  P(\mb{r}) = \pi G\rho_0^2\left[A_3 a_3^2 - A_1x^2 - A_2y^2 -
  A_3z^2\right] + \f{\rho_0}{2}\mb{v}^2,
   \eeq
 we calculate $U_{SI} = \int P(\mb{r})d^3\mb{r}$, which, for the Maclaurin
 spheroid with $A_1=A_2$, $a_1=a_2 \equiv a$, $a_3 \equiv c$,
 results in
  \beq \label{macpress}
   \int P(\mb{r})d^3\mb{r} = \f{8}{15}A_3(e)(1-e^2)^{2/3}\pi^2 G
   \rho_0^2R^5.
    \eeq
 Setting this expression equal to $U_{SI} = g\rho_0^2/(2m^2)\int d^3\mb{r}$ according to (\ref{internal}) and solving for
 the mean radius $R$, we get
\beq \label{nradius}
  R = \left(\f{15}{3A_3(e)(1-e^2)^{2/3}}\right)^{1/2}\left(\f{K_p}{2\pi
  G}\right)^{1/2}
 \eeq
  with $K_p$ in (\ref{selfpressure}).
In the spherical limit, $e \to 0$, this reduces to
 \beq \label{zerosphere}
  R_0 = \sqrt{\f{15}{2}}\left(\f{K_p}{2\pi
  G}\right)^{1/2},
   \eeq
which is a factor of about $1.15$ smaller than the radius of a
spherical ($n=1$)-polytropic halo given in (\ref{onesphere}).
Equ.(\ref{nradius}) connects the particle parameters $(m,g)$ via
$K_p$ to the mean radius $R$ such that \tx{virialized} haloes
fulfill a certain constraint as can be seen as follows: using the
above definitions for $K_p$ and (\ref{gmin})-(\ref{becdef}), we
rewrite
 \beq \label{first}
  \f{g}{g_H} = \f{m^2}{m_H^2}\f{2K_p}{(g_H/m_H^2)} ~~\mbox{ or }~~ y =
  \sqrt{\f{g_H}{2K_p m_H^2}}x
   \eeq
 and
  \beq \label{zwischen}
   \f{g_H}{m_H^2} = \f{\pi G}{2} R^2.
   \eeq
Inserting now the radius (\ref{nradius}), we arrive at the following
constraint, implied by imposing virial equilibrium,
 \beq \label{npoly}
  y(x) = \left(\f{5}{8A_3(e)(1-e^2)^{2/3}}\right)^{1/2}x.
   \eeq
   The corresponding dimensional relationship is accordingly
  \beq \label{npolydim}
  m(g) =
   \left(\f{5}{8A_3(e)(1-e^2)^{2/3}}\right)^{1/2}\f{m_H}{\sqrt{g_H}}
  \sqrt{g},
 \eeq
  i.e. virialized haloes as rotating, homogeneous Maclaurin spheroids lie on a line in
  ($\log m, \log g$)-space, whose slope is completely determined by the eccentricity $e$ and the mean
  radius $R$ of the halo.  We will make use of this result in \tx{Section 4}. \\
For now, we continue the calculation of $\lambda = \lambda(e)$.
Inserting (\ref{nradius}) into (\ref{vir}), we get
  \beq \label{lam4}
 \left|\f{E}{W}\right|^{1/2} =
 t^{1/2}\left(1+\f{e}{t}\f{A_3(e)(1-e^2)^{1/2}}{\arcsin(e)}\right)^{1/2}.
  \eeq
 In connection with (\ref{lam3}), the $\lambda$-spin parameter as a function of the eccentricity $e$ of the halo
 is finally given by
 \beq \label{lambdaincomp}
 \lambda =
 \f{6}{5\sqrt{5}}\f{\arcsin{e}}{e}t\left(1+\f{e}{t}\f{A_3(e)(1-e^2)^{1/2}}{\arcsin(e)}\right)^{1/2}
  \eeq
  with $t = t(e)$ in (\ref{tpam}). In our case, however, we fix $\lambda$ and solve for
  $e$. In what will follow in \tx{Section 4}, we will take three
representative values, $\lambda = (0.01, 0.05, 0.1)$, which
correspond to eccentricities $e = (0.062, 0.302, 0.550)$. A table of
spheroid parameters as a
  function of $\lambda$ can be found in \tx{Appendix A}. For given
values of the halo mass density and eccentricity, the angular
velocity is fixed. For example, for a mean dark matter density of
$1.6 \cdot 10^{-26}$ g/cm$^3$ for the Milky-Way and $\lambda =
0.05$, we have $\Omega \sim 1.3\cdot 10^{-17}$ rad/s, so
$\Omega/\Omega_{G} \sim 0.22$.

The halo model presented in this subsection has the advantage of
computational simplicity, however, the approximation of uniform
rotation means its velocity flow is not strictly irrotational in the
rest frame as required in the absence of vortices. The approximation
of uniform density, moreover, does not capture the effect of
compressibility, which results from the dependence of
self-interaction pressure on density in equ.(\ref{selfpressure}).
Therefore, we will consider a further model, which improves upon
those approximations, in the next subsection.

\subsection{BEC haloes as irrotational Riemann-S ellipsoids}

\subsubsection{General properties}

In this subsection, we relieve some of the simplifying assumptions
from above by allowing the density to vary in space and by imposing
\tx{strict} irrotationality in the rest frame prior to vortex
formation. This will allow us to take account of the compressibility
of the BEC fluid, which corresponds in the Thomas-Fermi regime to an
$(n=1)$-polytrope (see (\ref{selfpressure})). It is possible to
generalize the Maclaurin spheroids of \tx{Section 3.1} to account
for this compressibility in an approximate way, as shown by
\cite{LRS}. However, this would not allow us to impose the
additional constraint of irrotationality. In general, a rotating
ellipsoidal halo cannot be both axisymmetric and irrotational if it
is non-singular at the origin as can be seen as follows: Suppose
without lack of generality that the velocity field of the halo has
the form
 \beq \label{diffvelo}
 \mb{v} = \mb{\Omega}(r,z) \times \mb{r} = v_{\theta}(r,z)\mb{\hat \theta} = r\Omega(r,z)\mb{\hat \theta},
 \eeq
allowing for differential rotation in $r$ and $z$. The corresponding
vorticity is given by
 \beq \label{vortvelo}
  \nabla \times \mb{v} = \left(-r\f{\p \Omega}{\p z}, 0,
  \f{1}{r}\f{\p}{\p r}(r^2\Omega)\right)\mb{\hat r}.
  \eeq
Requiring irrotationality $\nabla \times \mb{v} = 0$ is now
equivalent to requiring $r^2\Omega(r,z) = constant$ in $r$ and
$\Omega = constant$ in $z$. This means that either $v_{\theta} =
\infty$ at $r=0$ or, if at $r=0$ we impose $v_{\theta} = 0$, then it
follows that $\Omega(r,z) = 0$. The assumption of axisymmetry along
with the constraint of irrotationality lead to diverging velocity
profiles in the centre of the halo or to a trivial
solution\footnote{For fully general-relativistic, rotating boson
stars, \cite{ryan} and \cite{YE} assumed axisymmetry but in their
case the velocity was singular at the centre.}.

It is thus necessary to consider non-axisymmetric objects if BEC
dark matter forces us to take into account the irrotationality of
its velocity field along with a bulk angular momentum \tx{prior} to
vortex formation. There exists indeed a 'classical' figure of
rotation which can serve this purpose, namely the irrotational
Riemann-S ellipsoid. The family of Riemann-S ellipsoids with
semi-axes $(a_1,a_2,a_3)$ along $(x,y,z)$ describes uniformly
rotating bodies, as in the Maclaurin case, but with internal
velocity fields \tx{superposed}, which combine with the uniform
rotation to yield a net flow with $\nabla \times \mb{v} = 0$, see
Fig.2. Exact solutions exist for Riemann-S ellipsoids only for the
case of uniform density, a limitation they share with the Maclaurin
spheroids (\cite{chandra}). Fortunately, as we shall see below,
\cite{LRS}, abbreviated LRS93 in the following, developed their
'ellipsoidal approximation' also for Riemann-S ellipsoids, so we
shall be able to consider the compressible case here as
well\footnote{As shown in LRS93, the approximate solutions for the
compressible case, which result from their ellipsoidal
approximation, agree well with the true equilibria.}.

Again, we assume the uniform rotation to be about the $z$-axis with
angular velocity $\Omega$. The internal velocity field having
angular velocity $\Lambda$ is required to have its uniform vorticity
parallel to $\Omega$ and to leave the ellipsoidal figure unchanged
(see \cite{chandra} and \tx{Appendix A} for more details).
  The mean radius of the ellipsoid is given by
   \beq \label{meanradius}
  R = (a_1a_2a_3)^{1/3} =
  a_1(1-e_1^2)^{1/6}(1-e_2^2)^{1/6},
   \eeq
   with eccentricities
     \beq
   e_1 = \sqrt{1-(a_2/a_1)^2} \mbox{ and } e_2 =
   \sqrt{1-(a_3/a_1)^2}.
   \eeq
   As in \cite{chandra} and LRS93, we write
the velocity field in the frame co-rotating with the figure as
 \beq \label{velo}
  \mb{v}' = C_1y\mb{x} + C_2x\mb{y}
\eeq
 with
  \beq \label{velo2}
  C_1 = -\f{a_1^2}{a_1^2+a_2^2}\zeta' = \f{a_1}{a_2}\Lambda,~~
 C_2 = \f{a_2^2}{a_1^2+a_2^2}\zeta' = -\f{a_2}{a_1}\Lambda.
  \eeq
 The vorticity in the rotating frame is defined as
  \beq
 \zeta' \equiv (\nabla' \times \mb{v}')_z  =
 -\f{a_1^2+a_2^2}{a_1a_2}\Lambda \equiv \Omega f_R,
 \eeq
and the vorticity in the rest frame is given by
 \beq \label{vortrie}
  \zeta \equiv (\nabla \times \mb{v})_z = \zeta' + 2\Omega = (f_R +
  2)\Omega.
   \eeq
 There is a whole class of Riemann-S ellipsoids, which differ in
 their value of $f_R$. However, only for those with $f_R = -2$
 does the vorticity vanish in the rest frame, the so-called
 \tx{irrotational} Riemann-S ellipsoids, and we will only consider
 those in the forthcoming analysis.
 It turns out that these ellipsoids fulfill $a_1 \geq a_3 \geq
a_2$, i.e. they are all prolate\footnote{As a curious note, we
remark that, interestingly, recent studies on standard CDM halo
formation suggest that most haloes are of prolate shape, see e.g.
\cite{plionis} and \cite{gottloeber}. The reason there, however, is
to be found in details of the tidal interactions, and not the
constraint of irrotationality.} bodies (see \cite{chandra}).
Furthermore, we can write the relationship between $\Omega$ and
$\Lambda$ for $f_R = -2$ according to LRS93 as
 \beq \label{velo3}
  \Omega = \f{1}{2}\left(\f{a_1}{a_2}+\f{a_2}{a_1}\right)\Lambda =
  \f{2-e_1^2}{2\sqrt{1-e_1^2}}\Lambda.
   \eeq
 The connection between angular velocity and axis ratios or
eccentricities, respectively, can again be stated in a concise form
as in (\ref{macomega}). We take the formula from LRS93 for $f_R =
-2$,
 \bdi
  \tilde{\Omega} \equiv \f{\Omega}{\Omega_G} =
  \left(\f{2B_{12}}{q_n}\right)^{1/2}\left(1 + \f{4a_1^2a_2^2}{(a_1^2+a_2^2)^2}\right)^{-1/2} =
   \edi
   \beq \label{rieomega}
  = \left(\f{2B_{12}}{q_n}\right)^{1/2}\left(1+\f{4(1-e_1^2)}{(2-e_1^2)^2}\right)^{-1/2},
  \eeq
 with $B_{12} = A_2 - a_1^2 (A_1-A_2)/(a_2^2-a_1^2)$ and the constant
 $q_n$, depending on the polytropic index $n$, can be found in
 (\ref{qup}) in \tx{Appendix A}.
In what follows, we will again take advantage of formulae which have
been derived in LRS93. These are again quantities entering the
Virial theorem: the gravitational potential energy
 \beq \label{riepoten}
  W = -\f{3}{5-n}\f{G M^2}{R}f(e_1,e_2)
   \eeq
    with
    \bdi
    f(e_1,e_2) =
    \f{1}{2}\left(A_1(1-e_1^2)^{-1/3}(1-e_2^2)^{-1/3} + \right.
   \edi
     \beq \label{aux}
     \left. + A_2(1-e_1^2)^{2/3}(1-e_2^2)^{-1/3} +
    A_3(1-e_1^2)^{-1/3}(1-e_2^2)^{2/3}\right)
     \eeq
      and $A_1, A_2, A_3$ as in \tx{Appendix A}. Furthermore, we will need the
total angular momentum
  \beq \label{angrie}
   \mb{L} = \f{\kappa_n}{5}M\left(\Omega(a_1^2+a_2^2) - 2
   a_1a_2\Lambda\right)\hat{\mb{z}}
    \eeq
 and the rotational kinetic energy
  \beq \label{rierot}
   T = \f{\kappa_n}{20}M(a_1-a_2)^2(\Omega + \Lambda)^2 +
   \f{\kappa_n}{20}M(a_1+a_2)^2(\Omega - \Lambda)^2,
    \eeq
 with $\kappa_n$ being another constant depending on the polytropic index $n$
(see \ref{kap}). Neglecting the quantum pressure by restricting to
the Thomas-Fermi regime will amount to setting $n=1$ in the above
formulae.

The irrotational Riemann-S ellipsoid experiences an internal motion
on top of the rigid rotation. Using (\ref{velo}) - (\ref{rieomega}),
its velocity field in the rotating frame can be derived as
  \bdi
   \mb{v}' = \f{2}{2-e_1^2}\Omega (y, -(1-e_1^2)x, 0) =
    \edi
   \beq \label{rievelo1}
   = 2\Omega_G\left(\f{2B_{12}}{q_n}\right)^{1/2}(8(1-e_1^2)+e_1^4)^{-1/2}(y,
   -(1-e_1^2)x, 0),
  \eeq
 while the velocity in the rest frame is accordingly
  \bdi
 \mb{v} = \mb{v}' + \mb{\Omega} \times \mb{r} = \Omega
 \f{e_1^2}{2-e_1^2}(y,x,0) =
  \edi
  \beq \label{rievelo2}
  = \Omega_G \left(\f{2B_{12}}{q_n}\right)^{1/2}(1+8(1-e_1^2)/e_1^4)^{-1/2}(y, x, 0).
 \eeq
Both expressions depend on the polytropic index $n$ via $q_n$.
Illustrative plots can be found in Fig.2. Note that the limit $e_1 =
0,~ e_2 \equiv e$ does \tx{not} reduce to the Maclaurin spheroid
case. Instead, the irrotational Riemann-S sequence bifurcates from
the non-rotating sphere. However, the sphere considered as the first
member of the irrotational sequence is viewed from a frame rotating
with angular velocity $\Omega$, which can be explicitly seen by
setting $e_1 = 0$ in the above expressions for the velocity,
resulting in $\mb{v} = \mb{0}$ and $\mb{v}' = -\mb{\Omega}\times
\mb{r}$ (see also LRS93 and their equ.(5.5) and (5.6), which reduce
to $L = 0$ and $T = 0$, hence $\mb{v} = \mb{0}$ in that case).

\begin{figure*}
     \begin{minipage}[b]{0.5\linewidth}
      \centering\includegraphics[angle=270,width=6cm]{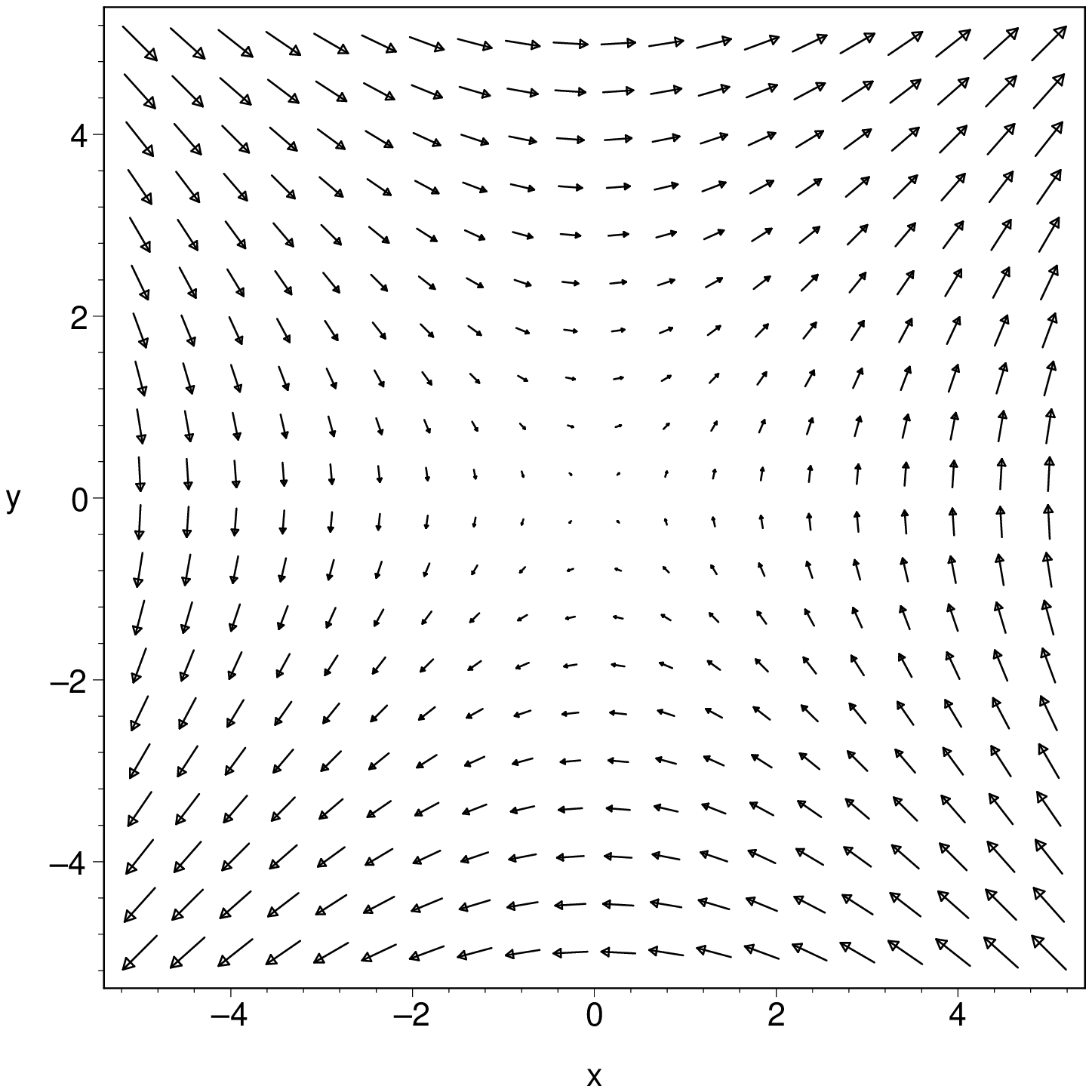}
     \hspace{0.1cm}
    \end{minipage}%
 \begin{minipage}[b]{0.5\linewidth}
      \centering\includegraphics[angle=270,width=6cm]{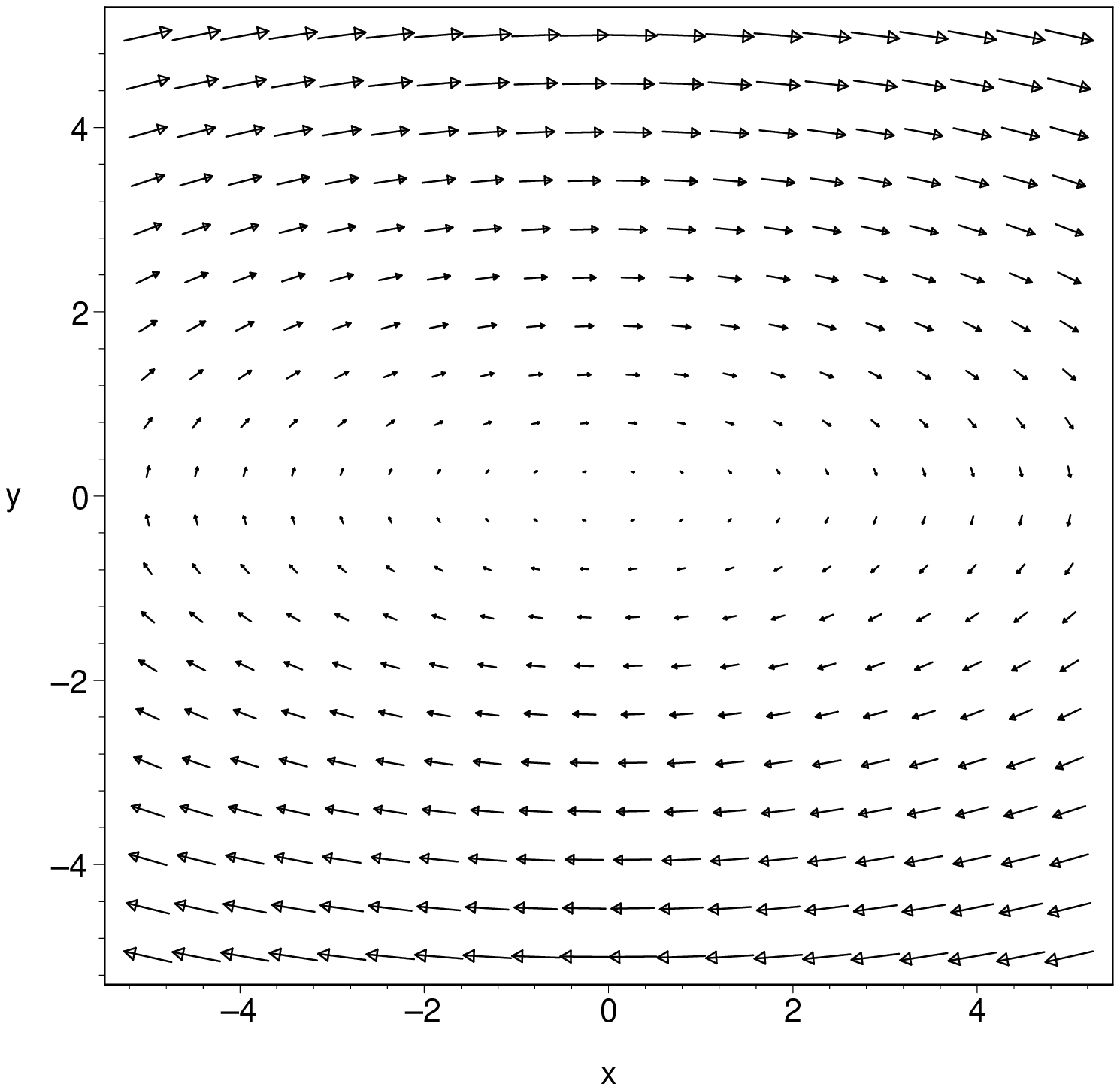}
     \hspace{0.1cm}
    \end{minipage}
 \caption{Illustrative velocity fields of an irrotational Riemann-S ellipsoid with $n=1$ and $\lambda = 0.05$ in the rest
 frame (\tx{left-hand-plot}) and in the co-rotating frame (\tx{right-hand-plot}).}
 \label{fig2}
\end{figure*}

\subsubsection{Comparison to CDM $\lambda$-spin parameter and virial
constraint}

In what will follow in the next section, we will also use the
Riemann-S ellipsoid as a model for rotating haloes in the
Thomas-Fermi regime. The condition of Virial equilibrium relates the
values of $y = m/m_H$ and $x = \sqrt{g/g_H}$ then to
($n=1$)-polytropes. For the general case of rotating haloes, we use
again
  (\ref{first}) and (\ref{zwischen}) along with formula (3.25) of LRS93, which
  relates the mean radius $R$ of the ellipsoid to the radius $R_0$ of the equilibrium spherical
  polytrope, equ.(\ref{onesphere}). For ($n=1$)\footnote{For ($n=0$), their formula (3.25) reduces to $R = R_0$. However,
  it would not have been correct to use this result for
     the BEC haloes we were considering in Section 3.1. due to our matching of the BEC pressure with the average pressure of an
     homogeneous spheroid, and so we needed to derive (\ref{nradius}).}, that formula becomes
 \beq \label{polyradius}
  R = R_0[f(e_1,e_2)(1-2t)]^{-1/2} \equiv R_0 g(e_1,e_2)^{-1/2},
  \eeq
 where the $t$-parameter and $g \equiv f(1-2t)$ depend on the eccentricities of
 the rotating figure only,
  \bdi
   t(e_1,e_2) = \f{\kappa_1}{5} \tilde \Omega^2
   (1-e_1^2)^{-1/3}(1-e_2^2)^{-1/3}\times
    \edi
    \beq \label{tparamrie}
    \times \left[\f{1}{2}(2-e_1^2) -
   \f{4(1-e_1^2)}{2-e_1^2} +
   \f{8(1-e_1^2)^2}{(2-e_1^2)^3}\right] |f(e_1,e_2)|^{-1}
   \eeq
    with $\tilde \Omega$ from (\ref{rieomega}), $f(e_1,e_2)$ in (\ref{aux}) and $\kappa_1$ in
    (\ref{kap}).
Inserting (\ref{polyradius}) into
   (\ref{zwischen}) and the result back into (\ref{first}) yields
   the Virial constraint
    \beq \label{compR}
  y(x) = \f{\pi}{\sqrt{8}}g(e_1,e_2)^{-1/2}x,
   \eeq
   with corresponding dimensional form
    \beq \label{compRdim}
     m(g) =
     \f{\pi}{\sqrt{8}}g(e_1,e_2)^{-1/2}\f{m_H}{\sqrt{g_H}}\sqrt{g},
     \eeq
  depending again only on the eccentricities $(e_1, e_2)$ and mean
  radius $R$ of the halo.
 Equ. (\ref{compR}) and (\ref{compRdim}) are the equivalent relations to (\ref{npoly}) and (\ref{npolydim}), constraining
  the BEC dark matter particle parameters such
 that rotating haloes as irrotational Riemann-S ellipsoids in the Thomas-Fermi regime fulfill virial equilibrium.
 Their formal equality does not come as a surprise, since the
 assumption of homogeneity in \tx{Section 3.1.2} results in the same
 neglect of $K_Q$ in (\ref{virial}) as in the Thomas-Fermi regime
 (albeit the latter does not neglect the spatial variation of the
 density, even though its Laplacian is likewise disregarded, as in the
 homogeneous case).

As in \tx{Section 3.1.2}, we shall now also derive a relationship
between the spin parameter $\lambda$ and the eccentricities of the
ellipsoid ($e_1, e_2$).  Using (\ref{riepoten}), (\ref{angrie}) and
(\ref{rierot}) for $n=1$,
 the corresponding ratio $2U_{SI}/T$ in (\ref{hilf}) is here given by
 \bdi
  \f{2U_{SI}}{T} = \f{2U_{SI}}{|W|t} =
   \edi
  \beq \label{ratio2}
  = \f{4}{3t}\f{g\rho_c^E}{2m^2}\f{R}{GM}|f(e_1,e_2)|^{-1} =
  \f{2}{3t}g(e_1,e_2)^{-1/2}f(e_1,e_2)^{-1},
   \eeq
   and we have also used (\ref{polyradius}) and (\ref{onesphere}).
   $\rho_c^E$ denotes the central density of the ellipsoid.
After some more algebra, we finally arrive at
 \bdi
  \lambda = \f{3\kappa_1}{20}(1-e_1^2)^{-1/3}(1-e_2^2)^{-1/3}(2-e_1^2)
  \left(1-\f{4(1-e_1^2)}{(2-e_1^2)^2}\right) \times
   \edi
    \beq \label{lambdarie}
   \times |f(e_1,e_2)|^{1/2}\tilde \Omega \left(t +
  \f{2f(e_1,e_2)^{-3/2}}{3(1-2t)^{1/2}}\right)^{1/2}
  \eeq
  with $\tilde \Omega$ in (\ref{rieomega}) and $t$ in (\ref{tparamrie}). We use this equation, along with
  (\ref{axisratios}), in order to solve for the eccentricities $(e_1,e_2)$ at given $\lambda$. Although highly
  nonlinear, involving many trigonometric and elliptic functions, equ.(\ref{lambdarie}) can be solved, in principle,
  as straightforwardly as the much simpler relationship (\ref{lambdaincomp}). The same values of
  $\lambda = (0.01, 0.05, 0.1)$ as in the previous section correspond now to eccentricities of $(e_1,e_2)=
 (0.707, 0.573), (e_1,e_2)=(0.881, 0.797)$ and $(e_1,e_2)=
 (0.934, 0.887)$, respectively. Note that even for small
  $\lambda$-values, the eccentricities are quite large, and surpass those in the
  case of the homogeneous spheroid.
 This is in part due to the fact that compressible bodies allow for
 higher eccentricities at a given angular momentum than
 incompressible ones do. A table of ellipsoid parameters as a
 function of $\lambda$ can be found in \tx{Appendix A}.

\section{Instability of rotating BEC haloes to vortex formation}

\subsection{Energy argument}

We shall use an energy argument in order to derive the critical
  angular velocity and energy for vortex creation in a rotating,
  self-gravitating BEC halo. Configurations rotating at an angular velocity $\Omega$ in the rest frame will be
  stationary solutions in the co-rotating frame. Therefore, we seek for vortex solutions which are (energetically)
  stable in the rotating frame. To this aim, we will compare the total energy with and without a vortex as measured in
   this frame. The equation of motion in the rotating frame is given by
equ.(\ref{gp}) with an additional operator $V_{rot} =
-\mathbf{\Omega}\cdot
 \mathbf{L}'$ on the right-hand-side under the brackets with angular momentum $\mathbf{L}' = -i\hbar
 \mathbf{r}'
 \times \nabla'$, and it is also understood that the primed
 variables and quantities are then those in the rotating frame.
 The GP energy functional is then
   \bdi
    \mathcal{E'}[\psi'] = \int_V \left[\frac{\hbar^2}{2m}
 |\nabla' \psi'|^2 + \frac{m}{2}\Phi |\psi'|^2 + \frac{g}{2}|\psi'|^4 - \right.
  \edi
   \begin{equation} \label{energie2}
  \left. - i\hbar \psi'^*\nabla' \psi' \cdot (\mb{\Omega}\times
 \mb{r}')\right]d^3\mb{r}'.
 \end{equation}
  In what follows, we shall use (\ref{energie2}) to determine at which
energy or angular velocity respectively, the presence of a vortex
starts to be energetically favoured, and to derive the vortex energy
as a function of the dark matter particle parameters.
\\
We will refer to a system without vortices (thus prior to vortex
formation) as the unperturbed system with
 $\psi'_0 = f'e^{iS_0'}$, and the unperturbed (vortex-free) halo density $|\psi'_0|^2=|f'|^2 = |f|^2$.
 The associated halo mass density and gravitational potential
 will be denoted as $\rho_0 = m|f|^2$ and $\Phi_0$, respectively. The
 corresponding phase $S_0'$ has no singularity in that case. These
 unperturbed haloes will be modelled by the ellipsoidal
 figures described in the previous section.
However, before we turn our attention to the energy analysis, we
highlight a necessary condition for vortex existence.

\subsection{Necessary minimum condition for vortex formation $L \geq L_{QM}$}

Applying a rotation with small enough, finite angular velocity to a
perfectly \tx{spherical-symmetric} BEC does not elevate its angular
momentum above zero. It is only above a critical value $\Omega_c$,
when a (singly-quantized) vortex starts to form, that the total
angular momentum is given by the amount necessary to sustain this
vortex, $L_{QM} = N \hbar$. However, non-spherical bodies carry
already a bulk angular momentum prior to vortex formation which is
responsible for their deformation, i.e. $L$ grows with $\Omega$ even
for $\Omega < \Omega_c$ (a very illustrative plot of such an
$L-\Omega$ relationship can be found in \cite{fetter}). We can
easily derive a relationship between the angular momentum $L$ of our
ellipsoidal haloes and the minimum angular momentum $L_{QM}$
necessary to sustain one vortex. In dividing equ.(\ref{macL}) by
$L_{QM}$ we have for the Maclaurin spheroid
 \bdi
  \f{L}{L_{QM}} = \f{m}{\hbar M}\f{2}{5}(1-e^2)^{-1/3}M R^2 \Omega
  = \f{2}{5}(1-e^2)^{-1/3} \f{m}{m_H}\f{\Omega}{\Omega_G}
  \edi
and by using (\ref{macomega}),
 \beq \label{macang}
   \f{L}{L_{QM}} = \f{m}{m_H}
   \f{2\sqrt{2}}{5(1-e^2)^{1/3}}\sqrt{A_1(e)-(1-e^2)A_3(e)}.
 \eeq
The equivalent relationship for the irrotational Riemann-S ellipsoid
using (\ref{angrie}) and (\ref{rieomega}) is
 \bdi
   \f{L}{L_{QM}} = \f{m}{m_H}
   \f{\kappa_n}{10}\f{2\tilde \Omega \sqrt{1-e_1^2}e_1^4}{(2-e_1^2)(1-e_1^2)^{5/6}(1-e_2^2)^{1/3}}
   \edi
   \bdi
    = \f{m}{m_H}\f{\kappa_n}{10}\times
     \edi
    \beq \label{rieang}
    \times \left(\f{2B_{12}}{q_n}\right)^{1/2}\left(2+\f{e_1^4}{4(1-e_1^2)}\right)^{-1/2}
    \f{e_1^4}{(1-e_1^2)^{5/6}(1-e_2^2)^{1/3}}.
 \eeq
 This shows that for fixed eccentricities (and fixed polytropic index in equ.(\ref{rieang})),
  the amount of angular momentum (in units of
 $L_{QM}$) of haloes depends only on the
 BEC-CDM particle mass (in units of $m_H$). That means that for a given shape
of the halo, the amount of angular momentum necessary to sustain a
vortex depends on the particle mass, such that only above a critical
mass has the system enough angular momentum to sustain the vortex.
At higher masses, even more angular momentum than needed can be
provided. We show in Fig.3, left-hand-plot, the ratio $m/m_{H}$ as a
function of $\lambda$ for the two figures considered in this paper,
for respective values of $L/L_{QM} = 1,10,100$. The lowest curve, $L
= L_{QM}$, establishes a lower bound on the particle mass for a
given halo's $\lambda$-value for vortex existence. We also show
$m/m_H$ as a function of $L/L_{QM}$ for fixed $\lambda = (0.01,
0.05, 0.1)$ (Fig.3, right-hand-plot). For a given $L/L_{QM}$, a
higher particle mass for vortex formation is required for haloes
with smaller spin-parameter.  \\
However, no bounds on the other BEC particle parameter, the
self-interaction strength $g/g_H$, can be determined from these
arguments, and only the energy analysis of the next subsection will
provide us with a sufficient condition on vortex formation,
constraining $g/g_H$ to be larger than a critical value
$(g/g_H)_{crit}$ for a given particle mass $m/m_H$, as we will show.

We will now investigate vortex formation for the two models for
rotating haloes described in \tx{Section 3}: \tx{Halo-Model A} is
referred to a halo which has much more angular momentum than the
minimum condition requires, $L \gg L_{QM}$. Since this is a regime
which can mimic solid-body rotation, we will use the Maclaurin
spheroid figures of \tx{Section 3.1}. The other model,
\tx{Halo-Model B}, has only enough angular momentum to support a
single vortex, i.e. the vortex takes up all of the angular momentum,
once it has formed. The underlying halo will be modelled by the
irrotational Riemann-S ellipsoids of \tx{Section 3.2} in this case.

\begin{figure*}
\begin{minipage}[b]{0.5\linewidth}
    \centering\includegraphics[width=8cm]{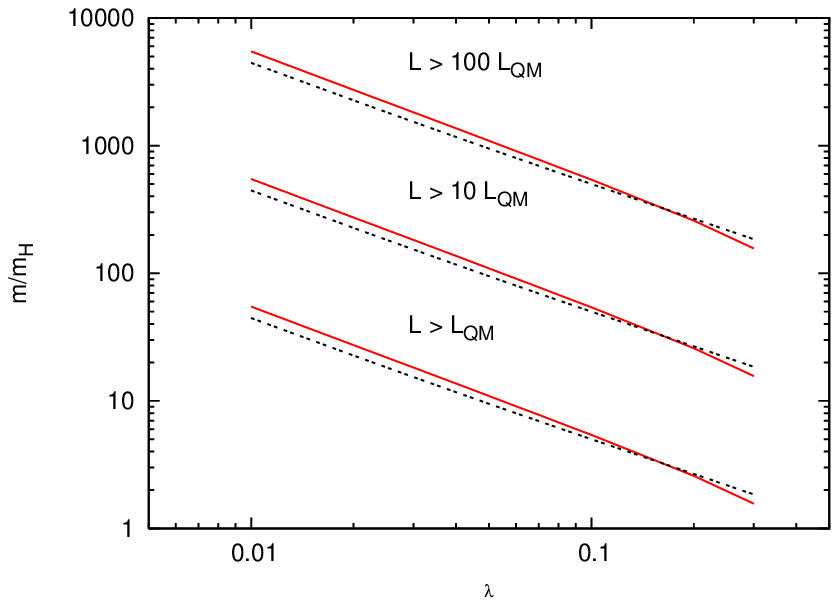}
     \hspace{0.1cm}
    \end{minipage}%
     \begin{minipage}[b]{0.5\linewidth}
      \centering\includegraphics[width=8cm]{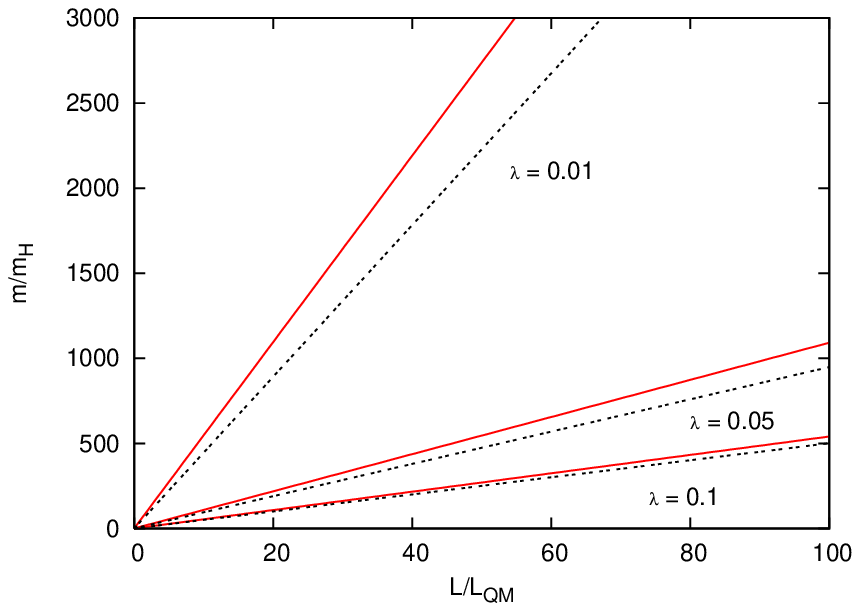}
     \hspace{0.1cm}
    \end{minipage}
 \caption{Dimensionless BEC-CDM particle mass $m/m_H$ and halo angular momentum:
 \tx{Left-hand-plot:} $m/m_H$ versus $\lambda$-spin parameter for
homogeneous Maclaurin spheroids (solid) and
 $(n=1)$-polytropic, irrotational Riemann-S ellipsoids (dashed) having $L/L_{QM} = 1,10,100$ (lower to upper curves).
 \tx{Right-hand-plot:} $m/m_H$ versus $L/L_{QM}$ for $\lambda = 0.01, 0.05, 0.1$ for Maclaurin
 spheroids (solid) and Riemann-S ellipsoids (dashed).}
 \label{fig3}
\end{figure*}

\subsection{Halo-Model A: $L \gg L_{QM}$}

\subsubsection{Energy splitting and vortex ansatz for the Maclaurin spheroid}

For brevity, we will omit the primes on variables in the forthcoming
sections \tx{except} for the phase functions and energies in order
to avoid a confusion between rotating and rest frames. In order to
pursue the energy analysis, we will decompose the halo wave function
$\psi$
  into a vortex-free part, $\psi_0 = fe^{iS_0'}$, and a
vortex part, $w = |w|e^{iS_1'}$, according to
 \beq \label{wsp}
 \psi = |\psi|e^{iS'} = \psi_0 w =
f|w|e^{i(S_0'+S_1')},
 \eeq
  where the corresponding amplitude $|\psi|=f|w|$ is a product state of unperturbed (vortex-free)
  and perturbed (vortex-carrying) parts, while the contributions in the phase are additive,
  $S'=S_0'+S_1'$. We will insert this ansatz for the perturbed dark
  matter halo into the energy functional in equ. (\ref{energie2}). This will lead to a convenient splitting of the
energy contributions, allowing us to compare them more easily. As a
prerequisite, we note that for vortex-free haloes, equ.(\ref{hd1})
and (\ref{hd2}) in the rotating frame are
 \bdi
  \Delta f - f(\nabla S_0')^2 -
\f{2m}{\hbar^2}f (m\Phi + g f^2 - \mu) +
 \edi
  \beq \label{fsa}
  + \f{2m}{\hbar}f \nabla S_0'
\cdot (\mb{\Omega} \times \mb{r}) = 0
 \eeq
   and
  \beq \label{conta}
\nabla \cdot \left[f^2\left(\nabla S_0' - \f{m}{\hbar}\mb{\Omega}
\times \mb{r}\right)\right] = 0 \eeq
 where $\mu$ is the associated chemical potential to the
 normalization $\int_V f^2 = N = \int_V |\psi|^2$.
 The corresponding energy functional for unperturbed haloes is given by
\bdi
 \gpf'[fe^{iS_0'}] = \int_V
\left\{\frac{\hbar^2}{2m}(\nabla f)^2 + f^2\left(\f{m}{2}\Phi +
\f{g}{2}f^2\right) + \right.
 \edi
  \beq \label{vofree}
  \left. + \frac{\hbar^2}{2m}f^2 \nabla S_0' \cdot
\left(\nabla S_0' - \f{2m}{\hbar}\mb{\Omega}\times \mb{r}\right)
\right\}d^3\mb{r}.
 \eeq
We refer the reader to \tx{Appendix B} for the derivation of the
energy splitting. It is shown there that for $\psi$ in (\ref{wsp}),
the associated Gross-Pitaevskii energy can be written as
 \beq \label{splitting}
\gpf'[\psi] = \gpf'[fe^{iS_0'}] + \mc{G}_f'[w] - \mc{R}_f'[w]
 \eeq
 with $\gpf'[fe^{iS_0'}]$ in (\ref{vofree}),
  \bdi
 \mc{G}_f'[w] \equiv \int \left(\frac{\hbar^2}{2m}f^2|\nabla w|^2 +
\frac{g}{2}f^4(1-|w|^2)^2\right)d^3\mb{r}  +
 \edi
  \beq \label{gl}
   + \int
\left(\f{m}{2}f^2\Phi_0 + \f{m}{2}f^2|w|^2\left[\Phi -
2\Phi_0\right]\right)d^3\mb{r}
 \eeq
 and
  \beq \label{rterm}
  \mc{R}_f'[w] \equiv \f{\hbar^2}{m}\int if^2 w^* \nabla w \cdot \left(\nabla S_0' -
\f{m}{\hbar}\mb{\Omega}\times \mb{r}\right)d^3\mb{r}.
 \eeq
 The terms apart from $\mc{E'}[fe^{iS_0'}]$
 describe the contribution of vortices to the energy. So, using the decomposition (\ref{wsp}) vortices of
$\psi$ (if present) are vortices of $w$ and they are described via
the energy functionals $\mc{G}_f'[w] - \mc{R}_f'[w]$ in
(\ref{splitting}). A similar splitting was deduced in \cite{aftdu}
and \cite{rindler} for laboratory BECs in the Thomas-Fermi regime
$R/\xi \gg 1$. However, it can be easily seen that this splitting
holds for any values of $R/\xi \geq 1$. There is, however, a notable
change in the form of an additional term in $\mc{G}_f'[w]$ which
stems from the fact that the external trap for atomic gases is
replaced here by the gravitational potential, which depends on the
density profile via Poisson's equation and is hence affected by the
presence of vortices. For laboratory BECs, on the other hand, the
trap potential is fixed from the outset by the adopted laser
configuration, and $\mc{G}_f'[w]$ does not contain the potential
explicitly.

We want to determine the critical angular velocity and energy above
which at least one vortex in the halo will form. In order to derive
an analytical result and to be as general as possible in the same
time, we will consider a $d$-quantized straight, axisymmetric
vortex, along the rotation-axis with core radius $s$. This core
radius is given by that (cylindrical) radius, where the density
recovers back from the inner vortex profile to its unperturbed bulk
value. The most general ansatz for the wavefunction of such a vortex
is $w = |w|(r)e^{id\phi}$, where the modulus depends only on the
radial variable. Inserting this ansatz into the Gross-Pitaevskii
equation, one may solve numerically for the density profile of the
halo in the presence of a vortex. However, it can be shown that for
gravitational potentials $\Phi$ falling off as $-1/r^b, b \leq 2$ at
infinity\footnote{More precisely, we require that if $O(\mu-\Phi)
\sim 1/r^b$, then $b \leq 2$.}, this profile goes like $r^{|d|}$ for
$r \to 0$, after imposing the constraint that it approaches the
unperturbed density for $r \to \infty$. This is the well-known
behaviour valid also for trap potentials used in atomic BEC gases.
So, the density of the halo will tend to zero towards the axis of
the vortex. For our purposes, it will be sufficient to catch this
behaviour by the simpler profile we are going to adopt. Our ansatz
for the vortex is
  \beq \label{testf}
   \tilde{w}(r,\phi) = |\tilde{w}|(r)e^{i S_1'}
    \eeq
  with amplitude
   \beq \label{vortexamplitude}
    |\tilde{w}|(r) =
\left\{\begin{array}{ll}
 1 & \textrm{for $r \geq s$}\\
 C_n\left(\frac{r}{s}\right)^d & \textrm{otherwise}
 \end{array} \right.
  \eeq
 and phase $S_1' = d\phi$.
 This amplitude reflects the fact that outside of the vortex, the halo (number) density
 is given by the unperturbed profile $|f|^2 = \rho_0/m$. Since $\tilde w$ is going
 to be multiplied by the unperturbed wave function according to the above
 decomposition (\ref{wsp}), $\tilde \psi = \psi_0\tilde{w}$, its amplitude is dimensionless and the
 $z$-dependence is trivial for the straight vortex. We think that the consideration of a bent vortex will not
 change our conclusions given the orders of magnitudes involved in the
 final results.
 The constant $C_n$ is determined by the normalization $\int
 |\tilde \psi|^2 = N$. Although the mass is conserved,
 the above ansatz for the vortex causes an unnatural steepening or 'overshooting' of
 the profile at the core radius, which would cause a singularity in
 the dynamical equations. However, this feature is invisible to the energy
 calculation we are going to perform.

\begin{figure*}
      \centering\includegraphics[width=4cm,height=5cm]{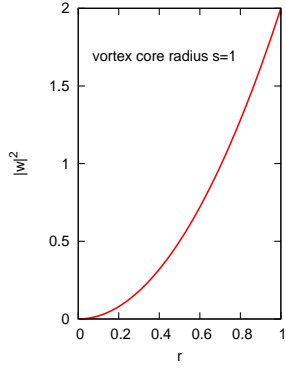}
 \caption{(Number) density profile of the vortex ansatz of equ.(\ref{vortexamplitude}) for $d=1$.}
 \label{fig4}
\end{figure*}

The vortex changes the gravitational potential of the smooth,
vortex-free halo (\ref{macpotential}). Owing to the linearity of the
Poisson equation, we may decompose the total halo potential and mass
density into unperturbed and perturbed parts,
 \bdi
  \Phi = \Phi_0 + \Phi_1,~~\rho = \rho_0 + \rho_1,
  \edi
 where $\Phi_1$ is the associated potential to the perturbation of the density due to the vortex,
 \bdi
  \rho_1 = \rho - \rho_0 = \rho_0(|\tilde w|^2-1) =
   \edi
   \beq \label{vortexansatz}
= \left\{\begin{array}{ll}
 0 & \textrm{outside the vortex}\\
 \rho_0\left(C_n^2\left(\frac{r}{s}\right)^{2d}-1\right) < 0 & \textrm{otherwise}
 \end{array} \right.
  \eeq
according to (\ref{testf}), and $\Delta \Phi = \Delta \Phi_0 +
\Delta \Phi_1 = 4\pi G (\rho_0 + \rho_1)$. So, we are left to solve
for the unknown potential of the vortex configuration,
 \bdi
  \Delta \Phi_1 = 4\pi G \rho_1 =
   \edi
  \beq \label{vortdiff}
 = \left\{\begin{array}{ll}
 0 & \textrm{outside the vortex}\\
 4\pi G \rho_0\left(C_n^2\left(\frac{r}{s}\right)^{2d}-1\right) < 0 &
 \textrm{otherwise}.
 \end{array} \right.
 \eeq
The first case amounts simply to solving the Laplace equation, while
the second case is an inhomogeneous extension. Since we only
encounter axisymmetric configurations, both cases can be solved
analytically. However, since the solution for $d > 1$ happens to be
a very cumbersome expression involving numerous hypergeometric
functions, we will in the forthcoming analysis restrict our
attention to singly-quantized vortices having $d=1$, whose
corresponding profile can be found in Fig.\ref{fig4}. In fact,
physical reasoning makes the ($d=1$)-vortex more interesting, since
multiply-quantized vortices are generally subject to splitting into
several singly-quantized vortices. The above differential equations
(\ref{vortdiff}) can be solved in a standard way, and the subtle
issue remaining is the choice of suitable boundary conditions. For
the outer-vortex solution, we impose that the potential approaches a
point-mass potential for large $r$ at fixed $z$, or large $z$ at
fixed $r$, respectively, resulting into
 \beq \label{outerphi}
  \Phi_1^{(o)}(r,z) = \f{2\pi G \rho_0 c s^2}{\sqrt{r^2+z^2}}.
  \eeq
For the inner-vortex solution, we require the gradient of the
potential to vanish, i.e. no net gravitational force, at the centre,
resulting in
  \beq  \label{innerphi}
 \Phi_1^{(i)}(r,z) = \pi G \rho_0 r^2
 \left(\f{1}{2}\left(\f{r}{s}\right)^2-1\right).
  \eeq
The total halo potential in the presence of this vortex is then
given by \beq \label{totpot}
 \Phi =
\left\{\begin{array}{ll}
 \Phi_0 + \Phi_1^{(o)} & \textrm{outside the vortex}\\
 \Phi_0 + \Phi_1^{(i)} & \textrm{otherwise}
 \end{array} \right.
 \eeq
 with the potentials given in (\ref{macpotential}),
 (\ref{outerphi}) and (\ref{innerphi}).

In \tx{Halo-Model A}, the vortex is essentially considered to be a
perturbation of the total angular momentum $L$ of the halo. Since
the latter is conserved, that part carried by the vortex, $L_{QM}$,
is assumed to be small compared to $L$, such that the total angular
momentum of the system before \tx{and} after vortex formation can be
given by $L$ in (\ref{macL}).

\subsubsection{Vortex energy and critical angular velocity}

\begin{figure*}
 \begin{minipage}[b]{0.5\linewidth}
  \centering\includegraphics[width=8.5cm]{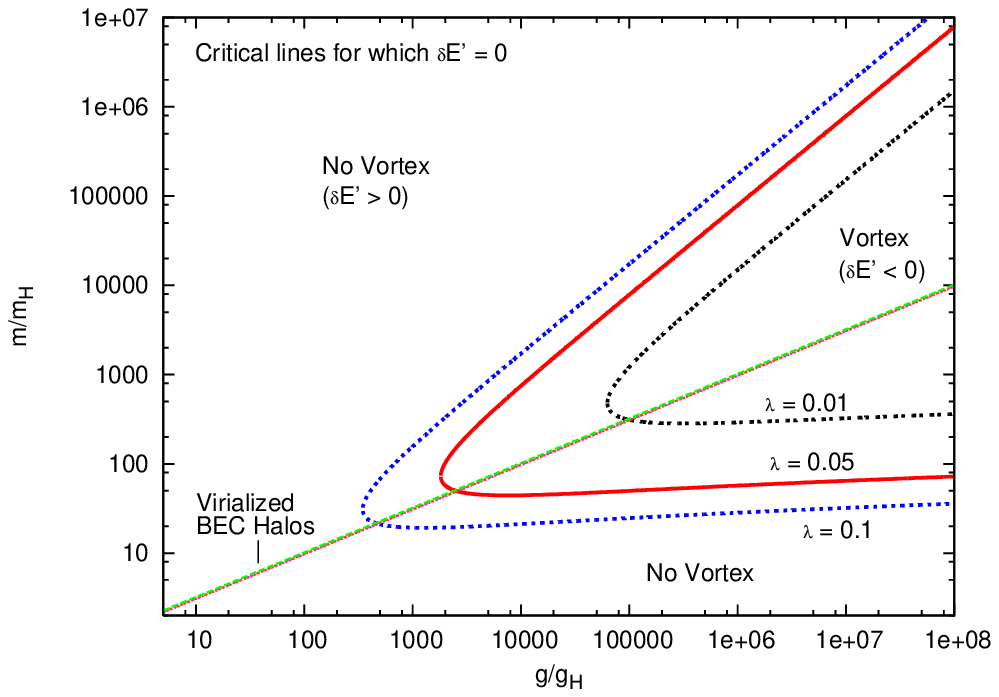}
   \hspace{0.1cm}
    \end{minipage}%
     \begin{minipage}[b]{0.5\linewidth}
      \centering\includegraphics[width=8.5cm]{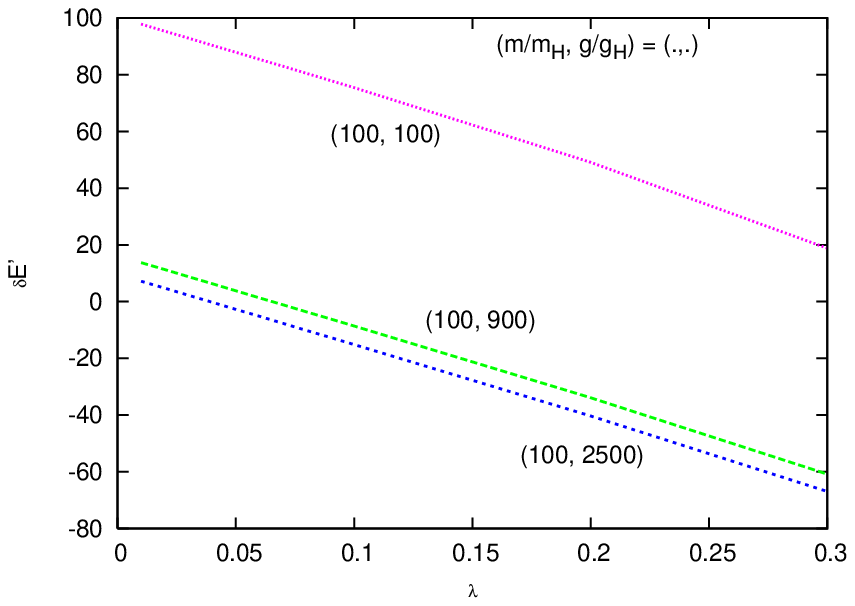}
     \hspace{0.1cm}
    \end{minipage}
\caption{BEC-CDM particle parameter space and vortex energy for
Halo-Model A: \tx{Left-hand-plot:} Critical curves $\delta E' = 0$
(or $\Omega = \Omega_c$) for vortex formation in the dimensionless
parameter space $(g/g_{H}, m/m_H)$ for $\lambda = 0.01 ~(e=0.062)$
(long-dashed black),
 $\lambda = 0.05 ~(e=0.302)$ (solid red), $\lambda = 0.1 ~(e=0.550)$
 (short-dashed blue);
 BEC haloes constrained by the Virial theorem: straight lines for the same $e$-values.
 \tx{Right-hand-plot:} Vortex energy in the rotating frame in units of $\Omega_{QM}L_{QM}$ versus $\lambda$-spin
 parameter for three different BEC-CDM particle models
$(m/m_H, g/g_H) = (100,100), (100,900), (100,2500)$.}
 \label{fig5}
\end{figure*}

Our prescription for the halo wave function is thus $\tilde \psi =
\psi_0\tilde w$ with the unperturbed wave function $\psi_0 =
fe^{iS_0'}$ having the above described geometry and rotational
properties of the Maclaurin spheroid and the
 vortex ansatz $\tilde w$ from (\ref{testf}). The normalization
 constant is then $C_n = \sqrt{d+1}$.
 Inserting this wave function ansatz into
  equ.(\ref{splitting}) will give us the critical angular velocity $\Omega_c$ above which the energy is lowered
  by the presence of that vortex. Since our ansatz leads to an energy greater than that
  which would result if the 'real' wave function were used in
place of the ansatz, this means
  vortex creation is energetically favoured, in general, if $\Omega >
  \Omega_c$. More precisely, we have the upper bound
   \bdi
    \gpf'[\psi] \leq \gpf'[\tilde{\psi}] = \gpf'[fe^{iS_0'}] +
    \glf'[\tilde w] - \rotf'[\tilde w] =
    \edi
   \bdi
   = \gpf'[fe^{iS_0'}] +
    \int \left(\frac{\hbar^2}{2m}f^2|\nabla \tilde{w}|^2 +
\frac{g}{2}f^4(1-|\tilde{w}|^2)^2 + \right.
 \edi
  \bdi
  \left. + \f{\rho_0}{2}\Phi_0 (1-|\tilde
w|^2) + \f{\rho_0 |\tilde w|^2}{2}\Phi_1\right)d^3\mb{r}-
 \edi
 \beq \label{preenergie}
   - \f{\hbar^2}{m}f^2 \int i \tilde{w}^* \nabla \tilde{w} \cdot \left(\nabla S_0' -
\f{m}{\hbar}\mb{\Omega}\times \mb{r}\right)d^3\mb{r}.
 \eeq
 Since $|f|^2 = \rho_0/m$, is constant, the integrals can be performed straightforwardly. Furthermore,
  $\nabla S_0' = 0$ since the spheroid has no net velocity in the rotating frame, $\mb{v}' = \mb{0}$.
 Let us consider each term separately, where we split the axisymmetric integration domain according to our ansatz:
 the quantum-kinetic term is affected by the presence of the vortex, so
  \bdi
   \int \f{\hbar^2}{2m}f^2 |\nabla \tilde{w}|^2 d^3\mb{r} =
   \edi
  \bdi
   = \f{\hbar^2}{2m}f^2 \int_{z=-c}^c \int_0^{2\pi} \int_{r=s}^{a\sqrt{1-\f{z^2}{c^2}}} |\nabla
   \tilde{w}|^2 r dr dz d\phi +
    \edi
   \bdi
   + \f{\hbar^2}{2m}f^2 \int_{z=-c}^c \int_0^{2\pi} \int_{r=0}^s |\nabla
   \tilde{w}|^2 r dr dz d\phi =
    \edi
 \beq \label{firstenergy}
  = \f{2\pi c f^2 \hbar^2 d^2}{m}\left[\ln \f{a}{s} + \ln 2 - 1 +
  \f{d+1}{d}\right],
  \eeq
   where the leading logarithmic term stems from the
   angular kinetic energy of the vortex. The term due to the self-interaction amounts to
 \beq \label{secondenergy}
 \int_{-c}^c \int_0^{2\pi} \int_0^s \f{g}{2}f^4 (1-|\tilde w|^2)^2 r dr d\phi dz = g f^4 \pi c s^2
  \f{d^2}{2d+1},
   \eeq
   while the rotation term becomes
  \bdi
  \rotf'[\tilde w] = \hbar f^2 \int_{-c}^c \int_0^{2\pi} \int_0^{a\sqrt{1-\f{z^2}{c^2}}} i \tilde{w}^* \nabla \tilde{w} \cdot
  (\mb{\Omega}\times \mb{r}) rdrd\phi dz =
   \edi
    \beq \label{thirdenergy}
    = \f{4}{3}\pi \hbar f^2  c a^2 d\Omega.
   \eeq
Inserting (\ref{outerphi}) - (\ref{totpot}) and (\ref{macpotential})
into the second integral of (\ref{gl}), the gravitational potential
energy due to the $(d=1)$-vortex perturbation is finally given by
 \bdi
  \int_{-c}^c \int_0^{2\pi} \int_{s}^{a\sqrt{1-\f{z^2}{c^2}}}
  \f{\rho_0}{2}\Phi_1^{(o)} r dr d\phi dz  +
   \edi
  \bdi
   + \int_{-c}^c\int_0^{2\pi}\int_{0}^s \left[\f{\rho_0}{2}\Phi_0
  \left(1-C_N^2\left(\f{r}{s}\right)^{2d}\right) + \right.
   \edi
   \bdi
  \left. + \f{\rho_0}{2}C_N^2\left(\f{r}{s}\right)^{2d}\Phi_1^{(i)}\right]r
  dr d\phi dz =
   \edi
  \bdi
   = 2\pi^2G\rho_0^2cs^2\left[c^2+\f{ac}{e}\arcsin(e) -
   c^2\sqrt{1+\left(\f{s}{c}\right)^2} - \right.
    \edi
    \beq \label{fourthenergy}
    \left. - s^2\ln
   \left(\f{c}{s}+\sqrt{\left(\f{c}{s}\right)^2+1}\right)\right] -
   \f{\pi^2G\rho_0^2}{6}s^4c(A_1(e)+5/2).
    \eeq
   The above terms constitute the total energy $\gpf'[\tilde \psi]$ of our ansatz for the wave function
 $\tilde \psi = \psi_0\tilde w$.
Before we collect the above terms, we set $d=1$ everywhere and
rewrite the semi-axes $a$ and $c$
   in terms of the mean radius $R$ and the eccentricity $e$. Also,
   we use (\ref{oqm}) in order to finally write the
   vortex energy in the rotating frame as
     \bdi
 \delta E' \equiv \mc{G}_f'[\tilde{w}] -
 \mc{R}_f'[\tilde{w}]
  \edi
  or
\bdi
 \f{\delta E'}{\Omega_{QM}L_{QM}} =
  \edi
   \bdi
   = \f{3}{2}(1-e^2)^{1/3}\left[\ln
 \left(\f{R}{\xi}\right) + \ln \left(2(1-e^2)^{-1/6}\right) +
 \f{13}{12}\right] +
  \edi
  \bdi
  +
  \f{3}{2}\left(\f{\Omega_G}{\Omega_{QM}}\right)^2\left(\f{\xi}{R}\right)^2(1-e^2)^{1/3}\times
   \edi
    \bdi
   \times \left[(1-e^2)^{2/3}+
  \f{\arcsin(e)}{e}(1-e^2)^{1/6}- \right.
   \edi
  \bdi
   \left. - \left(\f{\xi}{R}\right)(1-e^2)^{1/3}\sqrt{1+\left(\f{R}{\xi}\right)^2(1-e^2)^{2/3}}\right.
-
   \edi
    \bdi
  - \left.\left(\f{\xi}{R}\right)^2\ln\left(\f{R}{\xi}(1-e^2)^{1/3} +
  \sqrt{1+\left(\f{R}{\xi}\right)^2(1-e^2)^{2/3}}\right)\right] -
   \edi
 \beq \label{vortexenergie1}
   - \f{1}{8}\left(\f{\Omega_G}{\Omega_{QM}}\right)^2\left(\f{\xi}{R}\right)^4(1-e^2)^{1/3}\left(A_1(e)+\f{5}{2}\right)
   - \f{\Omega}{\Omega_{QM}}.
   \eeq
For singly-quantized vortices, the core radius $s$ is almost the
same size than the healing length $\xi$ (see e.g. \cite{PS}), and
can be very well approximated by it, so we have replaced $s$ by
$\xi$ in the above expressions altogether. We have also used
(\ref{healing}) to replace the coupling strength $g$ by the healing
length $\xi$.\\
 We can rewrite the variables appearing in (\ref{vortexenergie1}) in
   terms of the BEC particle parameters $m$ and $g$ using
   (\ref{mmin})-(\ref{freefall}) and
   (\ref{gmin}) - (\ref{becdef}) in order to arrive
  at
  \bdi
 \f{\delta E'}{\Omega_{QM}L_{QM}} =
  \edi
  \bdi
  = \f{3}{2}(1-e^2)^{1/3}\left[\ln
 \f{1}{2}\left(\f{g}{g_H}\right) + \ln \left(2(1-e^2)^{-1/6}\right) +
 \f{13}{12}\right] +
  \edi
  \bdi
  +
  \f{3}{2}\left(\f{m}{m_H}\right)^2\f{g_H}{g}(1-e^2)^{1/3}\left[(1-e^2)^{2/3}+
  \f{\arcsin(e)}{e}(1-e^2)^{1/6}- \right.
   \edi
   \bdi
   \left. - \left(\f{g_H}{g}\right)^{1/2}(1-e^2)^{1/3}\sqrt{1+\f{g}{g_H}(1-e^2)^{2/3}}
  - \right.
   \edi
     \bdi
   - \left.\f{g_H}{g}\ln\left(\sqrt{\f{g}{g_H}}(1-e^2)^{1/3} +
  \sqrt{1+\f{g}{g_H}(1-e^2)^{2/3}}\right)\right] -
   \edi
  \beq \label{vortexenergie2}
   - \f{1}{8}\left(\f{m}{m_H}\right)^2\left(\f{g_H}{g}\right)^2(1-e^2)^{1/3}\left(A_1(e)+\f{5}{2}\right)
   - \tilde \Omega(e)\f{m}{m_H}
   \eeq
with $\tilde \Omega$ in (\ref{macomega}). Now, the total energy
(\ref{preenergie}) will be \tx{lower} than the vortex-free,
unperturbed energy
 $\gpf'[f e^{iS_0'}]$ \tx{if} the vortex energy in the rotating frame $\delta E'$
  becomes smaller than zero, i.e. if the system with vortex is
  energetically favoured.
The critical condition amounts to setting $\delta E' = 0$ in
(\ref{vortexenergie2}) and finding the respective relationship
between the particle parameters $y = m/m_H$, $x = \sqrt{g/g_H}$ for
fixed halo eccentricity. We see that the critical curves are just
the two solution branches
 of a quadratic equation
  \beq \label{critcurve}
  y_{1,2}(x) = b/(2a(x))[1 \pm\sqrt{1-4a(x)c(x)/b^2}]
   \eeq
  with
\bdi
 a(x) \equiv \f{1}{x^2}\left[(1-e^2)^{2/3} + \f{\arcsin(e)}{e}(1-e^2)^{1/6}
 - \right.
  \edi
  \bdi
 \left. - \f{(1-e^2)^{1/3}}{x^2}\sqrt{1+x^2(1-e^2)^{2/3}}\right. -
  \edi
  \bdi
 - \left.\f{1}{x^2}\ln\left(x(1-e^2)^{1/3} +
 \sqrt{1+x^2(1-e^2)^{2/3}}\right)\right] -
  \edi
  \beq \label{c1}
 - \f{A_1(e)+5/2}{12}\f{1}{x^4},
  \eeq
 \beq \label{c2}
  b \equiv \f{2}{3}\tilde \Omega(1-e^2)^{-1/3},
   \eeq
and
 \beq \label{c3}
  c(x) \equiv \ln x + \ln \left(2(1-e^2)^{-1/6}\right) + \f{13}{12}.
   \eeq
 For a given eccentricity of the
 halo, these curves constrain the allowed space for the
 BEC-CDM particle parameters $(m/m_H, g/g_H)$ for vortex formation,
 for which there is a minimum allowed value of $g/g_{H} > 1$,
according to (\ref{gmin}) (see Fig.\ref{fig5}). This is reasonable
since it makes no sense for a vortex core radius $\xi$ to be
 indefinitely large. Equ.(\ref{critcurve}) also
describes the critical curve $\Omega = \Omega_c$ in this parameter
space\footnote{We can also solve for the critical angular velocity
of the halo, $\Omega_c$, above which a vortex is energetically
  favoured:
 \bdi
\Omega_c =
\Omega_{QM}\left\{\f{3}{2}(1-e^2)^{1/3}\left[\ln\left(\f{R}{\xi}\right)
+ \ln \left(2(1-e^2)^{-1/6}\right) + \f{13}{12}\right] + \right.
 \edi
  \bdi
\left. +
\f{3}{2}(1-e^2)^{1/3}\left(\f{\xi}{R}\right)^2\left(\f{\Omega_G}{\Omega_{QM}}\right)^2\left[(1-e^2)^{2/3}+
\right.\right.
 \edi
  \bdi
 \left. \left. + \f{\arcsin(e)}{e}(1-e^2)^{1/6} - \f{\xi}{R}(1-e^2)^{1/3}\sqrt{\left(\f{R}{\xi}\right)^2(1-e^2)^{2/3}+1}-
\right. \right.
 \edi
   \bdi
  \left.\left. - \left(\f{\xi}{R}\right)^2\ln\left(\f{R}{\xi}(1-e^2)^{1/3} +
\sqrt{\left(\f{R}{\xi}\right)^2(1-e^2)^{2/3}+1}\right)\right] -
\right.
 \edi
  \beq \label{omega1}
 \left. - \f{1}{8}(1-e^2)^{1/3}(A_1(e)+5/2)\left(\f{\Omega_G}{\Omega_{QM}}\right)^2\left(\f{\xi}{R}\right)^4\right\}.
  \eeq
This expression differs from the result given in \cite{RS}, since we
have taken here into account the gravitational potential of the
vortex more carefully. It can be shown that the general expression
for equ.(\ref{omega1}) is a monotonically increasing
  function of $d$, so our restriction to the lowest
  critical angular velocity for which there is $d=1$ is
  well-motivated, after all.
 In the Thomas-Fermi regime, as $R/\xi \to \infty$, the leading-order term in the critical angular velocity diverges
logarithmically, $\Omega_c \simeq \Omega_{QM} \ln(R/\xi)$,
 just as for laboratory condensates (see e.g. \cite{lundh} or
 \cite{PS}).}.

The results strongly suggest that BEC dark matter particles with
smaller coupling constant, especially those without
self-interaction, $g \equiv 0$, are \tx{not} able to form a vortex.

\begin{figure*}
\begin{minipage}[b]{0.5\linewidth}
      \centering\includegraphics[angle=270,width=7.8cm]{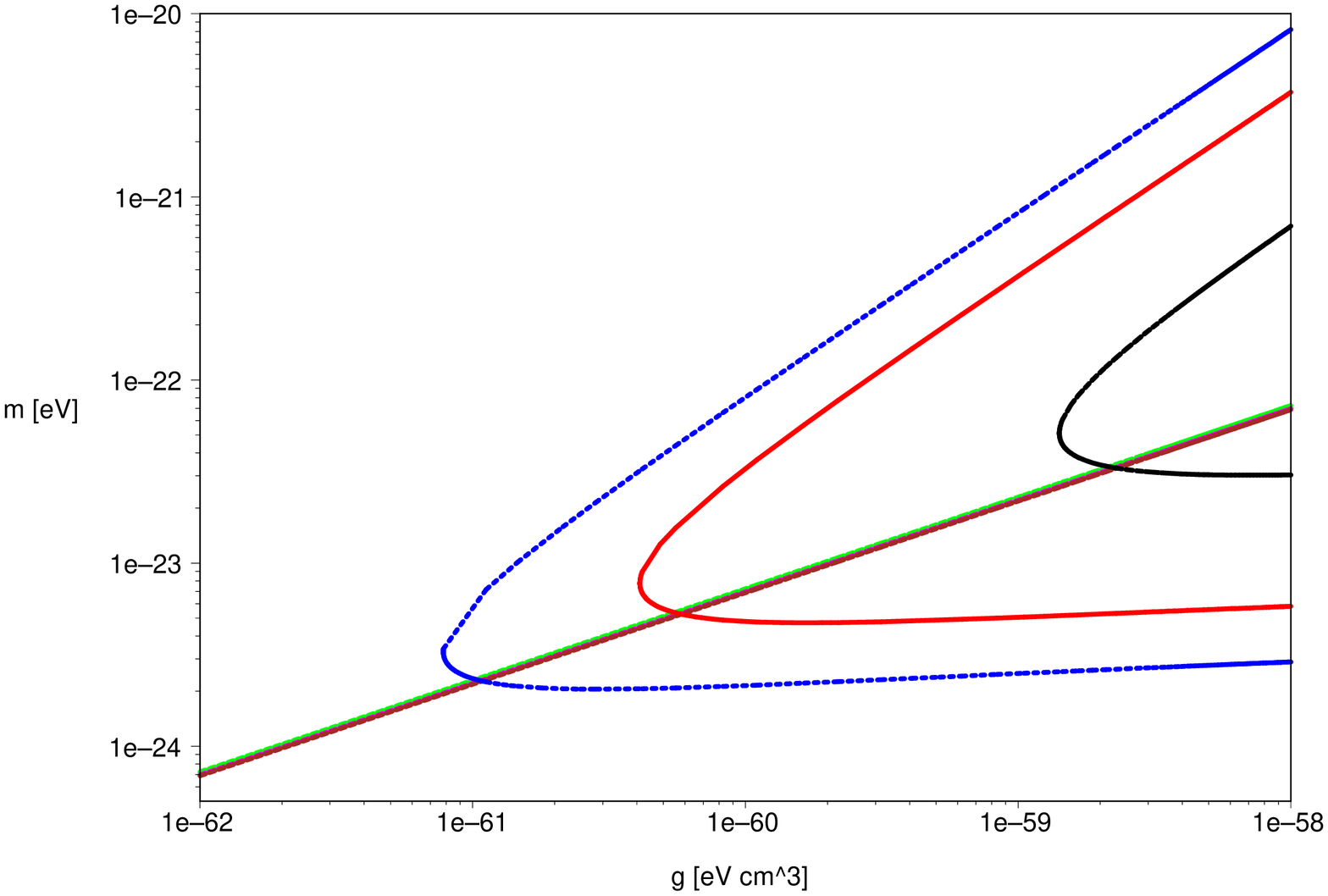}
     \hspace{0.1cm}
    \end{minipage}%
     \begin{minipage}[b]{0.5\linewidth}
      \centering\includegraphics[angle=270,width=7.8cm]{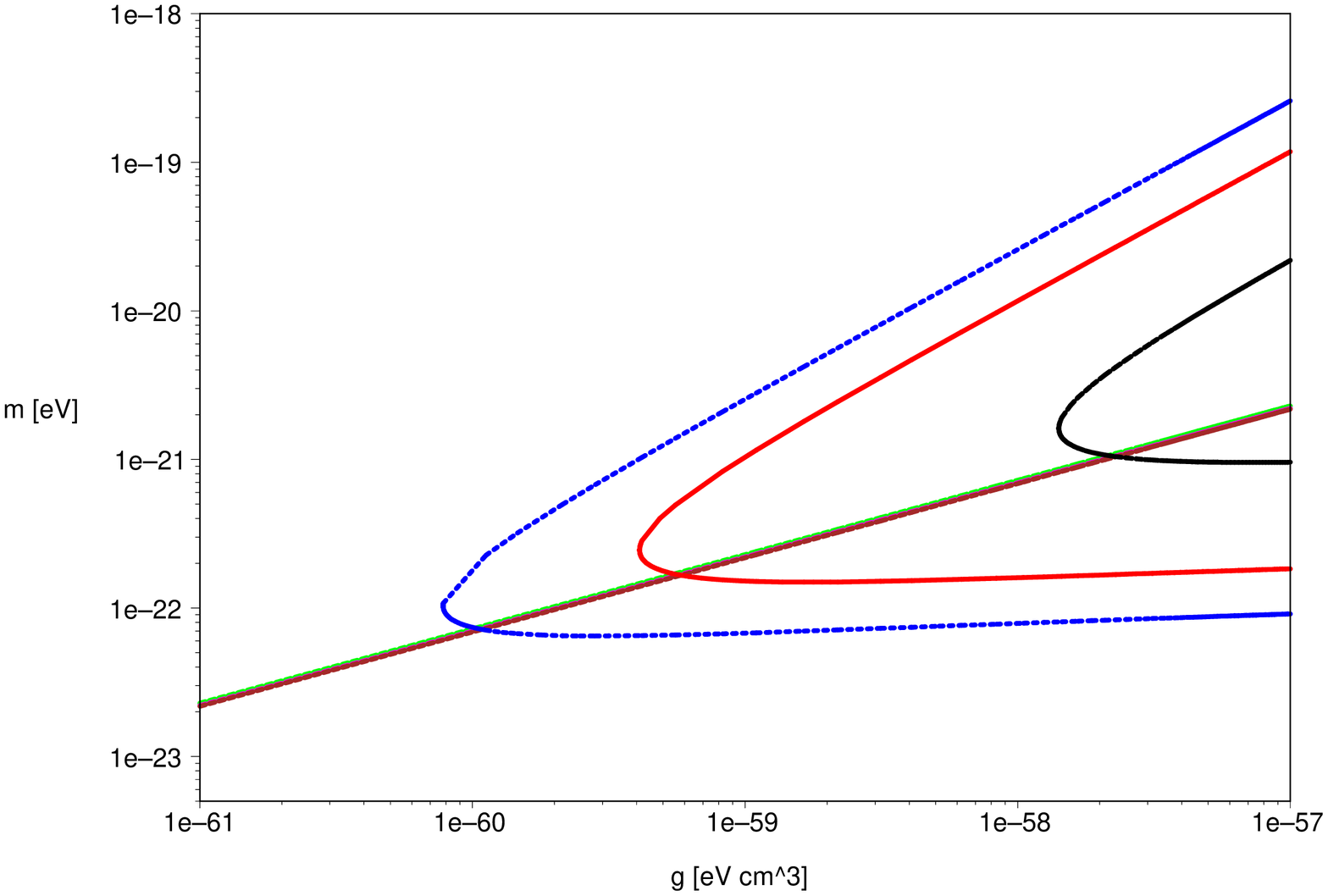}
     \hspace{0.1cm}
    \end{minipage}
 \caption{Critical curves $\delta E' = 0$ for vortex formation in the dimensional
 BEC-CDM particle
 parameter space
 for Halo-Model A: particle mass $m$ in eV versus $g$ in eV cm$^3$ for a Milky-Way-sized halo, $M = 10^{12}M_{\odot}, R = 100$
 kpc (\tx{left-hand plot}) and a dwarf-galaxy-sized halo, $M = 10^{10}M_{\odot}, R = 10$ kpc (\tx{right-hand plot});
 caption as in Fig.\ref{fig5} (lhs)}
 \label{fig6}
\end{figure*}

In Fig.\ref{fig5} (left-hand-plot), we show the relationship
(\ref{critcurve}) (i.e. $m/m_H$ versus $g/g_{H}$) \tx{independently}
of halo size, but for given halo eccentricity, i.e. given spin
parameter, $\lambda = (0.01, 0.05, 0.1)$. One can easily check on
which side of the critical curves vortex formation is favoured by
determining the sign of $\delta E'$. The result is that for each set
of solution branches, no vortex is allowed for parameters in the
space outside of the region bound by the critical curves, or with
other words, only inside that bound region, for a given $\lambda$,
is the vortex energy in the rotating frame negative. One can see
that the parameter space of vortex existence grows as $e$ or
$\lambda$ increases, as expected. However, the energy calculation
did not incorporate virial equilibrium of the halo, which we have
seen further constrains $(m/m_H, g/g_H)$ according to (\ref{npoly}).
We therefore also show this relationship in Fig.\ref{fig5} for the
same $e$-values. The sensitivity to the eccentricities is weak, so
the respective lines seem to lie on top of each other on the
double-logarithmic plot. In light of the discussion in \tx{Section
2.2}, we expect the virial relationship (\ref{npoly}) to be valid
only if $g/g_H \gg 2$. Now, for a given spin-parameter, virialized
BEC haloes will form vortices for those $(m/m_H, g/g_H)$-values, for
which the virial line lies inside the region bound by the critical
curves. The intersection of the virial line with the critical energy
curve defines thus a set of critical values $(m/m_H)_{crit},
(g/g_H)_{crit}$ for each $\lambda$, above which a vortex will form.
The results are shown in Table \ref{tab2}.
\\
In Fig.\ref{fig6}, we translate the bounds on vortex formation from
Fig.\ref{fig5} to the \tx{dimensional} BEC particle parameters $m$
and $g$. To this end, a halo mass and size have to be chosen, which
will fix $m_H$ and $g_H$ in (\ref{mfidu}) and (\ref{gfidu}),
respectively. We chose a Milky-Way halo and a dwarf-galaxy-sized
halo\footnote{Note again that the critical energy curves need $R$
and $M$ as input, while the virial constraint needs $R$ only to
specify the dimensional ($m, g$)-space, see equ. (\ref{zwischen})
and (\ref{npolydim}).}. For the latter, the parameter space of
vortex formation is shifted to higher values of particle mass as
compared to the Milky-Way halo. Generally, however, the parameter
space of vortex formation shifts to lower values of the particle
mass for haloes of the \tx{same} size but lower mean density.
\\
As a further illustration of the importance of self-interaction, we
plot the vortex energy $\delta E'$ of equ.(\ref{vortexenergie2}) as
a function of the spin parameter $\lambda$ for several chosen values
of BEC models $(m/m_H,g/g_H)$ in Fig.\ref{fig5}, right-hand-plot.
Again, we see the trend of the result in Fig.\ref{fig5} (lhs) at
play: for a given spin-parameter $\lambda$, the vortex is
increasingly favoured for larger coupling strength $g/g_H$. On the
other hand, for fixed $g/g_H$, a larger spin-parameter makes the
vortex favoured. So, the total energy of the system is lowered above
some critical $\lambda$, which is smaller for higher values of
$g/g_H$. For the depicted examples, we see that the BEC model of the
upper curve with $(m/m_H, g/g_H) = (100, 100)$ is never able to form
a vortex, since it would cross the ($\delta E' = 0$) -horizontal
only at a $\lambda$-value, which is beyond the stability limit of
the spheroidal halo, see Table \ref{tab4}.

We also note that the condition of a vortex to be energetically
favoured, $\delta E' < 0$, does not automatically fix the numerical
value of $L/L_{QM} > 1$. In fact, the value of $L/L_{QM}$ varies
along the critical energy curves in Fig.\ref{fig5} (lhs), being
larger for higher $m/m_H$, according to (\ref{macang}). The minimum
on those curves at the critical BEC-CDM particle parameters is a
factor of $L/L_{QM} \simeq 4-6$ for the $\lambda$-values considered,
see Table \ref{tab2}. This reflects the fact that the spheroidal
haloes carry a lot of excess angular momentum over $L_{QM}$ in all
of the relevant parameter space. It seems at first sight
counter-intuitive that $(L/L_{QM})_{crit}$ decreases with increasing
$\lambda$. However, one must bear in mind that this result comes
from critical conditions on the energy for vortex formation, i.e. if
$\lambda$ happens to be small, so must the ratio $L/L_{QM}$ be large
enough for a vortex to form, or, conversely, a higher $\lambda$
makes a vortex possible at lower $L/L_{QM}$. The spin-parameter,
after all, does not only depend on $L$, but also on the total energy
$E$ of the system, see equ.(\ref{lam}).

Table \ref{tab2} also shows the corresponding vortex core radii for
the chosen haloes, according to the relation in (\ref{becdef}).
Since the vortex core shrinks with increasing self-interaction
strength (see equ.(\ref{healing})), the values displayed are the
maximum possible ones.

\begin{table}
\caption{\tx{Halo-Model A:} Lower bounds on BEC-CDM particle mass
and self-interaction coupling strength for vortex formation in
haloes of given spin-parameter; the corresponding $\xi$-values are
upper bounds for the vortex core radius} \label{tab2}
\begin{center}
 Independent of halo size: \\
\begin{tabular}{l|l|l|l|l}
\hline
          & $\lambda$   & $(m/m_H)_{crit}$ & $(g/g_H)_{crit}$ & $(L/L_{QM})_{crit}$     \\
\hline
          & $0.01$ & $309.41$  & $1.02\cdot 10^{5}$   & $5.65$ \\
          & $0.05$ & $49.52$  & $2549.24$  &  $4.53$  \\
          & $0.10$ & $21.73$  & $454.54$  &  $4.02$  \\
\hline
\end{tabular}
\end{center}
\begin{center}
 Milky-Way-sized halo:  $M = 10^{12}M_{\odot}$, $R = 100$ kpc \\
\begin{tabular}{l|l|l|l|l}
\hline
          & $\lambda$   & $m_{crit}$ [eV] & $g_{crit}$ [eV cm$^{3}$]  & $\xi_{max}$ [kpc]     \\
\hline
          & $0.01$ & $3.30 \cdot 10^{-23}$  & $2.30 \cdot 10^{-59}$   & $0.31$ \\
          & $0.05$ & $5.28 \cdot 10^{-24}$  & $5.74 \cdot 10^{-61}$   & $1.98$ \\
          & $0.10$ & $2.32 \cdot 10^{-24}$  & $1.02 \cdot 10^{-61}$   & $4.69$ \\
\hline
\end{tabular}
\end{center}
\begin{center}
 Dwarf-galaxy-sized halo:  $M = 10^{10}M_{\odot}$, $R = 10$ kpc \\
\begin{tabular}{l|l|l|l|l}
\hline
          & $\lambda$   & $m_{crit}$ [eV] & $g_{crit}$ [eV cm$^{3}$]  & $\xi_{max}$ [kpc]     \\
\hline
          & $0.01$ & $1.04 \cdot 10^{-21}$  & $2.30 \cdot 10^{-58}$   & $0.03$  \\
          & $0.05$ & $1.67 \cdot 10^{-22}$  & $5.74 \cdot 10^{-60}$   & $0.20$  \\
          & $0.10$ & $7.33 \cdot 10^{-23}$  & $1.02 \cdot 10^{-60}$   & $0.47$  \\
 \hline
\end{tabular}
\end{center}
\end{table}

\subsection{Halo-Model B: $L = L_{QM}$}

\subsubsection{Energy and density profiles of the irrotational Riemann-S ellipsoid}

To account for the effects of compressibility and irrotationality,
we take as our unperturbed halo with no vortex an ($n=1$)-polytropic
Riemann-S ellipsoid having $L=L_{QM}$, just enough angular momentum
for one quantum vortex. We will perform a similar energy analysis as
in \tx{Section 4.3} to determine if a vortex, which carries all of
the angular momentum, is energetically favoured. Since the vortex
takes up all of the angular momentum in this model, however, the
final shape will be a spherical halo with a cylindrical vortex in
its centre.

Before we study this in more detail, we first consider the energy of
the vortex-free ellipsoidal halo model. The energy of the
irrotational Riemann-S ellipsoid with $(n=1)$ is given by
(\ref{sumenerg}), where we set $K_Q=0$ and use
equ.(\ref{riepoten})-(\ref{rierot}) for the remaining terms in order
to arrive at
  \bdi
  E_R = U_{SI} + W + T =
   \edi
    \beq \label{rieenergie}
    = \f{\rho_c^E}{2}M\f{g}{2m^2} - \f{3}{4}\f{GM^2}{R}f(e_1,e_2) +
  \f{\kappa_1}{20}\mc{K}(e_1,e_2)MR^2\bar \Omega_G^2,
   \eeq
    where
     \bdi
      \mc{K}(e_1,e_2) \equiv
       \edi
       \bdi
       \equiv \left((1-e_1^2)^{-1/6}(1-e_2^2)^{-1/6}
      - (1-e_1^2)^{1/3}(1-e_2^2)^{-1/6}\right)^2 \times
       \edi
       \bdi
       \times(\tilde \Omega +
      \tilde \Lambda)^2  +
     \edi
      \bdi
        + \left((1-e_1^2)^{-1/6}(1-e_2^2)^{-1/6}
      + (1-e_1^2)^{1/3}(1-e_2^2)^{-1/6}\right)^2\times
       \edi
       \beq \label{auxK}
       \times(\tilde \Omega -
      \tilde \Lambda)^2
       \eeq
       is a function of the eccentricities only. Here we defined $\bar \Omega_G \equiv \sqrt{\pi G \bar
       \rho^E}$, $\tilde \Lambda = \Lambda/\Omega_G$ and $\bar \rho^E, \rho_c^E$ are the mean and central densities of
        the Riemann-S ellipsoid, respectively (see also \tx{Appendix
        A.3}). The total angular momentum is given in (\ref{angrie})
        and we will force it to be equal to $L_{QM}$ in the
        calculation of the vortex energy in the next subsection. Although we will only make
        use of the global energies as stated in
\tx{Section 3.2} for the calculation which follows, we find it
useful to plot the actual density profiles of the Riemann-S
ellipsoids in the absence of a vortex, see Fig.\ref{fig7}
\tx{left-hand-plot}. In fact, as we show in \tx{Appendix A3},
analytic profiles can be derived if the ellipsoidal approximation is
assumed.

\subsubsection{Vortex ansatz and vortex energy}

Once formed, the vortex is assumed to carry all of the angular
momentum $L=L_{QM}$ of the halo. That means we anticipate a
transition of the Riemann-S ellipsoid to a sphere, but with a vortex
in the centre. A spherical halo with central vortex has been studied
e.g. in \cite{kain}, where an approximate calculation of the density
profile in the presence of a vortex has been presented. For
convenience, we will thus in the following take advantage of the
result for the approximate energy of a spherical halo with
cylindrical vortex as presented
 there\footnote{As we indicated in \tx{Section 1}, the approach in \cite{kain} to calculate the
 energy of the halo with vortex neglects angular momentum in the first place, making
 the very appearance of the vortex unmotivated. However, since we start from a
halo as a Riemann-S ellipsoid \tx{with} minimum angular momentum
$L_{QM}$, provided by the $\lambda$-spin parameter, it is meaningful
to ask the question of whether it is energetically favoured to drive
the system to a state having a vortex in the center.}.

\begin{figure*}
\begin{minipage}[b]{0.5\linewidth}
      \centering\includegraphics[width=8.3cm]{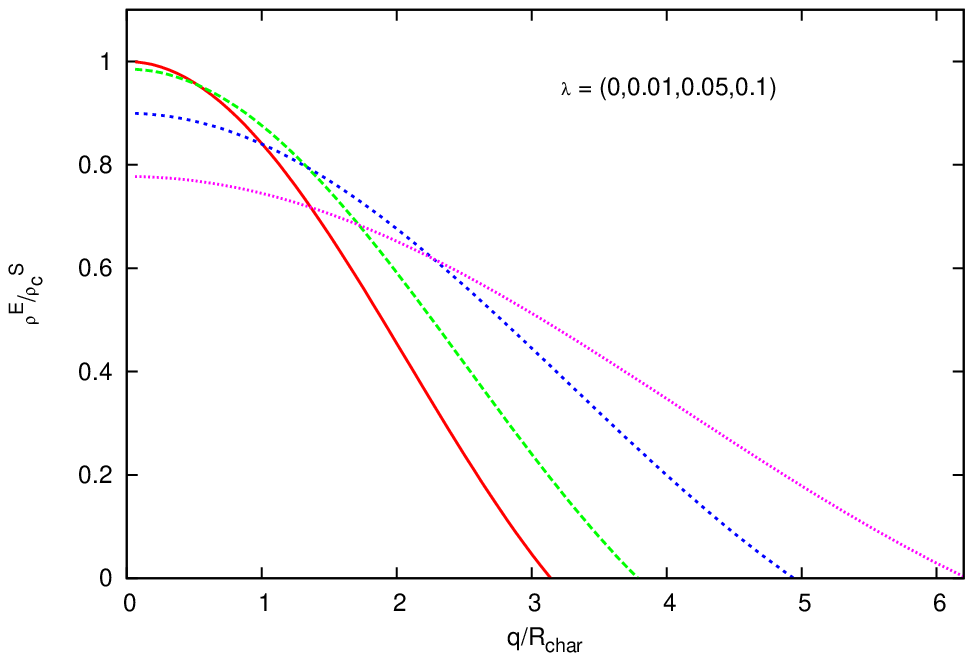}
     \hspace{0.1cm}
    \end{minipage}%
     \begin{minipage}[b]{0.5\linewidth}
 \centering\includegraphics[width=8.3cm]{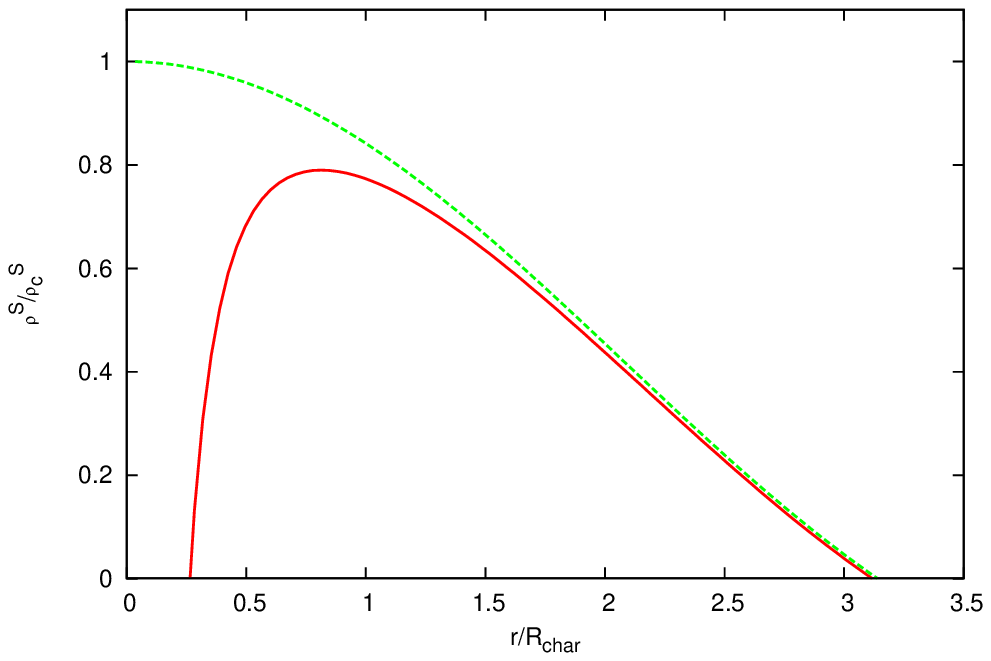}
  \hspace{0.1cm}
    \end{minipage}
 \caption{Density profiles for Halo-Model B: \tx{Left-hand-plot}:
 Profiles of the (vortex-free) ($n=1$)-polytropic Riemann-S ellipsoidal haloes having
 $\lambda = (0.01, 0.05, 0.1)$, employing the ellipsoidal approximation, according to equ.(\ref{ellipdensity}).
  The solid curve is the ($n=1$)-polytropic sphere, equ.(\ref{tfprofile}), and is added for comparison. The densities are
  all normalized to $\rho_c^S = 1$. The locus of
  the outer surface where the density vanishes increases with $\lambda$;
  \tx{Right-hand-plot}: Profile of the spherical halo with vortex in the
 centre (solid) using equ.(4.16) of Kain \& Ling (2010)
 with $\Theta_0 = 1$ and $R_c/\xi$ for $\lambda = 0.05$.
 The unperturbed sphere (dashed) is added for comparison.}
 \label{fig7}
\end{figure*}

The usual ansatz for a singly-quantized, axisymmetric vortex is used
in \cite{kain}, in our notation $\tilde w = |w|e^{i\phi}$. Then, the
GP energy of a halo with such a vortex on top of the spherical
background density (\ref{tfprofile}) is calculated in \cite{kain}.
Using our notation, the energy of the spherical halo in the presence
of the vortex given in equ.(6.22) of \cite{kain} is
  \bdi
  E_S = U_{SI} + W + T = \f{\rho_c^S}{2}M\f{g}{2m^2} - \f{3}{4}\f{GM^2}{R_0} +
   \edi
   \beq \label{kl}
  + \f{1}{4}L_{QM}\Omega_{QM}(1 + \pi Si(\pi)(\ln 2 + \ln (R_0/R_c)) + \pi\Gamma),
  \eeq
 where $Si(\pi) \approx
 1.852$ is the sine integral evaluated at $\pi$, $\Gamma \approx
 -2.658$, and $\rho_c^S$ is the central density of the halo, which is an ($n=1$)-polytropic
 \tx{sphere}
 with radius $R_0$ given in (\ref{onesphere}).
Some shortcomings enter their calculation of the angular kinetic
part of the vortex energy: the actual vortex profile, as for
instance given by $|w|^2$ in equ.(\ref{vortexansatz}), is replaced
by a finitely thick, empty cylinder with radius $R_c$, in our
notation $R_c = \xi \sqrt{\bar \rho^E/\rho_c^S}$. Also, the change
in the gravitational potential due to the vortex is neglected as can
be seen in the derivation of equ.(6.22) in \cite{kain}\footnote{More
precisely, the gravitational potentials $V_G$ in equ.(6.8) and
(6.13) of \cite{kain} are simply set equal, whereas our analysis in
\tx{Section 4.3.1} clearly distinguishes between the unperturbed
halo potential $\Phi_0$ and the perturbed one, $\Phi$.}. Using their
result for the energy of the spherical halo in the presence of the
vortex provides us thus with a rougher upper bound estimate of the
conditions for vortex formation, than if we had used a better
implementation of the vortex ansatz of \tx{Section 4.3.1} in the
Riemann-S ellipsoid. However, given the orders of magnitudes
involved in the resulting bounds on the BEC particle parameters, we
do not expect these effects to be very significant. We stress that
the above expression for the energy, equ. (\ref{kl}), assumes that
all of the angular momentum is in the singly-quantized
 vortex, whose amount is $L_{QM} = N\hbar$. Equ.(\ref{kl}) is the energy as measured in the \tx{rest
 frame}, therefore $E_S' = E_S - \Omega L_{QM}$ is the energy in the
 frame rotating rigidly with $\Omega$. The energy in (\ref{rieenergie})
 is also the one given in the rest frame, so again $E_R' = E_R - \Omega L$ is
 the energy in the rotating frame. Now, in order to compare the
 two states on an equal footing, we force the Riemann-S ellipsoid to have
 the same amount of angular momentum as the sphere with vortex,
 i.e. all of the angular momentum of the ellipsoid shall be
 transferred to the vortex, leaving the bulk of the halo spherical. We will thus
 set $L = L_{QM}$ in $E_R'$. Moreover, since the total mass $M$ of the halo shall not change across the
transition to the vortex state, the mean and central densities,
$\bar \rho^S, \bar \rho^E$ and $\rho_c^S, \rho_c^E$,
      for the spherical and the ellipsoidal halo shape, respectively, will differ but be related according to
  \beq \label{ratden}
   \bar \rho^E = \bar \rho^S g(e_1,e_2)^{3/2} \mbox{ and } \rho_c^E =
   \rho_c^S g(e_1,e_2)^{3/2},
   \eeq
   where we used $\bar \rho^E = 3M/(4\pi R^3)$ and
   (\ref{polyradius}) (see also equ.(3.10) in LRS93). Furthermore, by using (\ref{ratden}) and (\ref{kreuz}), we
   can write
   $\bar \rho^E/\rho_c^S = \f{3}{\pi^2}g(e_1,e_2)^{3/2}$, which we
   use in (\ref{kl}) in order to rewrite
    \beq
    \ln (R_0/R_c) = \ln R/\xi + \f{1}{2} \ln (\pi^2/3g(e_1,e_2)^{1/2}).
     \eeq
 Calculating the energy difference and rearranging terms,
 we get for the (dimensionless) vortex energy,
  \bdi
   \f{\delta E'}{\Omega_{QM}L_{QM}} \equiv
   \f{E_S'-E_R'}{\Omega_{QM}L_{QM}} = \f{E_S-E_R}{\Omega_{QM}L_{QM}} =
   \edi
    \bdi
     =
   \f{\pi^2}{24}(f(e_1,e_2)^{-3/2}-1)\left(\f{g}{g_H}\right) +
   \f{\pi Si(\pi)}{8}f(e_1,e_2)^{-1}\ln \f{g}{g_H} -
    \edi
     \bdi
    - \left[g(e_1,e_2)^{-1/2}-f(e_1,e_2)\right]\left(\f{m}{m_H}\right)^2 +
   \f{1}{4}g(e_1,e_2)^{-1}\times
     \edi
     \bdi
   \times \left[1+\pi Si(\pi)\ln 2 + \pi \Gamma +
   \f{\pi Si(\pi)}{2}\ln \left(\f{\pi^2}{3g(e_1,e_2)^{1/2}}\right) -
   \right.
    \edi
 \beq \label{engdiff}
   \left. - \f{\kappa_1}{5}\mc{K}(e_1,e_2)g(e_1,e_2)\left(\f{m}{m_H}\right)^2\right].
    \eeq
We stress again that formula (\ref{engdiff}) has been derived by
imposing $L = L_{QM}$. In Fig.\ref{fig7}, \tx{right-hand-plot}, we
show a spherical halo with central vortex which transitioned from a
Riemann-S ellipsoid having $\lambda = 0.05$.

\begin{figure*}
\begin{minipage}[b]{0.5\linewidth}
      \centering\includegraphics[width=8.3cm]{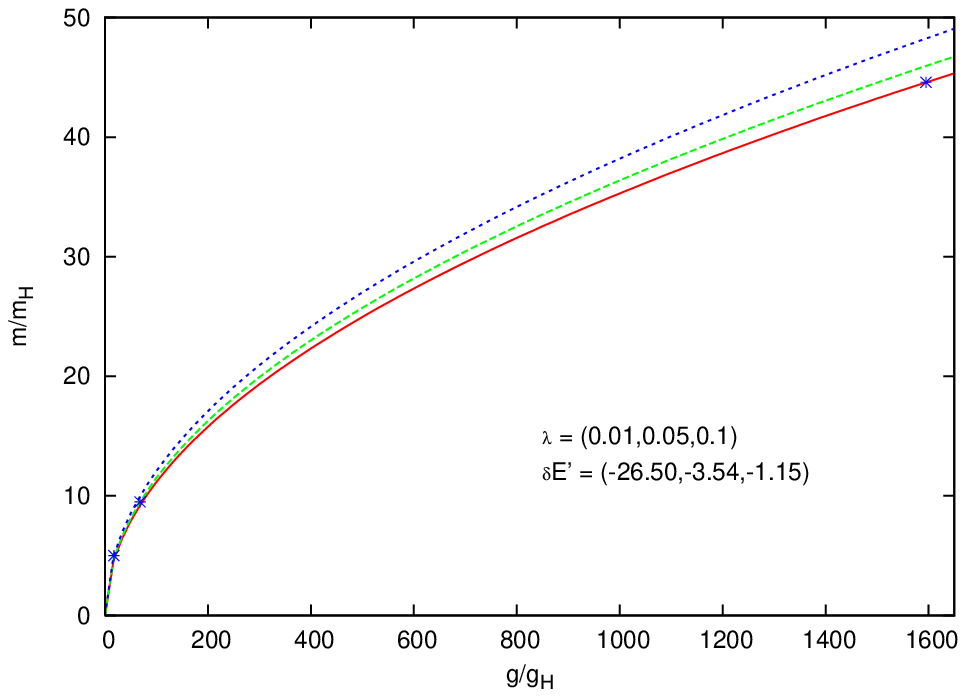}
     \hspace{0.1cm}
    \end{minipage}%
     \begin{minipage}[b]{0.5\linewidth}
 \centering\includegraphics[width=8.3cm]{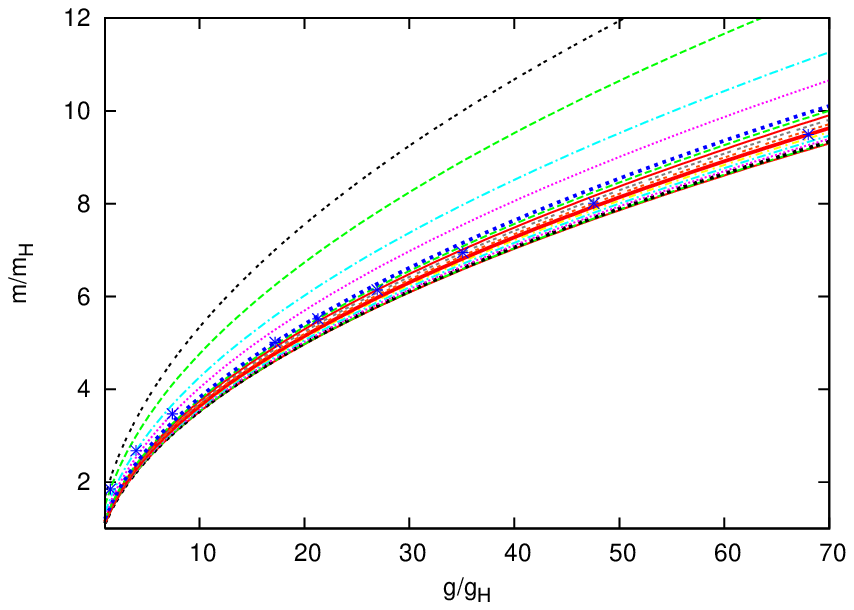}
  \hspace{0.1cm}
    \end{minipage}
 \caption{BEC-CDM particle parameter space for Halo-Model B (no logarithmic scales here !): \tx{Left-hand-plot}: Virialized haloes according to (\ref{compR})
 for $\lambda = (0.01, 0.05, 0.1)$ (lower to upper curves) with
 depicted points $(m/m_H, g/g_H)$, according to Table \ref{tab3}, at which $L=L_{QM}$ and corresponding vortex
 energy $\delta E'$; \tx{Right-hand-plot:} the same but including more curves in
 $\lambda$ according to Table \ref{tab5}, and zoomed-in closer to the origin; dots denote BEC haloes having $L=L_{QM}$.}
 \label{fig8}
\end{figure*}

\begin{figure*}
\begin{minipage}[b]{0.5\linewidth}
      \centering\includegraphics[width=8cm]{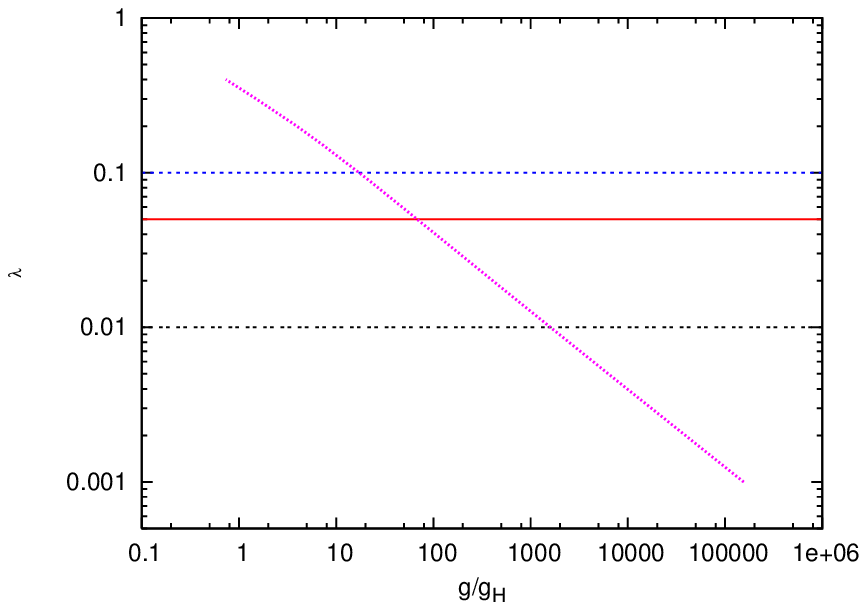}
     \hspace{0.1cm}
    \end{minipage}%
     \begin{minipage}[b]{0.5\linewidth}
 \centering\includegraphics[width=8cm]{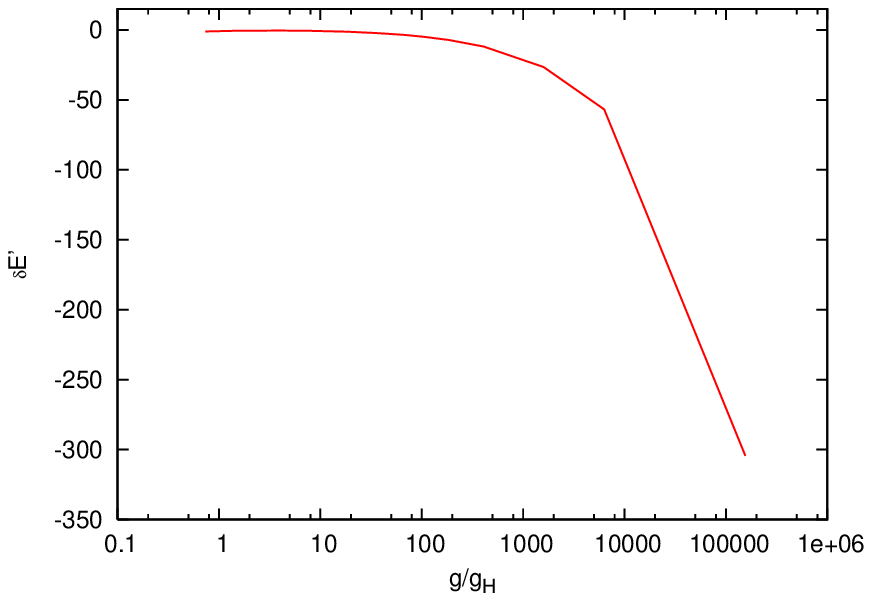}
     \hspace{0.1cm}
    \end{minipage}
 \caption{Halo-Model B: Virialized BEC haloes according to (\ref{compR}) for which $L = L_{QM}$: \tx{Left-hand plot}:
  spin parameter vs. self-interaction coupling
 strength;
  \tx{Right-hand-plot}: Vortex energy (\ref{engdiff}) in units of
 $\Omega_{QM}L_{QM}$ vs. self-interaction strength.}
 \label{fig9}
\end{figure*}

In Fig.\ref{fig8}, left-hand-plot, we show virialized haloes which
are irrotational Riemann-S ellipsoids according to (\ref{compR}) for
$\lambda = (0.01, 0.05, 0.1)$: Thanks to the relationship in
(\ref{rieang}), the constraint of having minimum angular momentum
for vortex formation, $L = L_{QM}$, means to fix the value of
$m/m_H$. For given $\lambda$ and $m/m_H$, the Virial constraint
(\ref{compR}) fixes then the self-interaction $g/g_H$. This means
that the condition $L=L_{QM}$ can be met at only \tx{one} point (for
each $\lambda$) in the BEC-CDM particle parameter space. These
points are also depicted in Fig.\ref{fig8}. Inserting those values
for the particle parameters in (\ref{engdiff}) shows that the
corresponding vortex energy is negative, i.e. the vortex is favoured
for all $\lambda$-values of interest. Again, the vortex is favoured
at high enough particle mass and self-interaction strength for a
given $\lambda$. To strengthen this picture by adding more 'data
points', we plot (\ref{compR}) for a larger range in $\lambda$
according to Table \ref{tab5} and the corresponding values $(m/m_H,
g/g_H$) at which $L = L_{QM}$ (see Fig.\ref{fig8}, right-hand-plot).
A vortex can be formed in all cases considered. In addition,
Fig.\ref{fig9} confirms the previous result of favouring vortices at
high enough self-interaction for a given $\lambda$ and vice versa.
Since vortex formation generally requires $L \geq L_{QM}$, the above
calculation provides us thus again with the \tx{critical} BEC-CDM
particle parameters, above which vortex formation is favoured, along
with the maximum vortex core sizes. Table \ref{tab3} summarizes
those parameters in dimensionless and physical units. Owing to the
constraint $L=L_{QM}$, we do not calculate the critical condition
$\delta E' = 0$ in \tx{Halo-Model B}, since that condition is
fulfilled at another $L ~(\not= L_{QM})$. Apart from the
consideration, that this model may be more realistic than the
rigidly rotating haloes of \tx{Model A} in \tx{Section 4.3}, the
purpose of this model was also to show that the results gained in
\tx{Section 4.3} are consistent with those derived here, despite the
simplifying assumptions on the unperturbed (vortex-free) halo, on
which the former were based. The same trend can be seen here as in
\tx{Model A} of \tx{Section 4.3}: vortex formation requires a high
enough mass $m$ and positive self-interaction $g$ of the BEC-CDM
particles, where smaller $\lambda$-spin parameters require higher
values for $m$ and $g$. Small-mass haloes set tighter constraints on
the particle parameters. Both halo models are thus qualitatively
consistent with each other. For the same halo with fixed $\lambda$,
\tx{Model A}, however, requires in general larger values of $m$ and
$g$ then \tx{Model B}. This is due to the fact, that \tx{Model A}
has more angular momentum $L/L_{QM} \gg 1$, resulting in higher
$m/m_H$ via (\ref{macang}), which in turn requires higher $g/g_H$
due to the virial constraint (\ref{npoly}). By the same token, the
maximum vortex core radii can be substantially larger in
\tx{Halo-Model B}, since the critical self-interaction strengths for
vortex formation are smaller than for \tx{Halo-Model A}.

We also see from Fig.\ref{fig8} that a significant portion of the
particle parameter space remains, where vortices will not form. In
fact, for all values of $(g/g_H, m/m_H)$ which lie left to the
depicted points on a given virial curve, depending on $\lambda$,
haloes fulfill $L < L_{QM}$. For coupling strengths which are high
enough for them to be still in the polytropic Thomas-Fermi regime,
$g/g_H \gg 2$, those haloes will just remain compressible,
irrotational Riemann-S ellipsoids, and will not undergo vortex
formation.

\begin{table}
\caption{\tx{Halo-Model B:} Lower bounds on BEC-CDM particle mass
and self-interaction coupling strength for vortex formation in
haloes of given spin-parameter; the corresponding $\xi$-values are
upper bounds for the vortex core radius} \label{tab3}
\begin{center}
 Independent of halo size: \\
\begin{tabular}{l|l|l|l|l}
\hline
          & $\lambda$   & $(m/m_H)_{crit}$ & $(g/g_H)_{crit}$ & $L/L_{QM}$     \\
\hline
          & $0.01$ & $44.58$  & $1595.07$   & $1$ \\
          & $0.05$ & $9.49$  & $68.00$  &  $1$  \\
          & $0.10$ & $5.01$  & $17.20$  &  $1$  \\
\hline
\end{tabular}
\end{center}
\begin{center}
 Milky-Way-sized halo:  $M = 10^{12}M_{\odot}$, $R = 100$ kpc \\
\begin{tabular}{l|l|l|l|l}
\hline
          & $\lambda$   & $m_{crit}$ [eV] & $g_{crit}$ [eV cm$^{3}$]    & $\xi_{max}$ [kpc]  \\
\hline
          & $0.01$ & $4.75 \cdot 10^{-24}$  & $3.59 \cdot 10^{-61}$    & $2.50$ \\
          & $0.05$ & $1.01 \cdot 10^{-24}$  & $1.53 \cdot 10^{-62}$    & $12.13$ \\
          & $0.10$ & $5.34 \cdot 10^{-25}$  & $3.87 \cdot 10^{-63}$    & $24.11$ \\
\hline
\end{tabular}
\end{center}
\begin{center}
 Dwarf-galaxy-sized halo:  $M = 10^{10}M_{\odot}$, $R = 10$ kpc \\
\begin{tabular}{l|l|l|l|l}
\hline
          & $\lambda$   & $m_{crit}$ [eV] & $g_{crit}$ [eV cm$^{3}$]   & $\xi_{max}$ [kpc]   \\
\hline
          & $0.01$ & $1.50 \cdot 10^{-22}$  & $3.59 \cdot 10^{-60}$    & $0.25$ \\
          & $0.05$ & $3.20 \cdot 10^{-23}$  & $1.53 \cdot 10^{-61}$    & $1.21$ \\
          & $0.10$ & $1.69 \cdot 10^{-23}$  & $3.87 \cdot 10^{-62}$    & $2.41$ \\
 \hline
\end{tabular}
\end{center}
\end{table}

\section{Conclusions and Discussion}

We have studied in this paper the conditions of vortex formation in
galactic haloes composed of dark matter particles which are able to
form a Bose-Einstein condensate. Haloes can then be described as
fluids, obeying quantum-mechanical fluid equations.
Quantum-mechanical effects make this form of dark matter behave
differently from standard CDM, resulting in new effects with
potentially observable consequences. There are essentially two
limiting cases one may consider. First, for quantum-coherence to be
relevant on the scale of a halo of radius $R$, we must either
require the particle de-Broglie wavelength (\ref{deB}) to be of the
order of the halo size,
 $\lambda_{deB} \lesssim R$, or else require $\lambda_{deB} \ll R$
 but with a strong enough repulsive self-interaction to hold the
 halo up against gravity.
 In the former case, if $v \simeq v_{vir}$ for the halo, this translates
 into a condition for the dark matter particle mass to be small
 enough, $m \gtrsim m_H = 1.066 \cdot 10^{-25}(R/100
~\rm{kpc})^{-1/2}(M/10^{12}~ M_{\odot})^{-1/2}$ eV. In the latter
case, in order for the repulsive self-interaction pressure force
greatly to exceed that due to quantum pressure (i.e. the
Thomas-Fermi regime), we must require $g \gg
   g_H = 2.252 \cdot 10^{-64}(R/100~
   \rm{kpc})(M/10^{12}~M_{\odot})^{-1}$ eV
   cm$^3$. BEC
haloes can then be approximated by polytropes with index $n=1$. If
we take $R$ to be the radius of the virialized object supported
   against gravity by the dominant, repulsive self-interaction, this
   condition becomes a condition on the particle mass given by $m
   \gtrsim \f{m_H}{4}\sqrt{15 g/g_H}$.

These results apply generally to the global halo structure. However,
rotating BEC haloes add new phenomenology, and the possibility to
distinguish this form of dark matter from other candidates.
 To this aim, we have studied here the
question of whether an angular velocity sufficient to create
vortices occurs in BEC-CDM cosmologies. As quantum fluid systems,
BEC haloes can be modelled as uniformly rotating ellipsoids, with
and without internal motions superposed. To this aim, we have
derived equations which relate the eccentricities of haloes to their
$\lambda$-spin parameter. Once the latter is fixed, the
eccentricities can be uniquely determined. Then, we have
analytically studied necessary and sufficient conditions for vortex
formation. Vortex formation requires as a necessary minimum
condition that the halo angular momentum fulfills $L \geq L_{QM} =
N\hbar$, which implies a lower bound on $m/m_H$, i.e. on the dark
matter particle mass. However, a sufficient condition for vortex
formation can be established by an energy analysis, which aims to
find the conditions when a vortex becomes energetically favoured.
This results, in addition, to lower bounds on the positive
self-interaction coupling strength of the dark matter particle,
$g/g_H$. While the sufficient condition also requires $L \geq
L_{QM}$, the amount necessary is determined by the model.

We have studied two classes of models for rotating haloes in order
to analyze stability with respect to vortex formation in two limits,
that for $L/L_{QM} \gg 1$ (\tx{Halo-Model A}) and for $L/L_{QM} = 1$
(\tx{Halo-Model B}), respectively. In \tx{Halo-Model A}, haloes are
modelled as homogeneous Maclaurin spheroids. The minimum angular
momenta for vortex formation are then $(L/L_{QM})_{crit} = (5.65,
4.53, 4.02)$ for $\lambda = (0.01, 0.05, 0.1)$, respectively, which
correspond to a constraint on the particle mass $m/m_H \geq
(m/m_H)_{crit}$, where $(m/m_H)_{crit} = (309.41, 49.52, 21.73)$,
respectively. As long as $m/m_H$ satisfies this condition, the
strength of the self-interaction must also then satisfy the
condition that $g/g_H \geq (g/g_H)_{crit}$, where $(g/g_H)_{crit} =
(1.02\cdot 10^5, 2549.24, 454.54)$ for these same $\lambda$-values,
respectively.

However, for \tx{Halo-Model B}, which is an $(n=1)$-polytropic
Riemann-S ellipsoid, strictly irrotational prior to vortex
formation, even $L/L_{QM} = 1$ can be sufficient for vortex
formation, \tx{if} the self-interaction strength is large enough.
The condition $L/L_{QM}=1$ fixes the value of $m/m_H$ for each
$\lambda$ according to (\ref{rieang}), and the condition of Virial
equilibrium (\ref{compR}) thereby also fixes $g/g_H$. For $\lambda =
(0.01, 0.05, 0.1)$, these values are given by $m/m_H = (44.58, 9.49,
5.01)$ and $g/g_H = (1595.07, 68.00, 17.20)$, respectively.
According to equ.(\ref{engdiff}), \tx{Halo-Model B} makes vortex
formation energetically favourable for those values of $m/m_H$ and
$g/g_H$. We interpret this to mean that, for $L/L_{QM} > 1$ (i.e.
$m/m_H > (m/m_H)(L=L_{QM})$), vortex formation will \tx{also} be
favoured, \tx{as long as} $g/g_H > (g/g_H)(L=L_{QM})$. Furthermore,
any values of $m/m_H$ and $g/g_H$ which satisfy the condition for
vortex formation in \tx{Halo-Model A} will automatically satisfy
that found by \tx{Halo-Model B}, which is less stringent, although
more accurate.

We can thus imagine vortex formation in BEC haloes composed of
repulsively interacting particles as follows: If the angular
momentum of a rotating BEC halo fulfills $L < L_{QM}$ (i.e. if
$(m/m_H) < (m/m_H)_{crit}$ for a given $\lambda$, according to
\tx{Halo-Model B}), no vortex will form, and the halo can be
modelled by a mildly compressible, irrotational Riemann-S ellipsoid,
which has a polytropic index of $n=1$. For $L = L_{QM}$ (i.e. if
$m/m_H = (m/m_H)_{crit}$), the irrotational Riemann-S ellipsoidal
halo can make a transition to a non-rotating, spherical halo with a
vortex at the center if the self-interaction is strong enough (i.e.
$g/g_H = (g/g_H)_{crit}$). For a range of angular momenta fulfilling
$L_{QM} < L \leq 2L_{QM}$, we may still expect a central vortex but
now with the excess angular momentum deforming the halo such that
again a Riemann-S ellipsoid forms. Finally, if $L \gg L_{QM}$,
oblate haloes described as Maclaurin spheroids can have a central
vortex if $m/m_H \geq (m/m_H)_{crit}$ and $g/g_H \geq
(g/g_H)_{crit}$ with the critical values now given by \tx{Halo-Model
A}. Those critical values determine as of when a single vortex is
energetically favoured, but since $L/L_{QM} \gg 1$ [i.e. ($L$ per
particle) $\gg \hbar$], it is also possible that multiple vortices
will form, even a lattice of vortices\footnote{It is known that
laboratory BEC quantum gases confined by a wide range of trap
potentials favour multiple vortices that are singly-quantized over a
single vortex that is multiply-quantized, see e.g. \cite{aftdu,
rindler}.}. For a dense lattice of vortices, the halo's vorticity
will approach that of a rigidly rotating body, according to the
quantum-mechanical correspondence principle argument, applied to the
analogous problem in superfluid helium by \cite{feynman}. A suitable
generalization of the homogeneous \tx{Halo-Model A} of \tx{Section
3.1} to compressible spheroidal haloes, according to the ellipsoidal
approximation of \cite{LRS}, will then constitute a viable model for
this high-angular momentum regime.

Generally, we have shown that BEC-CDM haloes in the polytropic
Thomas-Fermi regime \tx{will} typically form vortices, since this
regime requires $m \gg m_H$ and $g \gg g_H$, which largely overlaps
the region of parameter space for vortex formation. By comparing the
characteristic parameters of \tx{Section 2.3} with the constraints
derived for vortex formation, we see that, apparently, for these
values of particle parameters, the angular momentum of CDM haloes is
typically of the order of the minimum value required for quantum
vortex formation. Since vortex formation happens then in a large
part of the particle parameter space, this is important to take into
account when BEC-CDM models are fitted to galactic rotation curves
and density profiles, especially in the very centers. While it is
true that the vortex becomes more and more favoured for increasing
self-interaction strengths $g/g_H$, its size, and hence its
dynamical importance, \tx{decreases}. However, at the critical
values for vortex formation and beyond, the size of the vortex in
both halo models is large enough to be expected to be able to
imprint a notable effect on the central halo dynamics, which can be
seen from Tables \ref{tab2} and \ref{tab3}.

The appearance of vortices in the central parts of BEC-CDM haloes,
whose core regions are depleted of dark matter mass, will change the
gravitational coupling to baryons, as compared to a smooth or
nearly-smooth dark matter distribution in the halo. As a result of
depleted dark matter in the vortex cores, the subsequent collapse of
baryonic matter can be delayed and so can star formation. A detailed
analysis is necessary, however, to be able to quantify those
effects.

Our results show furthermore that haloes with particles of high
enough mass to satisfy the minimum condition that $L \geq L_{QM}$
are nevertheless vortex-free, \tx{unless} their self-interaction
strength is high enough. In particular, vortices will not form in
BEC dark matter which has no self-interaction, $g=0$. Axions have
often been modelled without self-interaction, but in doing so they
will not be able to form vortices in galactic haloes, contrary to
what has been proposed in \cite{SY}. In order to have vortices mimic
a net rotational component, a dense lattice of vortices is needed.
Both halo models studied here are in principle able to sustain these
structures. However, not only is it required that $L \gg L_{QM}$,
but so is positive self-interaction. On the other hand, our analysis
does not simply extend to cases when $g < 0$, such that the small
attractive interaction of axions can be taken into account. While we
do still not expect quantum vortices to appear in this case, other
defect structures, like bright solitons, have been shown to being
able to form not only in laboratory BECs with attractive
self-interaction, but also in some models of axion haloes, see
\cite{MP, MP2}.

Our results also indicate that there remain notable regions in
$(m,g)$-parameter space where vortices are \tx{not} favoured. In
these cases, rotating haloes may be modelled as irrotational,
compressible ($n=1$)-polytropic Riemann-S ellipsoids, as we have
shown.

We have considered here the case of BEC-CDM which is a pure, i.e.
zero-temperature, condensate. In the Thomas-Fermi regime, the
characteristic size of virialized objects is in that case the radius
of an ($n=1$)-polytrope, fixed by the ratio $g/m^2$ of particle
parameters. As mentioned in \tx{Section 2.2}, this can be
interpreted either as the full size of a halo or as the size of the
core region of a larger halo. Recently, \cite{slepian} have replaced
the assumption of zero-temperature condensate by that of
thermodynamic equilibrium in isothermal haloes at the virial
temperature. The haloes which result contain an ($n=1$)-polytropic
core of condensate surrounded by an 'atmosphere' of non-condensed
bosons. They claim that such haloes are incompatible with
astronomical constraints. However, as they point out, their
assumption of thermodynamic equilibrium at finite temperature breaks
down if the 2-body elastic scattering collision time exceeds a
Hubble time, depending upon the ratio $\sigma_s/m$ with the
scattering cross section in equ.(\ref{scattcross}). In order to
guarantee local thermodynamic equilibrium, they must assume
$\sigma_s/m \approx 1$ cm$^{2}/$g. In fact, this condition is
\tx{not} met by the ultra-light bosons considered here, for which
$\sigma_s/m$ is much smaller. Using equ.(\ref{scattcross}) and
(\ref{coupst}), along with (\ref{onesphere}) and (\ref{mfidu}), we
can write
 \bdi
  \f{\sigma_s}{m} =
   2.094\cdot
  10^{-102}\left(\f{m}{m_H}\right)^5 \times
   \edi
   \beq
   \times \left(\f{R}{100~\rm{kpc}}\right)^{3/2}
  \left(\f{M}{10^{12}~M_{\odot}}\right)^{-5/2}~\f{\rm{cm}^2}{\rm{g}},
  \eeq
which is valid as long as $m/m_H \gg 1$. Even for large $m/m_H$, and
for that matter, also for the critical values for the dark matter
particle mass for vortex formation calculated above, the value of
$\sigma_s/m$ is much, much smaller than 1 cm$^2/$g.
\\
\\
\\
{\large \tb{Acknowledgements}}\\
\\
We thank Pierre Sikivie, and Steven Weinberg, Eiichiro Komatsu and
other members of the Texas Cosmology Center for stimulating
discussion. We are especially grateful to Stuart Shapiro for his
helpful comments on the effects of compressibility on rotating
figures of equilibrium. This work was supported in part by U.S. NSF
grants AST-0708176 and AST-1009799, NASA grants NNX07AH09G,
NNG04G177G, NNX11AE09G, and Chandra grant SAO TM8-9009X to PRS. TRD
also acknowledges support by the Texas Cosmology Center of the
University of Texas at Austin and by the German DFG through FG960.

\appendix

\section{Approximate equilibrium rotating figures as halo models}

\subsection{General relationships}

In this appendix, we summarize some of the properties of the
homogeneous Maclaurin spheroid and irrotational, $(n=1)$-polytropic
Riemann-S ellipsoid used in this paper. A thorough treatment of
these figures of rotation can be found in the works of
\cite{chandra} and \cite{LRS} (LRS93). We make use of uniformly
rotating figures as models for haloes with non-vanishing angular
momentum. We assume haloes to rotate about their $z$-axis with
angular velocity $\mb{\Omega} = (0,0,\Omega)$. Their mass density is
$\rho_0 = m|f|^2$ with $f$ the corresponding (vortex-free) halo BEC
wavefunction and their semi-axes $(a_1,a_2,a_3)$ are along
$(x,y,z)$. The mean radius of the ellipsoids is defined as $R =
(a_1a_2a_3)^{1/3}$, which differs from the equilibrium radius $R_0$
of the spherical polytrope according to
equ.(\ref{nradius})-(\ref{zerosphere}) (or (\ref{rtilde})) for the
Maclaurin spheroid, and equ.(\ref{polyradius}) \& (\ref{onesphere})
(or (\ref{rtilde2})) for the irrotational Riemann-S ellipsoid,
respectively.

The gravitational potential inside a homogeneous ellipsoidal body is
given by
  \bdi
  \Phi_0(x,y,z) =
   \edi
   \beq \label{riepotential}
   = \pi G \rho_0 (A_1x^2+A_2y^2+A_3z^2 -
  A_1a_1^2-A_2a_2^2-A_3a_3^2)
  \eeq
(see e.g. \cite{chandra}), where the functions $A_1, A_2, A_3$,
depending on the axis ratios $a_2/a_1$, $a_3/a_1$ or eccentricities
 \bdi
  e_1 = \sqrt{1-(a_2/a_1)^2},~
e_2 = \sqrt{1-(a_3/a_1)^2},
 \edi
  respectively, are given by
 \beq \label{aa1}
  A_1 =
  2\f{a_2}{a_1}\f{a_3}{a_1}\f{F(\theta,\phi)-E(\theta,\phi)}{\sin^3 \phi
  \sin^2\theta},
  \eeq
\beq \label{aa2}
  A_2 =
  2\f{a_2}{a_1}\f{a_3}{a_1}\f{E(\theta,\phi)-F(\theta,\phi)\cos^2 \theta - \f{a_3}{a_2}\sin^2 \theta \sin \phi}{\sin^3 \phi
  \sin^2\theta \cos^2 \theta},
  \eeq
\beq \label{aa3}
  A_3 =
  2\f{a_2}{a_1}\f{a_3}{a_1}\f{\f{a_2}{a_3}\sin \phi - E(\theta,\phi)}{\sin^3 \phi
  \cos^2\theta}
  \eeq
 with $\cos \phi = a_3/a_1$, $\sin \theta =
 \sqrt{\f{1-(a_2/a_1)^2}{1-(a_3/a_1)^2}}$ and the standard
 incomplete elliptic integrals
  \bdi
   E(\theta, \phi) = \int_0^{\phi} (1-\sin^2 \theta
   \sin^2\phi')^{1/2}d\phi',
    \edi
    \bdi
    F(\theta, \phi) = \int_0^{\phi} (1-\sin^2 \theta
   \sin^2\phi')^{-1/2}d\phi'.
 \edi
The functions $A_1,A_2,A_3$ are generally defined this way, however,
equ.(\ref{riepotential}) is only valid for homogeneous bodies. While
we use $\Phi_0$ for the homogeneous Maclaurin spheroid explicitly in
the calculations for \tx{Halo-Model A}, we do not need to use the
gravitational potential of the compressible Riemann-S ellipsoid for
\tx{Halo-Model B}, since we make use of the global energy
quantities, like for instance $W$, as provided by LRS93.

\subsection{Homogeneous Maclaurin spheroids}

\begin{figure*}
     \begin{minipage}[b]{0.5\linewidth}
      \centering\includegraphics[width=6cm]{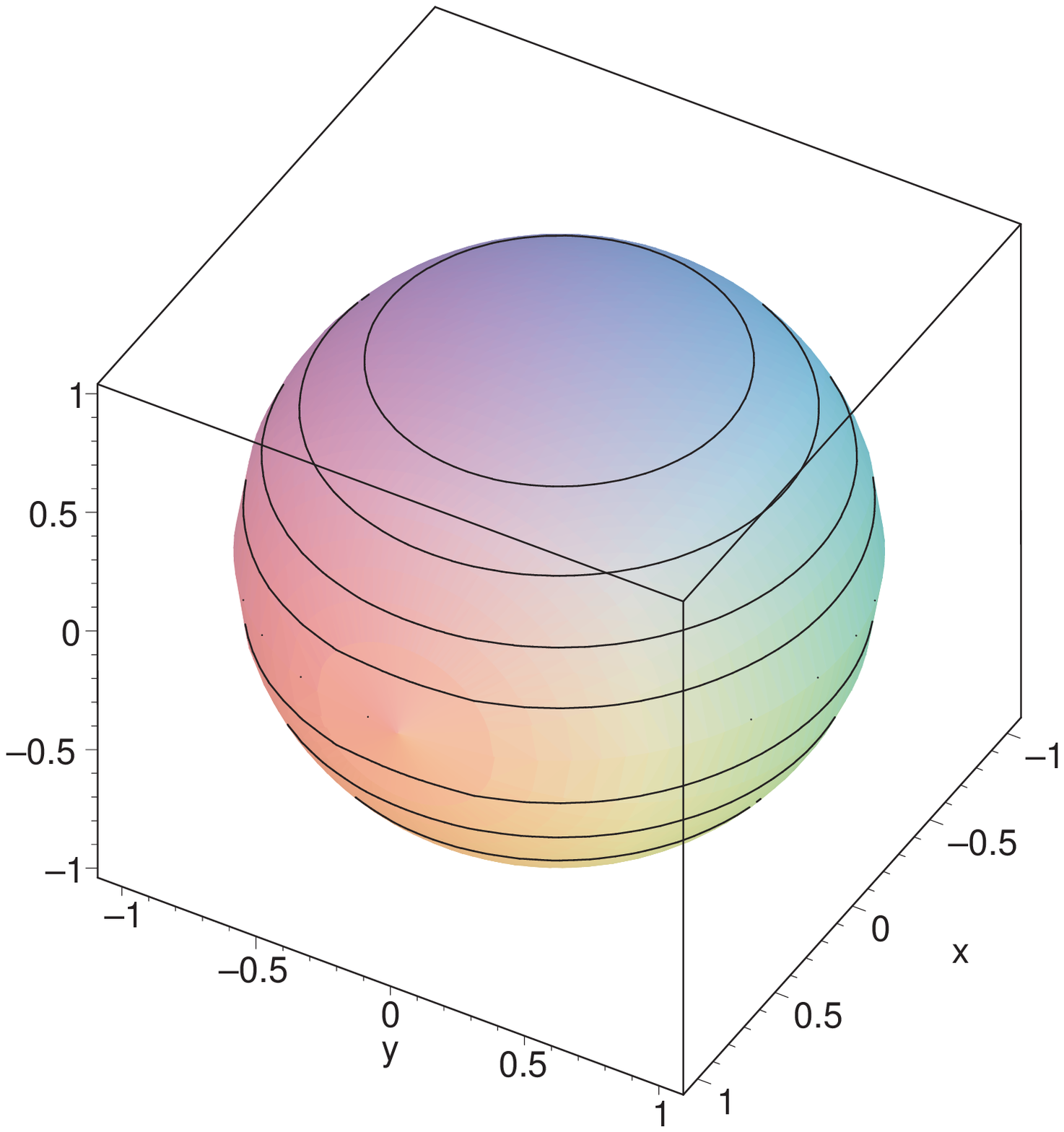}
     \hspace{0.1cm}
    \end{minipage}%
 \begin{minipage}[b]{0.5\linewidth}
      \centering\includegraphics[width=6cm]{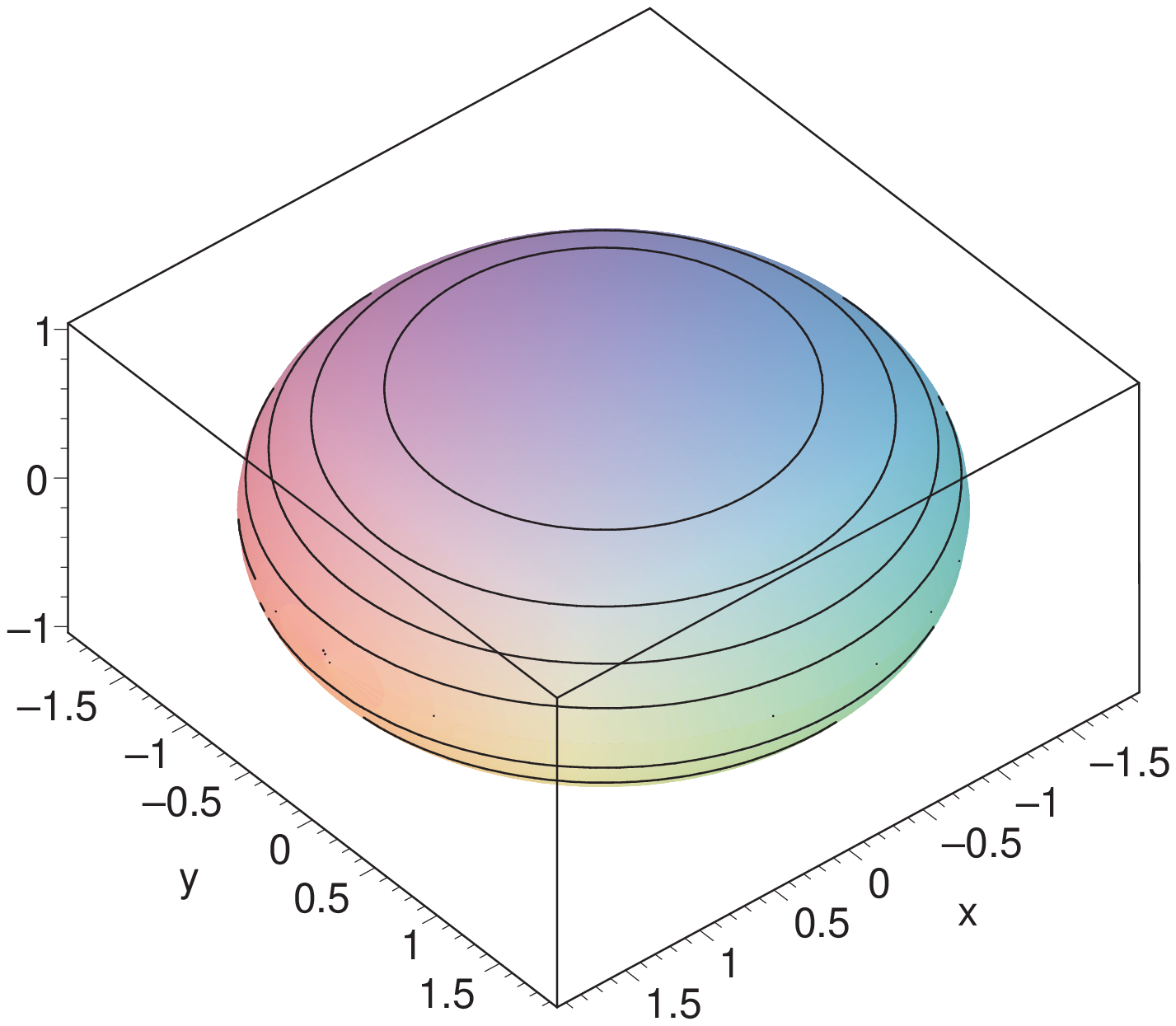}
     \hspace{0.1cm}
    \end{minipage}
 \caption{Maclaurin spheroidal halo rotating about the $z$-axis having $a_3 = 1$ and
$\lambda = 0.05$ (\tx{left-hand-plot}) and $\lambda = 0.2$
(\tx{right-hand-plot}), respectively.}
 \label{fig10}
\end{figure*}

For the Maclaurin spheroid of \tx{Section 3.1}, $a_1 = a_2 \equiv
a$, $a_3 \equiv c$ and (\ref{riepotential}) reduces to
(\ref{macpotential}). Most of the needed relationships can be
already found in the main text. Similarly to LRS93, we can write the
dimensionless forms of $W$ (\ref{macpot}) and $R$ (\ref{nradius}) as
well as $L$ (\ref{macL}) as functions of the eccentricity only:
 \beq \label{wtilde}
  |\tilde W| \equiv \f{5}{3}\f{|W|}{(GM^2/R)} =
  (1-e^2)^{1/6}\f{\arcsin(e)}{e},
   \eeq
    \beq \label{rtilde}
   \tilde R \equiv \f{R}{R_0} =
   \left(\f{2}{3A_3(e)(1-e^2)^{2/3}}\right)^{1/2}
    \eeq
    and
 \beq \label{ltilde}
 \tilde L^2 = \f{L^2}{GM^3R} = \f{3}{25}(1-e^2)^{-2/3}\tilde
 \Omega^2
 \eeq
  with $\tilde \Omega$ in (\ref{macomega}).
On the other hand, we expressed also the $\lambda$-spin parameter of
haloes (\ref{lam}) as a function of the eccentricity only in
\tx{Section 3.1.2}, equ.(\ref{lambdaincomp}). Thus, fixing $\lambda$
determines the geometric and energetic quantities of the spheroidal
halo unambiguously and independently of the BEC particle parameters.
Table \ref{tab4} summarizes them for a range of $\lambda$-values
which encompass those relevant for CDM haloes. Also, Fig.\ref{fig10}
shows two examples of halo shapes for different spin-parameters.

\begin{table*}
\begin{minipage}{13cm}
 \caption{Parameters of the homogeneous Maclaurin spheroid as a
function of $\lambda$} \label{tab4}
\begin{center}
\tiny {\begin{tabular}{l|l|l|l|l|l|l|l}
          & $\lambda$   &  $e$  &  $t$  &  $\tilde \Omega$   &  $\tilde L^2$    &  $|\tilde W|$    &  $\tilde R$        \\
\hline
          & $0.01$     & $.6240952030(-1)$    &  $.5204(-3)$   & $.4558481107(-1)$  &
          $.2500065974(-3)$  & $.9999996616$  &
           $1.000520746$ \\
          & $0.02$     & $.1242809380$     & $.207472(-2)$    &  $.9086159255(-1)$  & $.1001034014(-2)$
           & $.9999946162$  &
          $1.002083877$ \\
   & $0.03$   & $.1850994726$    & $.464440(-2)$   & $.1355078928$  & $.2255299174(-2)$  & $.9999730069$  &
   $1.004690543$ \\
   & $0.04$  & $.2443945077$   & $.819899(-2)$   & $.1792395408$  & $.4016798968(-2)$  & $.9999158013$  &
   $1.008343664$ \\
   & $0.05$   & $.3017559569$   & $.1269872(-1)$  & $.2217968643$  & $.6291209890(-2)$ & $.9997977879$  &
   $1.013048263$ \\
   & $0.06$   & $.3568449744$   & $.1809622(-1)$  & $.2629514670$   & $.9085911668(-2)$ & $.9995887660$ &
   $1.018812234$ \\
 & $0.07$   & $.4093987579$    & $.24338998(-1)$
  & $.3025107370$  & $.1241015884(-1)$  & $.9992548016$  &
  $1.025647448$  \\
 & $0.08$   & $.4592299927$   & $.31371503(-1)$  & $.3403173945$  & $.1627516169(-1)$
 & $.9987594192$  & $1.033570749$ \\
 & $0.09$   & $.5062220297$   & $.39137176(-1)$
 & $.3762479967$  & $.2069422508(-1)$  & $.9980646348$ &
 $1.042605066$  \\
  & $0.10$    & $.5503211259$   & $.47580078(-1)$  & $.4102099657$
  & $.2568290981(-1)$  & $.9971317748$  &  $1.052780577$ \\
  & $0.20$   & $.8470361891$    & $.1590675677$ & $.6351541756$    &
  $.1124337842$  & $.9662629271$  &  $1.231978660$ \\
  &  $0.30$   & $.9658444276$  & $.3004590826$  & $.6495893878$   &  $.3065151508$   &
  $.8638313134$  & $1.703156470$  \\
 \hline
\end{tabular} }
\end{center}
\end{minipage}
\end{table*}

\subsection{Compressible, irrotational Riemann-S ellipsoids}

The Riemann-S ellipsoid rotates rigidly with angular velocity
$\mb{\Omega} = (0,0,\Omega)$ like the Maclaurin spheroid. However,
on this background rigid rotation, an internal velocity field is
superposed specified by two requirements: (i) it shall have a
uniform vorticity parallel to $\mb{\Omega}$; (ii) it shall leave the
ellipsoidal figure unchanged, that is, the velocity vector at any
point in the fluid shall be tangent to the isodensity surface
passing through that point.
 The fluid circulation along the equator is given by
  \beq \label{riecirc}
   \oint_{\mc{C}} \mb{v}\cdot d\mb{l} = \pi (2+f_R)a_1a_2\Omega
    \eeq
  with $f_R \equiv \zeta'/\Omega$. The flow is irrotational in the rest frame only for $f_R = -2$.

LRS93 have extended the analysis of \cite{chandra} from
incompressible bodies to compressible ones by exploiting the
so-called \tx{ellipsoidal approximation} which assumes the
following: i) the surfaces of constant density are self-similar
ellipsoids (i.e. the axis ratios $a_2/a_1$ and $a_3/a_1$ are the
same for all interior isodensity surfaces), ii) the density profile
$\rho^E(m)$, with $m$ denoting here the mass interior to an
isodensity surface, is assumed to be identical to that of a
spherical polytrope of same $K_p$ and $n$, but with radius $R =
(a_1a_2a_3)^{1/3}$. Both assumptions are strictly valid only in the
incompressible limit ($n=0$), but in the general case, where $n
\not= 0$, it provides an approximation to the true equilibrium
solution.

Specifying the spin-parameter $\lambda$, we can determine the axis
ratios (or eccentricities, respectively) using (\ref{lambdarie}) and
equation (5.16) in LRS93, which is
 \bdi
 \f{4(a_2/a_1)^2}{(1+(a_2/a_1)^2)^2} -
  \edi
  \beq \label{axisratios}
 - \f{4B_{12}(a_2/a_1)^2}{\left[(a_3/a_1)^2A_3 -
 (a_2/a_1)^2\f{A_1-A_2}{(a_2/a_1)^2-1}\right]}\f{1}{1+(a_2/a_1)}+1
 = 0.
  \eeq
   The constants $\kappa_n$ and $q_n$ in the energy terms appearing in \tx{Section 3.2},
   depend on the polytropic index $n$ via
      \beq \label{kap}
    \kappa_n \equiv \f{5}{3}\f{\int_0^{\chi_1}\theta^n\chi^4
    d\chi}{\chi_1^4 |\theta_1'|} \left\{\begin{array}{ll}
 = 1 & \textrm{for $n = 0$}\\
 = \f{5}{3}\left(1-\f{6}{\pi^2}\right) \approx 0.653 & \textrm{for $n = 1$}
 \end{array} \right.
  \eeq
   and
  \beq \label{qup}
   q_n \equiv \kappa_n \left(1-\f{n}{5}\right) \left\{\begin{array}{ll}
 = 1 & \textrm{for $n = 0$}\\
 = \f{4}{3}\left(1-\f{6}{\pi^2}\right) \approx 0.523 & \textrm{for $n = 1$}.
 \end{array} \right.
  \eeq
 $\theta = \rho^S/\rho_c^S$ and $\chi = r/R_{char}$, with $R_{char} = \sqrt{K_p/(2\pi G)}$, are the dimensionless Lane-Emden variables for the density and radius, respectively, of a
 polytrope. We denote the first zero of the density profile as $\chi_1$ (i.e. $\chi_1 =
 R_0/R_{char}$) and $\theta_1 = \theta (\chi_1)$.

\begin{figure*}
     \begin{minipage}[b]{0.5\linewidth}
      \centering\includegraphics[width=6cm]{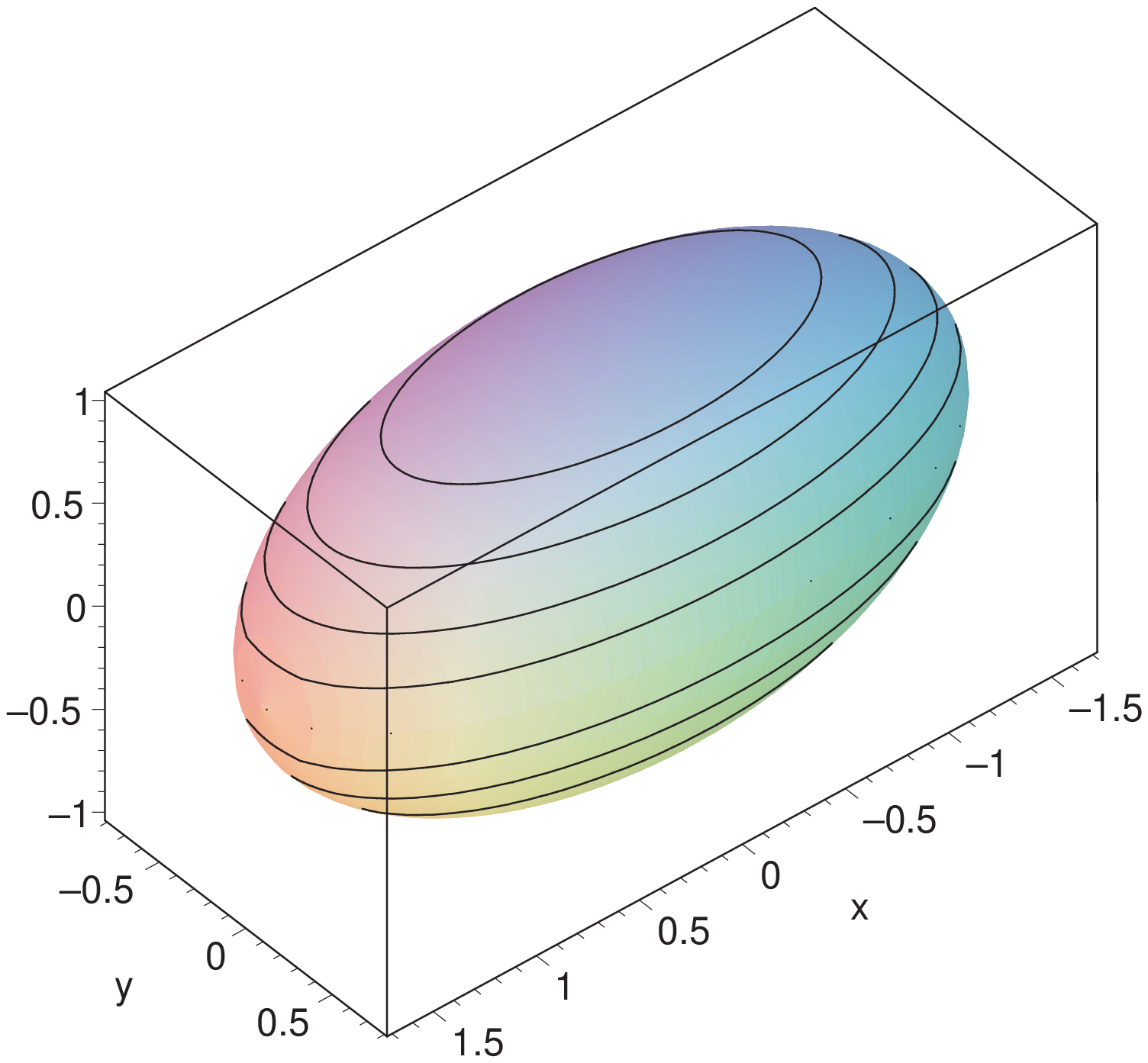}
     \hspace{0.1cm}
    \end{minipage}%
 \begin{minipage}[b]{0.5\linewidth}
      \centering\includegraphics[width=6cm]{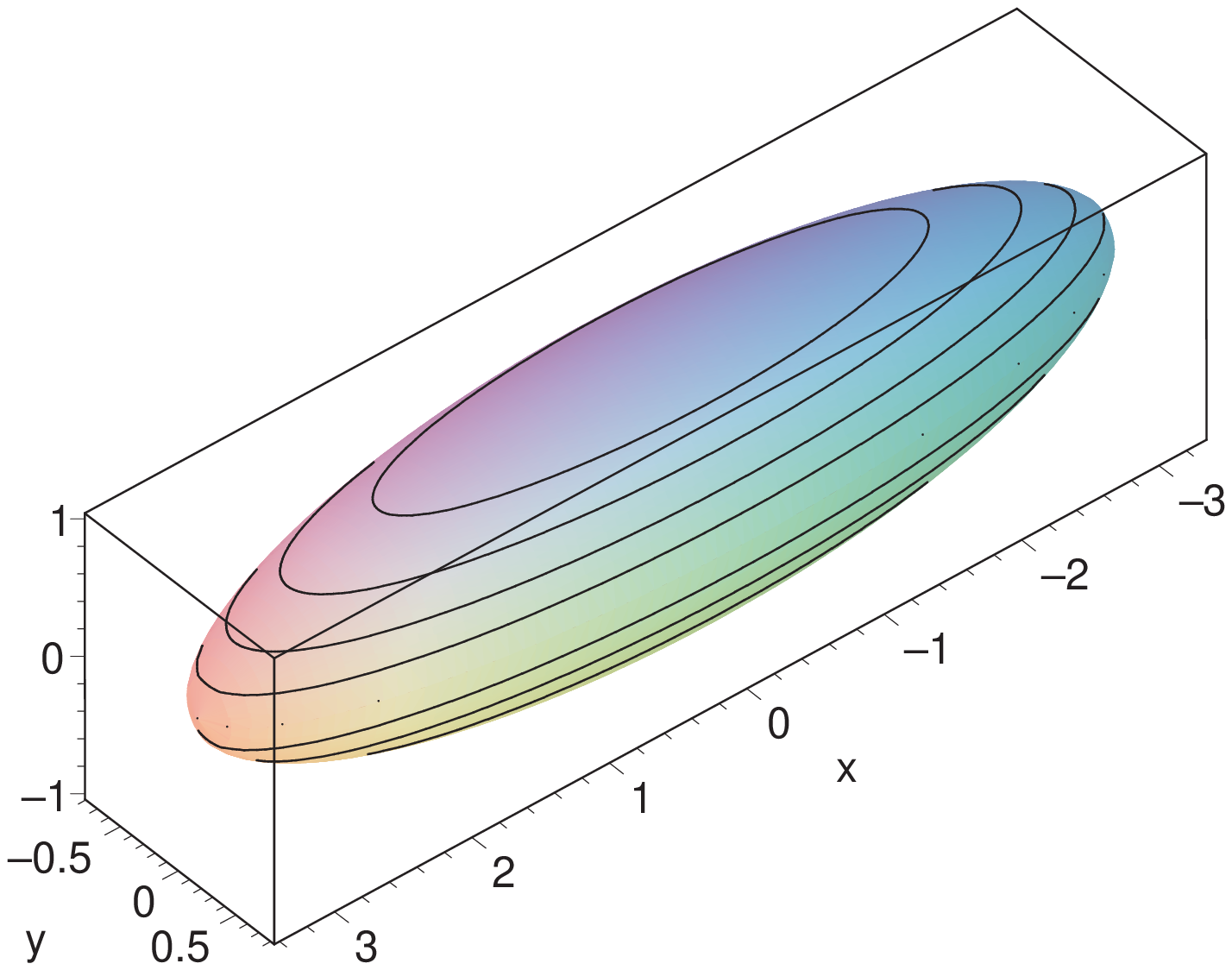}
     \hspace{0.1cm}
    \end{minipage}
 \caption{Irrotational Riemann-S ellipsoidal halo rotating about the $z$-axis having $a_3 = 1$ and
$\lambda = 0.05$ (\tx{left-hand-plot}) and $\lambda = 0.2$
(\tx{right-hand-plot}), respectively.}
 \label{fig11}
\end{figure*}

Now, specializing to $n=1$, the dimensionless versions of $W, R, L$
in (\ref{riepoten}), (\ref{polyradius}), (\ref{angrie}) can again be
 given as functions of the eccentricities only:
 \beq \label{wtilde2}
 |\tilde W| \equiv \f{4}{3}\f{|W|}{(GM^2/R)} = f(e_1,e_2)
  \eeq
  with $f(e_1,e_2)$ in (\ref{aux}),
   \beq \label{rtilde2}
  \tilde R \equiv \f{R}{R_0} = [f(e_1,e_2)(1-2t)]^{-1/2}
   \eeq
   (see (\ref{polyradius})), and
  \beq \label{ltilde2}
   \tilde L^2 \equiv \f{L^2}{GM^3R} =
   \f{3}{4}\left(\f{\kappa_1}{5}\right)^2\gamma^2(e_1,e_2)\tilde
   \Omega^2
    \eeq
   where
    \bdi
     \gamma(e_1,e_2) \equiv
      \edi
     \bdi
     \left[(1-e_1^2)^{-1/3}(1-e_2^2)^{-1/3} +
     (1-e_1^2)^{2/3}(1-e_2^2)^{-1/3} - \right.
      \edi
      \beq \label{auxZ}
     \left. - 4\f{(1-e_1^2)^{1/3}(1-e_2^2)^{-2/3}}{(1-e_1^2)^{-1/3}(1-e_2^2)^{-1/3}
     + (1-e_1^2)^{2/3}(1-e_2^2)^{-1/3}}\right]
     \eeq
      and $\tilde \Omega$ in (\ref{rieomega}). The angular velocity
      of the internal motions $\Lambda$ is related to $\tilde \Omega$ via
      (\ref{velo3}), defining $\tilde \Lambda \equiv
      \Lambda/\Omega_G$.
The $\lambda$-spin parameter determines the above quantities
unambiguously via equ.(\ref{lambdarie}). Table \ref{tab5} summarizes
them for a range of $\lambda$-values, encompassing those relevant
for CDM haloes. Note that $\tilde \Lambda$ is a monotonically
decreasing function of $\lambda$, while $\tilde \Omega$ is \tx{not}
monotonic, in contrast to the case of the Maclaurin spheroid. Two
illustrative examples of halo shapes can be found in
Fig.\ref{fig11}.

\subsubsection{Density profile for $n=1$}

LRS93 confirm their results by numerical comparison of the density
profiles of rotating versus non-rotating configurations for
different polytropic indices $n$. However, we observe that the
ellipsoidal approximation in conjunction with the known analytic
density profile of the equilibrium ($n=1$)-sphere,
equ.(\ref{tfprofile}), makes possible the derivation of an analytic
profile for a rotating, ellipsoidal ($n=1$)-polytrope, as follows.

According to the ellipsoidal approximation, $\rho^S(m)/\bar \rho^S$
and $\rho^E(m)/\bar \rho^E$ have the same dependence on $m/M$. From
this follows that the shapes of the cumulative mass profiles are the
same. The ellipsoid has a larger volume, however, than the
equilibrium sphere according to (\ref{polyradius}). Since we
consider the case where the total mass $M$ shall be the same, this
means that the mean densities (and hence central densities) are
different according to (\ref{ratden}).

Now, the equation of an ellipsoidal isodensity surface can be
written as
 \beq \label{qober}
 q^2 = x^2 + \f{y^2}{1-e_1^2} + \f{z^2}{1-e_2^2},
  \eeq
 whose dimensionless form, after dividing by $R_{char}$,
 will be denoted carrying a 'tilde'-sign as
 \beq \label{qoberdimless}
  \tilde q \equiv q/R_{char}.
   \eeq
The outer surface of the ellipsoid is then given by
 \beq \label{outerober}
  \tilde q_{max}^2 = \tilde x^2 + \f{\tilde y^2}{1-e_1^2} + \f{\tilde
  z^2}{1-e_2^2} = \f{a_1^2}{R_{char}^2}.
\eeq
 Writing the semi-major axis in terms of mean radius and
 eccentricities equ.(\ref{meanradius}), and using (\ref{polyradius}), we see that
  \beq \label{qmax}
   \tilde q_{max} = \f{\pi
   g(e_1,e_2)^{-1/2}}{(1-e_1^2)^{1/6}(1-e_2^2)^{1/6}}.
   \eeq
Since the isodensity surfaces are self-similar ellipsoids, it
follows more generally that
 \beq
 \tilde q = \f{\chi
 g(e_1,e_2)^{-1/2}}{(1-e_1^2)^{1/6}(1-e_2^2)^{1/6}},
   \eeq
 with $\chi$ from above.
  Because of (\ref{polyradius}), the volumes differ according to
   $V^E = V^S g(e_1,e_2)^{-3/2}$
   and correspondingly we have for the volume elements that
    \bdi
     dV^E = dV^S g(e_1,e_2)^{-3/2}
      \edi
     \beq
     = 4\pi R_{char}^3 (1-e_1^2)^{1/2}(1-e_2^2)^{1/2} \tilde q^2 d\tilde q.
   \eeq
      Using now the fact that the shapes of the cumulative mass
   profiles are the same, we can derive the following density profile of the
   ($n=1$)-polytropic ellipsoids,
    \beq \label{ellipdensity}
     \rho^E (\tilde q) = \rho_c^E \f{\sin \left[\tilde q
     (1-e_1^2)^{1/6}(1-e_2^2)^{1/6}g(e_1,e_2)^{1/2}\right]}{\tilde q
     (1-e_1^2)^{1/6}(1-e_2^2)^{1/6}g(e_1,e_2)^{1/2}}
     \eeq
with $\tilde q$ from above and with $\rho_c^E$, as usual, denoting
the central density of the ellipsoid. Using this density profile,
the total mass can be calculated to be
 \beq
  M^E = 4\pi^2 R_{char}^3\rho_c^E g(e_1,e_2)^{-3/2} = 4\pi^2
  R_{char}^3\rho_c^S = M^S,
  \eeq
  i.e. the mass of the ($n=1$)-polytropic ellipsoid is the same as that for the
  equilibrium ($n=1$)-sphere, as we have demanded.

\begin{table*}
\begin{minipage}{13cm}
  \caption{Parameters of the irrotational, $(n=1)$-polytropic Riemann-S ellipsoid as a
function of $\lambda$} \label{tab5}
\begin{center}
\tiny {\begin{tabular}{l|l|l|l|l|l|l}
          & $\lambda$   &  $a_2/a_1$  &  $a_3/a_1$  &  $e_1$   & $e_2$   &
          $t$ \\
 \hline
    & $0.01$   &  $.7073606352$    & $.8191876671$  &  $.7068528360$   & $.5735255584$  & $.8990308186(-3)$  \\
     & $0.02$  & $.6157806530$     & $.7444171156$  &
     $.7879176273$  & $.6677148778$   & $.3256446859(-2)$  \\
     & $0.03$  & $.5551042656$ & $.6884452492$ & $.8317807730$ &
     $.7252883143$  & $.6676951963(-2)$ \\
     & $0.04$ & $.5095561821$ & $.6425906919$  & $.8604373872$ & $.7662096337$ &
     $.1087731987(-1)$ \\
    &  $0.05$  & $.4731628114$ & $.6033952054$ & $.8809750019$ &
    $.7974423027$
 & $.1565115011(-1)$ \\
  & $0.06$ & $.4429370293$ & $.5690393124$ & $.8965527247$ & $.8223103191$ &
  $.2084571079(-1)$ \\
   & $0.07$ & $.4171522426$ & $.5384160866$ & $.9088366226$ & $.8426791309$ &
   $.2634641496(-1)$ \\
   & $0.08$ & $.3947154535$ & $.5107881879$ & $.9188034125$  & $.8597065936$ &
   $.3206621692(-1)$ \\
  & $0.09$ & $.3748897504$ & $.4856329988$ & $.9270694014$  & $.8741627940$ &
  $.3793824738(-1)$ \\
  & $0.10$  & $.3571546372$ & $.4625629986$  & $.9340452693$ & $.8865864156$ &
  $.4391062656(-1)$ \\
  & $0.20$ & $.2417900586$ & $.3025300680$ & $.9703285874$ & $.9531398418$ &
  $.1032663262$ \\
  & $0.30$ & $.1766421365$ & $.2101855698$ & $.9842751422$ & $.9776615090$ &
  $.1550787050$ \\
 \hline
\end{tabular} }
\end{center}
\begin{center}
\tiny {\begin{tabular}{l|l|l|l|l|l|l}
          & $\lambda$   &  $\tilde \Omega$  &  $\tilde \Lambda$  &  $\tilde L^2$   & $|\tilde W|$   &
          $\tilde R$ \\
 \hline
   & $0.01$  & $.7169963591$ & $.6760714962$ & $.8424269299(-6)$ & $.9919850657$ & $1.004935593$ \\
   & $0.02$ & $.7176346622$ & $.6408208912$ & $.3253671535(-5)$ & $.9842263532$ &
   $1.011279929$ \\
   & $0.03$ & $.7165548594$ & $.6081343468$ & $.7088561436(-5)$ & $.9766699520$ & $1.018697841$ \\
   & $0.04$ & $.7140826379$ & $.5777254698$ & $.1223320264(-4)$ & $.9692743288$ &
   $1.026958151$ \\
   & $0.05$ & $.7104728779$ & $.5493488048$ & $.1859786170(-4)$ & $.9620070989$ & $1.035897363$ \\
    & $0.06$ & $.7059271906$ & $.5227939592$ & $.2611162661(-4)$ & $.9548427521$ &
    $1.045397324$ \\
   & $0.07$ & $.7006067868$ & $.4978802570$ & $.3471838378(-4)$ & $.9477609961$ & $1.055371038$ \\
   & $0.08$ & $.6946418722$ & $.4744520038$ & $.4437376265(-4)$ & $.9407455674$ &
   $1.065753362$ \\
   & $0.09$ & $.6881386566$ & $.4523744952$ & $.5504279856(-4)$ & $.9337833787$ & $1.076494765$ \\
    & $0.10$ & $.6811846403$ & $.4315306946$ & $.6669814241(-4)$ & $.9268637895$ &
    $1.087557099$ \\
   & $0.20$ & $.5984642204$ & $.2734205668$ & $.2348312504(-3)$ & $.8587224921$ & $1.211459922$ \\
    & $0.30$ & $.5100096167$ & $.1747264833$ & $.49746267360(-3)$ & $.7911442386$ &
    $1.353621249$ \\
 \hline
\end{tabular} }
\end{center}
\end{minipage}
\end{table*}

\section{Splitting of the Gross-Pitaevskii energy functional}

In this appendix, equ. (\ref{splitting}) to (\ref{rterm}) are
derived. All energies and wave functions, $\psi, f, w$, as well as
the respective amplitudes and phases, $S, S_0, S_w$, are those in
the rotating frame of reference, so primes are omitted in this
section altogether. The integrations are over the whole space.

We insert (\ref{wsp}) into the energy functional (\ref{energie2})
considering each term separately: the kinetic energy term has the
following form,
 \bdi
\int \f{\hbar^2}{2m}|\nabla \psi|^2 = \int \f{\hbar^2}{2m}\left\{f^2
|\nabla w|^2 + |w|^2\left[f^2(\nabla S_0)^2 + (\nabla f)^2\right]
\right. \edi
 \beq \label{appenergy}
 \left. + \f{1}{2}\nabla(f^2)\cdot \nabla(|w|^2) + f^2 \nabla S_0 \cdot
(iw\nabla w^* - iw^* \nabla w)\right\}.
 \eeq
The rotation term is given by
 \bdi
 -i\hbar \int \psi^* \mb{\Omega} \cdot (\nabla \psi \times \mb{r}) =
  \edi
   \beq \label{approt}
 = i\hbar \int f^2 w^*\nabla w \cdot (\mb{\Omega} \times \mb{r}) -
 \hbar \int f^2 |w|^2 \nabla S_0 \cdot (\mb{\Omega} \times \mb{r}),
 \eeq
  while the potential energies due to gravitation and self-interaction
  are simply
   \beq \label{apppot}
    \int \left(\f{m}{2}\Phi|\psi|^2 + \f{g}{2}|\psi|^4\right) =
  \int \left(\f{m}{2}\Phi f^2 |w|^2 + \f{g}{2} f^4 |w|^4\right).
 \eeq
  It is advantageous to write the energy such that the vortex-free
  contribution is clearly separated from the other terms.
   Using the above expressions, we may thus recast the functional into the following form
   \bdi
   \gpf[\psi] = \gpf[fe^{iS_0}] + \int
   \f{\hbar^2}{2m}(|w|^2-1) \times
   \edi
    \bdi
     \times \left[(\nabla f)^2 + f^2(\nabla S_0)^2 -
   \f{2m}{\hbar}f^2\nabla S_0 \cdot (\mb{\Omega}\times
   \mb{r})\right] +
    \edi
     \bdi
      + \int \f{m}{2} f^2 \left[\Phi |w|^2 -
    \Phi_0\right] +
   \int \f{\hbar^2}{4m}\nabla(f^2)\cdot \nabla(|w|^2)
    \edi
     \bdi
      + \int
   \f{\hbar^2}{2m}f^2|\nabla w|^2 + \f{g}{2}\int f^4(|w|^4-1)
   +
     \edi
     \bdi
      + \int \f{\hbar^2}{2m} \left[f^2\nabla S_0 \cdot (iw\nabla
   w^* - iw^* \nabla w) + \right.
   \edi
   \beq \label{appenergy2}
   \left. i\f{2m}{\hbar}f^2w^*\nabla w \cdot
   (\mb{\Omega}\times \mb{r})\right]
   \eeq
    with $\mc{E}[fe^{iS_0}]$ being the vortex-free energy
    of equ.(\ref{vofree}).
  Evaluating the second term in line 3 results in
   \bdi
   \int \nabla(f^2)\cdot \nabla(|w|^2) = \int
    \nabla(f^2)\cdot \nabla(|w|^2-1) =
     \edi
    \beq \label{spec3}
     -\int
    (|w|^2-1)\Delta(f^2) = -2 \int (|w|^2-1) (f\Delta f + (\nabla
    f)^2),
     \eeq
    which, by using (\ref{fsa}), becomes
   \bdi
   \int \f{\hbar^2}{4m}\nabla(f^2)\cdot \nabla(|w|^2) = \int
   \f{\hbar^2}{2m}(|w|^2-1) \times
    \edi
     \bdi
     \times \left[\f{2m}{\hbar}f^2\nabla S_0 \cdot
   (\mb{\Omega}\times \mb{r}) - f^2(\nabla S_0)^2 - \right.
    \edi
    \beq \label{spec4}
    \left. - \f{2m}{\hbar^2}f^2 (m\Phi_0 + gf^2 - \nu) - (\nabla
   f)^2\right],
    \eeq
    assuming that the halo mass density either goes to zero as $|\mb{r}| \to \infty$
   or is identically zero beyond some finite radius.
   Furthermore, we take advantage of rewriting the following terms:
\beq \label{spec5}
   \f{g}{2}\int f^4(|w|^4-1) \equiv g \int f^4(|w|^2-1)
   + \f{g}{2}\int f^4 (1-|w|^2)^2
  \eeq
   and
    \bdi
 \int \left[\f{\hbar^2}{2m} f^2 \nabla S_0 \cdot (iw\nabla w^* - iw^*\nabla w) + i\hbar f^2 w^* \nabla w \cdot
 (\mb{\Omega}\times
 \mb{r})\right] =
  \edi
   \beq \label{spec6}
   = -\f{\hbar^2}{m} \int if^2w^* \nabla w \cdot \left(\nabla S_0 - \f{m}{\hbar}\mb{\Omega}
 \times \mb{r}\right),
  \eeq
  respectively,
   using the conservation of particle number, $\int (|w|^2-1)f^2\mu = 0$.
  Collecting all of the above expressions, we shall finally arrive at the following splitting of the energy
  functional:
 \beq \label{appsplitting}
\gpf[\psi] = \gpf[fe^{iS_0}] + \mc{G}_f[w] - \mc{R}_f[w],
 \eeq
 where
  \bdi
 \mc{G}_f[w] \equiv \int \left(\frac{\hbar^2}{2m}f^2|\nabla w|^2 +
\frac{g}{2}f^4(1-|w|^2)^2\right) +
 \edi
  \beq \label{appG}
  + \int \left(\f{m}{2}f^2\Phi_0 +
\f{m}{2}f^2|w|^2\left[\Phi - 2\Phi_0\right]\right)
 \eeq
 and
  \beq \label{appR}
  \mc{R}_f[w] \equiv \f{\hbar^2}{m}\int if^2 w^* \nabla w \cdot \left(\nabla S_0 -
\f{m}{\hbar}\mb{\Omega}\times \mb{r}\right).
 \eeq

\section{The chemical potential of BEC haloes}

We have dealt in this paper mostly with the energy of a given BEC
halo. The equation of motion for stationary systems in
equ.(\ref{stat}) also involve the chemical potential $\mu$, which is
fixed by the conservation of particles. For a static ground state
without vortex, (\ref{stat}) easily leads to the following
   expression for the equilibrium condensate chemical potential
 \beq \label{chempot}
 \mu = -\f{\hbar^2}{2m}\f{\Delta \sqrt{n}}{\sqrt{n}} + m
 \Phi + g n
  \eeq
   with corresponding time-independent particle number density $n = |\psi|^2$.
  Multiplying (\ref{variation}) with $\p \psi^*/\p N$ results
 in the well-known thermodynamic relationship, $\mu = \p E/\p N$, where $E$ is the total energy of
 the BEC halo wave function under
consideration. In the TF regime, for instance, (\ref{chempot})
reduces to
  $\mu = g \rho/m + m\Phi$.
 Multiplying by $\rho$
 and integrating results in $\mu = 2mE/M$,
 where $E$ is the energy in the Thomas-Fermi regime, i.e.
 (\ref{sumenerg}) with $K_Q=0$ (and $T=0$ if $\mb{v}=0$). The chemical potential for ($n=1$)-polytropic BEC haloes is
 thus given by
  \beq \label{mupoly}
   \mu = 2\f{E}{N}.
    \eeq
We can derive this relationship also by using the energy expressions
of LRS93. The total energy of the polytropic sphere with index $n$
according to LRS93 is
 \beq \label{polyenergy}
  E = U + W = k_1K(\rho_c^S)^{1/n}M - k_2(\rho_c^S)^{1/3}GM^{5/3},
   \eeq
 where the constants $k_1, k_2$ depend on $n$,
 $\chi_1$ and $\theta_1'$.
 Using the formula for the central density of the sphere (see also
 (\ref{kreuz}))
  \beq \label{ratden2}
   \rho^S_c = \f{1}{3}\f{\chi_1}{|\theta_1'|}\bar \rho^S = \f{\chi_1}{4\pi|\theta_1'|}\f{M}{R^3} \equiv
   g_1\f{M}{R^3},
    \eeq
 we rewrite
  \beq \label{polyenergy2}
   E = k_1Kg_1^{1/n}\f{(Nm)^{1/n+1}}{R^{3/n}} -
   k_2g_1^{1/3}\f{G}{R}(Nm)^2.
    \eeq
The chemical potential of this sphere is thus
 \beq \label{polychem}
  \mu = \f{\p E}{\p N} = \left(\f{1}{n}+1\right)\f{U}{N} + 2\f{W}{N}
  = \f{1}{N}\left(E + \f{U}{n} + W\right),
   \eeq
 which reduces to $\mu = 2 E/N$ for $n=1$, in accordance with (\ref{mupoly}).

\label{lastpage}

\end{document}